%% file: main.tex
\def\isarxiv{}

\documentclass[a4paper,11pt]{article}
\usepackage[margin=1.05in]{geometry}
\usepackage{setspace}
\usepackage{graphicx}%
\usepackage{multirow}%
\usepackage{amsmath,amssymb,amsfonts}%
\usepackage{amsthm}%
\usepackage{mathrsfs}%
\usepackage[title]{appendix}%
\usepackage[dvipsnames]{xcolor}%
\usepackage{textcomp}%
\usepackage{manyfoot}%
\usepackage{booktabs}%
\usepackage{algorithm}%
\usepackage{algorithmicx}%
\usepackage{algpseudocode}%
\usepackage{listings}%
\usepackage{xspace}%
\usepackage{subcaption}
\usepackage{tikz}
\usepackage{lineno}
\usepackage{cmtt}

\usepackage{abstract}

\usepackage[colorlinks=true,linkcolor=blue,citecolor=blue,urlcolor=blue]{hyperref}

\input{src/preamble}

\theoremstyle{thmstyleone}%
\theoremstyle{thmstyletwo}%

\theoremstyle{thmstylethree}%

\input{src/macro}

\input{figures/figure_preamble}

\usetikzlibrary{external}

\begin{document}

\title{Robust and Generalizable Background Subtraction on Images of Calorimeter Jets using Unsupervised Generative Learning}

\author{
Yeonju Go$^{1,}$\thanks{These authors contributed equally to this work. \\ \hspace*{1.5em} Corresponding authors: \texttt{ygo@bnl.gov}, \texttt{dtorbunov@bnl.gov}.}\,\ ,\ 
Dmitrii Torbunov$^{2,\,*}$,\,
Yi Huang$^{2}$,\,
Shuhang Li$^{3}$,\\\,
Timothy Rinn$^{1}$,\,
Haiwang Yu$^{1}$,\,
Brett Viren$^{1}$,\,
Meifeng Lin$^{2}$,\\\,
Yihui Ren$^{2}$,\,
Dennis Perepelitsa$^{4}$,\,
Jin Huang$^{1}$%
}

\date{
\small
$^{1}$Physics Department, Brookhaven National Laboratory, Upton, NY 11973, USA\\
$^{2}$Computational Science Initiative, Brookhaven National Laboratory, Upton, NY 11973, USA\\
$^{3}$Physics Department, Columbia University, New York, NY 10027, USA\\
$^{4}$Physics Department, University of Colorado Boulder, Boulder, CO 80309, USA
}

\maketitle

\begin{abstract}
Accurate separation of signal from background is one of the main challenges for precision measurements across high-energy and nuclear physics.
Conventional supervised learning methods are insufficient here because the required paired signal and background examples are impossible to acquire in real experiments. 
Here, we introduce an unsupervised unpaired image-to-image translation neural network that learns to separate the signal and background from the input experimental data using cycle-consistency principles. We demonstrate the efficacy of this approach using images composed of simulated calorimeter data from the sPHENIX experiment, where physics signals (jets) are immersed in the extremely dense and fluctuating heavy-ion collision environment. Our method outperforms conventional subtraction algorithms in fidelity and overcomes the limitations of supervised methods. Furthermore, we evaluated the model’s robustness in an out-of-distribution test scenario designed to emulate modified jets as in real experimental data. The model, trained on a simpler dataset, maintained its high fidelity on a more realistic, highly modified jet signal. This work represents the first use of unsupervised unpaired generative models for full detector jet background subtraction and offers a path for novel applications in real experimental data, enabling high-precision analyses across a wide range of imaging-based experiments.
\par\vspace{0.5em}
\noindent\textbf{Keywords:} Background subtraction; Signal extraction; Generative AI; High-energy nuclear physics

\end{abstract}

\sloppy

\input{src/intro}
\input{src/result}
\input{src/discussion}
\input{src/methods}

\section*{Acknowledgment}
We thank the sPHENIX collaboration for their generous support to this work by providing access to high fidelity full detector simulation dataset. We also thank for valuable feedback and discussion with the sPHENIX collaboration on this work. This work is supported by the LDRD Program at Brookhaven National Laboratory and US Department of Energy under Contract DE-SC0012704. Shuhang Li was partially supported by DOE-SC through the Office of Nuclear Physics under Award No.~DE‐FG02‐86ER40281. Dennis Perepelitsa acknowledges support from the DOE Office of Science grant DE-FG02-03ER41244.

\clearpage
\bibliographystyle{unsrt}
\bibliography{main}

\clearpage
\begin{appendices}
\input{src/appendix}

\end{appendices}

\end{document}

%% file: src/preamble.tex
\usepackage{xspace}

%% file: src/macro.tex
\newcommand{\ourmodel}{UVCGAN-S\xspace}

\newcommand{\deta}{\ensuremath{\Delta\eta}\xspace}
\newcommand{\dphi}{\ensuremath{\Delta\phi}\xspace}

\newcommand{\pt}{\ensuremath{p_\mathrm{T}}\xspace}
\newcommand{\ptratio}{\ensuremath{p_\mathrm{T}^\mathrm{sub}/p_\mathrm{T}^\mathrm{real}}\xspace}
\newcommand{\ptreal}{\ensuremath{p_\mathrm{T}^\mathrm{real}}\xspace}

\newcommand{\hijing}{\textsc{Hijing}\xspace}
\newcommand{\pythia}{\textsc{Pythia}\xspace}
\newcommand{\fastjet}{\textsc{Fastjet}\xspace}
\newcommand{\geant}{\textsc{Geant4}\xspace}
\newcommand{\jewel}{\textsc{Jewel}\xspace}
\newcommand{\pythiahijing}{\textsc{Pythia+Hijing}\xspace}
\newcommand{\jewelhijing}{\textsc{Jewel+Hijing}\xspace}

\newcommand{\zg}{\ensuremath{z_{g}}\xspace}
\newcommand{\zleading}{\ensuremath{z_\mathrm{_{leading}}}\xspace}
\newcommand{\rg}{\ensuremath{r_\mathrm{g}}\xspace}

\newcommand{\auau}{\mbox{$Au$+$Au$}\xspace}

\newcommand{\thename}{UVCGAN-S\xspace}

%% file: figures/figure_preamble.tex
\usepackage{tikz}
\usepackage{pgfplots}
\usepackage{ifthen}
\usetikzlibrary{
    fit,
    math,
    calc,
    arrows,
    shapes,
    shadows,
    fadings,
    patterns,
    external,
    plotmarks,
    arrows.meta,
    backgrounds,
    positioning,
    shadows.blur,
    intersections,
    pgfplots.groupplots,
    pgfplots.statistics,
    pgfplots.fillbetween,
    decorations.pathreplacing,
}
\pgfplotsset{compat=1.18}

\definecolor{b1}{HTML}{099CF0} 
\definecolor{b2}{HTML}{0FA5C7} 
\definecolor{or}{HTML}{F8A300} 

\definecolor{yg}{HTML}{A1CB2E} 
\definecolor{pg}{HTML}{5AAA4F} 

\definecolor{pr}{HTML}{7000AD} 
\definecolor{gr}{HTML}{EAEAF2} 

\pgfdeclarelayer{back}
\pgfsetlayers{back,main}

\makeatletter
\pgfkeys{%
  /tikz/on layer/.code={
    \def\tikz@path@do@at@end{\endpgfonlayer\endgroup\tikz@path@do@at@end}%
    \pgfonlayer{#1}\begingroup%
  }%
}
\makeatother

\makeatletter
\pgfplotsset{
    boxplot/hide outliers/.code={
        \def\pgfplotsplothandlerboxplot@outlier{}%
    }
}
\makeatother

%% file: src/intro.tex
\section{Introduction}\label{sec:intro}
High-energy heavy-ion collisions at the Relativistic Heavy Ion Collider (RHIC) and the Large Hadron Collider (LHC) recreate the extreme conditions that existed in the universe a microsecond after the Big Bang. In this regime ordinary hadrons dissolve into a deconfined medium of quarks and gluons known as the quark-gluon plasma (QGP). Studying this state tests Quantum Chromodynamics at high temperature and density and clarifies how quarks and gluons interact in a deconfined medium~\cite{Busza:2018rrf}. Among the most incisive probes of the QGP are jets, collimated sprays of hadrons created from hard-scattered quarks and gluons. As jets traverse the dense QGP medium, they interact and lose energy and their internal structure is modified, a phenomenon known as \textit{jet quenching}~\cite{Bjorken:1982tu,Qin:2015srf,Connors:2017ptx}.

\begin{figure}[ht!]
    \centering
    \ifdefined\isarxiv
        \includegraphics[width=0.80\textwidth]{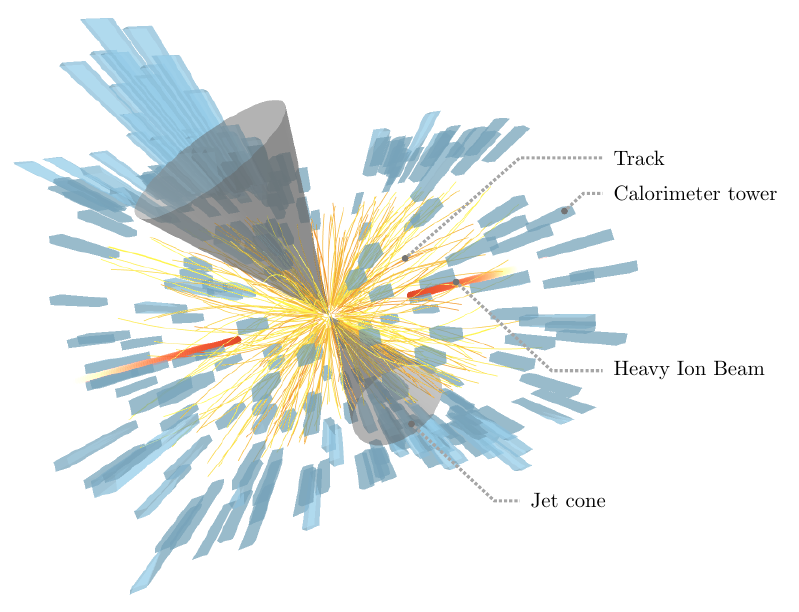}
    \else
        \tikzsetnextfilename{jet_bkg_diagram}
        \resizebox{.90\textwidth}{!}{
            \input{figures/jet_bkg_diagram/jet_bkg_diagram}
        }
    \fi
    \caption{
    Schematic of jet production and detection in heavy-ion collisions. Red lines show heavy-ion beams. Yellow lines are charged particle tracks. Blue boxes are calorimeters measuring energy. Light-gray cones show the jet signals, which are difficult to identify due to the large underlying event activity (background) in these collisions.
    }
    \label{fig:jet_bkg_diagram}
\end{figure}

However, the precision measurements of jets are challenging because the jet signal is superposed on a large underlying event generated by the medium itself as depicted in \autoref{fig:jet_bkg_diagram}. This underlying event adds soft particles across the detector and varies from event to event as well as within an event across rapidity and azimuth. Traditional background subtraction strategies, such as area-based (``Area Method'')~\cite{Soyez:2009cw} and iterative method~\cite{Hanks:2012wv} with constituent subtraction~\cite{Berta:2014eza} (``Iterative Constituent Subtraction, ICS Method''), estimate an average event-level background density to remove contamination. These approaches struggle when fluctuations are large and spatial correlations are strong such as in central heavy-ion collisions when the two nuclei collide head-on. This restricts precision jet measurements, particularly for large jet radii and low transverse momentum (\pt).

Machine learning (ML) has emerged as a promising avenue to address these limitations. Early ML studies~\cite{ALICE:2023waz, Mengel:2023mnw} trained supervised models to predict a corrected scalar such as the true particle-level jet \pt from reconstructed jet, jet constituents, and event information. These models are typically trained from simulation as a regression task by mapping the reconstructed values to the known truth values. These efforts improved access to jets at lower \pt and larger jet radius relative to the conventional methods (i.e., Area Method). Yet, such supervised models carry two major caveats: First, supervised predictors can inherit biases when the training domain is not realistic or does not match the domain of application. For example, a model trained on unquenched jets can bias final physics observables when applied to quenched jets because the internal jet structure presented to the model differs between the training and application domains~\cite{Stewart:2024mkx}. Training on quenched simulations can mitigate this effect~\cite{Li:2024fzn}, but performance then depends heavily on how faithfully the generator reproduces the real experimental data. Second, scalar regressions compress rich image-level information into a single number or a finite set of numbers, which limits downstream physics analyses that rely on jet substructure and correlations.

Image-based learning provides a more natural representation for detector response data. Recent work~\cite{Qureshi:2025ylv} has applied transformer-based models to translate images of truth particle energies towards background-removed jet images. Since the output image preserves full-dimensional information at the event level, analysts can construct any desired observable from it. However, these studies remain supervised and require simulation-generated labeled pairs, which are inherently physical-model dependent and never perfectly realistic. While some supervised methods have demonstrated robustness across different Monte Carlo (MC) event generators, the transferability to real experimental data — where conditions may differ from any available simulation — remains an open question. \textit{Unsupervised unpaired} image-to-image translation offers a different path. Cycle-consistent generative adversarial networks (CycleGAN)~\cite{8237506} learn mappings between two domains and enforce that translating from one domain to the other and back returns the original image. Modern variants such as UVCGANv2~\cite{torbunov2023} replace the original convolutional generators with hybrid architectures that combine Vision Transformer encoders~\cite{Dosovitskiy:2020qjv} with Style-Modulated U-Net decoders, enabling the model to capture both global context and fine-grained local structure in the translated images. Together with improved optimization strategies including self-supervised pre-training of the encoders, these advances yield significantly better training stability and image fidelity compared to the original CycleGAN, making the framework suitable for complex scientific applications.

Here, we present an unsupervised, unpaired image-to-image translation framework for background subtraction that operates directly on full detector images. In our formulation, one domain contains images of pure jet signals and pure background separately, and the other domain contains their sum as it appears in a heavy-ion collision event. The network learns both to separate and to recombine. We demonstrate this approach on simulated calorimeter images from the sPHENIX~\cite{Belmont:2023fau} experiment at RHIC. Without labeled targets, the model recovers jet kinematics and internal structure with higher accuracy than conventional methods (i.e., Area Method and ICS Method), and it improves over scalar supervised regressions by preserving full-dimensional information.
The unsupervised framing opens a route to data-driven training without paired labels and reduces reliance on specific generators for target outputs, which makes the approach portable across experiments, collision systems, and analysis tasks. 

%% file: figures/jet_bkg_diagram/jet_bkg_diagram.tex




    \begin{tikzpicture}
        \tikzset{
            shadowed path/.style={
                postaction={
                    copy shadow={shadow xshift=0.2em, 
                                 shadow yshift=-0.2em, 
                                 opacity=0.2, 
                                 fill opacity=0}
                }
            },
            flow/.style={
                -{Latex[length=4mm, width=4mm]}, 
                rounded corners=10pt, 
                line width=4 * \lw, 
                draw=black!55, 
                shadowed path,
                shorten >= 2pt,
                shorten <= 2pt
            },
            marker/.style={
                circle, 
                fill=black!55, 
                draw=black!55,
                inner sep=1pt,
            },
            legend_line/.style={
                black!35, 
                densely dotted, 
                line width=1.5pt,
            }
        }
        
        \node[inner sep=5] {
            \begin{tikzpicture}
                \node[inner sep=0] (detector){\includegraphics[scale=.45,trim=0 0 0 0, clip]{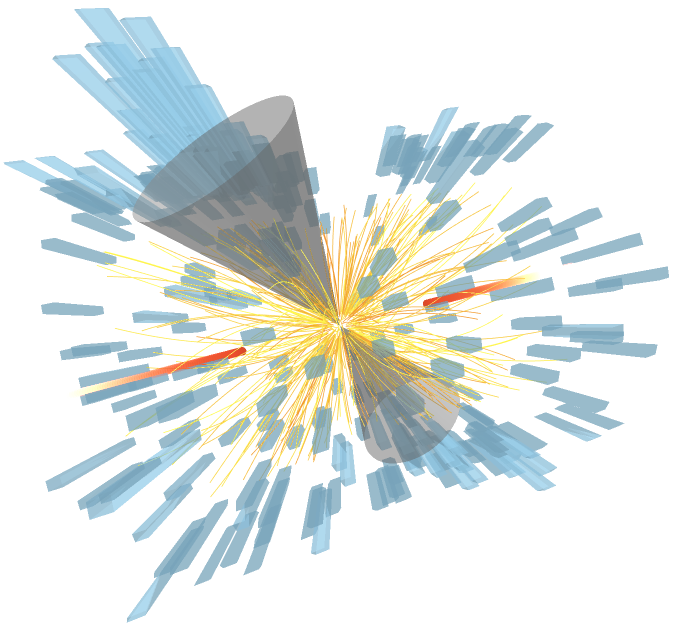}};
                
                \tikzmath{coordinate \C; \C=(detector.north east)-(detector.south west);}
                

                \coordinate (small_cone_x) at ($(detector.west)!.63!(detector.east)$);
                \coordinate (small_cone_y) at ($(detector.north)!.70!(detector.south)$);
                \node[marker] (small_cone) at (small_cone_x |- small_cone_y) {};


                \coordinate (right_beam_x) at ($(detector.west)!.70!(detector.east)$);
                \coordinate (right_beam_y) at ($(detector.north)!.46!(detector.south)$);
                \node[marker] (right_beam) at (right_beam_x |- right_beam_y) {};

                \coordinate (tower_x) at ($(detector.west)!.87!(detector.east)$);
                \coordinate (tower_y) at ($(detector.north)!.34!(detector.south)$);
                \node[marker] (tower) at (tower_x |- tower_y) {};
                
                \coordinate (track_x) at ($(detector.west)!.62!(detector.east)$);
                \coordinate (track_y) at ($(detector.north)!.42!(detector.south)$);
                \node[marker] (track) at (track_x |- track_y) {};

                \coordinate (legend_x) at ($(detector.west)!0.8!(detector.east)$);
                \coordinate (legend_x_tower) at ($(detector.west)!0.93!(detector.east)$);


                \coordinate (small_cone_turn) at ([xshift=.13 * \Cx, yshift=-.13 * \Cy]small_cone);
                \draw[legend_line] (small_cone) -- (small_cone_turn) -- (legend_x |- small_cone_turn);
                \node[anchor=west] at (small_cone_turn -| legend_x) {Jet cone};


                \coordinate (right_beam_turn) at ([xshift=.15 * \Cx, yshift=-.15 * \Cy]right_beam);
                \draw[legend_line] (right_beam) -- (right_beam_turn) -- (legend_x_tower |- right_beam_turn);
                \node[anchor=west] at (right_beam_turn -| legend_x_tower) {Heavy Ion Beam};


                \coordinate (tower_turn) at ([xshift=.03 * \Cx, yshift=.03 * \Cy]tower);
                \draw[legend_line] (tower) -- (tower_turn) -- (legend_x_tower |- tower_turn);
                \node[anchor=west] at (tower_turn -| legend_x_tower) {Calorimeter tower};

                \coordinate (track_turn) at ([xshift=.18 * \Cx, yshift=.17 * \Cy]track);
                \draw[legend_line] (track) -- (track_turn) -- (legend_x_tower |- track_turn);
                \node[anchor=west] at (track_turn -| legend_x_tower) {Track};
                
            \end{tikzpicture}
        };
    \end{tikzpicture}

%% file: src/result.tex

\section{Results}\label{sec:result}
The performance of our unsupervised unpaired image-to-image translation framework, \ourmodel (where ``S'' denotes ``Stratified,'' reflecting the decomposition of the translation domain into separate signal and background strata, as detailed in Section~\ref{sec:stratifiedCycleGAN}), was evaluated and compared to widely-used conventional background subtraction methods (i.e., Area Method, ICS Method). All background subtraction methods, including the Area and ICS Methods, are applied at the reconstructed calorimeter energy level, ensuring a consistent and fair comparison. For the ICS method, we use two iterations with a maximum particle-background matching distance of $\Delta R_{\mathrm{max}} = 0.3$ and a momentum-weighting parameter of $\alpha = 1$. We quantify gains in momentum response, position response, and substructure observables. 

\subsection{Training samples and conditions}

The training and validation data were generated using MC event generators followed by a full detector simulation. 
High-\pt particles coming from hard-scattering processes were generated to produce jets (signal) via \pythia8~\cite{Sjostrand:2014zea}, and soft particles of minimum-bias (i.e., events recorded with minimal trigger selection, designed to represent the inclusive distribution of heavy-ion collisions without favoring any particular process) \auau collision events (background) were generated using \hijing v1.383~\cite{Wang:1991hta} at a center-of-mass energy per nucleon pair $\sqrt{s_\mathrm{_{NN}}}=200~\text{GeV}$. To test the subtraction under the most challenging conditions, we selected the most central $0-10\%$ \hijing\ events, where the underlying event (UE) is largest. For a realistic description of the underlying event, collective flow effects are included as an afterburner to \hijing, introducing azimuthal anisotropies that modulate the background. \hijing also includes event-by-event fluctuations in particle multiplicity and energy deposition. No explicit minimum \pt cut is applied at the particle level.

The particles generated by \pythia and \hijing were propagated through a \geant~\cite{GEANT4:2002zbu} simulation with the full sPHENIX detector geometry, including the 1.4 T solenoidal magnetic field, to model a realistic detector response. The sPHENIX detector includes a tracking system, an electromagnetic calorimeter, and inner and outer hadronic calorimeters. In this study, jets are reconstructed using calorimeter information only, without incorporating tracking data, as our image-based method operates directly on calorimeter tower energy maps. For each event, the deposited energy across all calorimeter layers was projected onto a common grid in pseudorapidity ($\eta$) and azimuthal angle ($\phi$) in cylindrical coordinates with the beam along the $z$ axis. The energy was aggregated per cell on a grid with a fixed granularity of $\Delta\eta=\Delta\phi=0.1$ (referred to as a tower), yielding a two-dimensional \textit{calorimeter tower energy map} that serves as the network's input image. The image consists of $64\times24$ pixels, corresponding to calorimeter towers in $\phi$ and $\eta$, respectively.

\begin{figure}[ht!]
    \centering
    \ifdefined\isarxiv
        \includegraphics[width=0.99\textwidth]{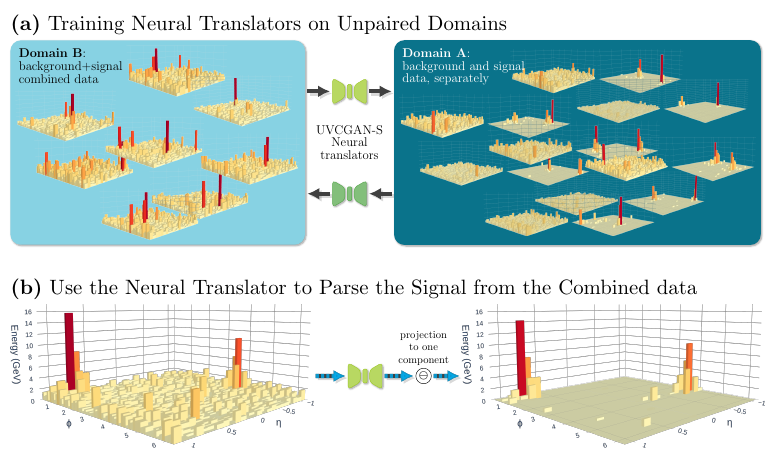}
    \else
        \tikzsetnextfilename{cyclegan}
        \input{figures/cyclegan/cyclegan}
    \fi
    \caption{Schematic diagram of the \ourmodel framework (a) training and (b) signal extraction. \ourmodel translates a jet plus background calorimeter image (Domain $B$) into its two separated components (Domain $A$): the isolated jet signal and the underlying background. 
    See \autoref{fig:stratified_arch} for more detail about the \ourmodel's training and inference procedure.
    }
    \label{fig:cyclegan}
\end{figure}

We constructed three datasets to train \ourmodel: (1) \pythia pure jet images, (2) \hijing pure background images, and (3) \pythiahijing formed by embedding \pythia jets into \hijing events by summing the calorimeter tower energies. Within the \ourmodel framework, domain $A$ contains the separate jet and background images, and domain $B$ contains the combined images as depicted in \autoref{fig:cyclegan}. The model was trained in an unsupervised manner using $1,000,000$ unpaired images drawn from these two domains. For validation and testing, we reserved $200,000$ Domain $B$ images with known ground truth components. Performance is strictly evaluated by comparing the model's outputs for jet and background images to the corresponding ground truth images used to construct the combined input. \autoref{fig:cyclegan} illustrates a representative translation from domain $B$ to domain $A$, where a single combined image is separated into jet signal and UE background components, whose sum reproduces the input within tolerance.

Jets were reconstructed from the calorimeter energies using the anti-$k_T$ algorithm~\cite{Cacciari:2008gp} via the \fastjet package~\cite{Cacciari:2011ma}. The jet radius ($R = \sqrt{(\Delta\eta)^2 + (\Delta\phi)^2}$), defining the area over which energies are aggregated to form jets, was varied from $R=0.2$ to $R=0.5$ within the detector acceptance of $|\eta| < 0.6$. 
In \pythia, the hard-scattering transverse momentum transfer was set to $\hat{p}_\mathrm{T} > 30$~GeV, ensuring that the leading jet in each event has a truth-particle level \pt above 30~GeV. Subleading jets with lower truth-level \pt are also present and included in this study. After accounting for detector response effects, the calorimeter-level reconstructed jet \pt is typically 20--30\% lower than the truth-particle level value. The performance of the subtraction for various jet observables is evaluated for all reconstructed (detector-level) jets, including both leading and subleading jets, in the \pt range 14 to 50~GeV. The background-subtracted jet observables, denoted by ``sub'', are compared with the corresponding detector-level signal jet observables, denoted by ``real''. Here, ``real'' refers to jets reconstructed from the simulated signal component alone after detector simulation, rather than to truth-particle level jets before detector simulation.

\subsection{Position Response}
\begin{figure}[ht!]
    \centering
    \includegraphics[width=0.45\textwidth]{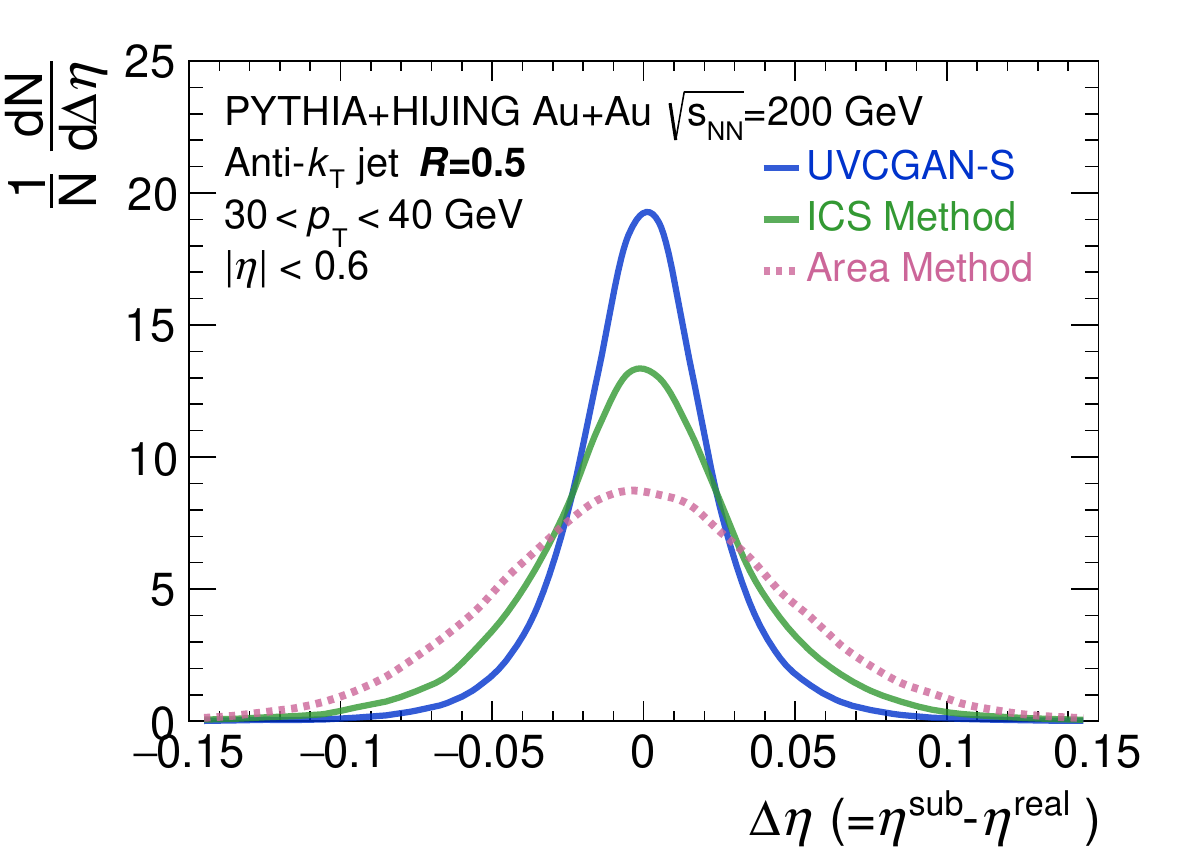}%
    \includegraphics[width=0.45\textwidth]{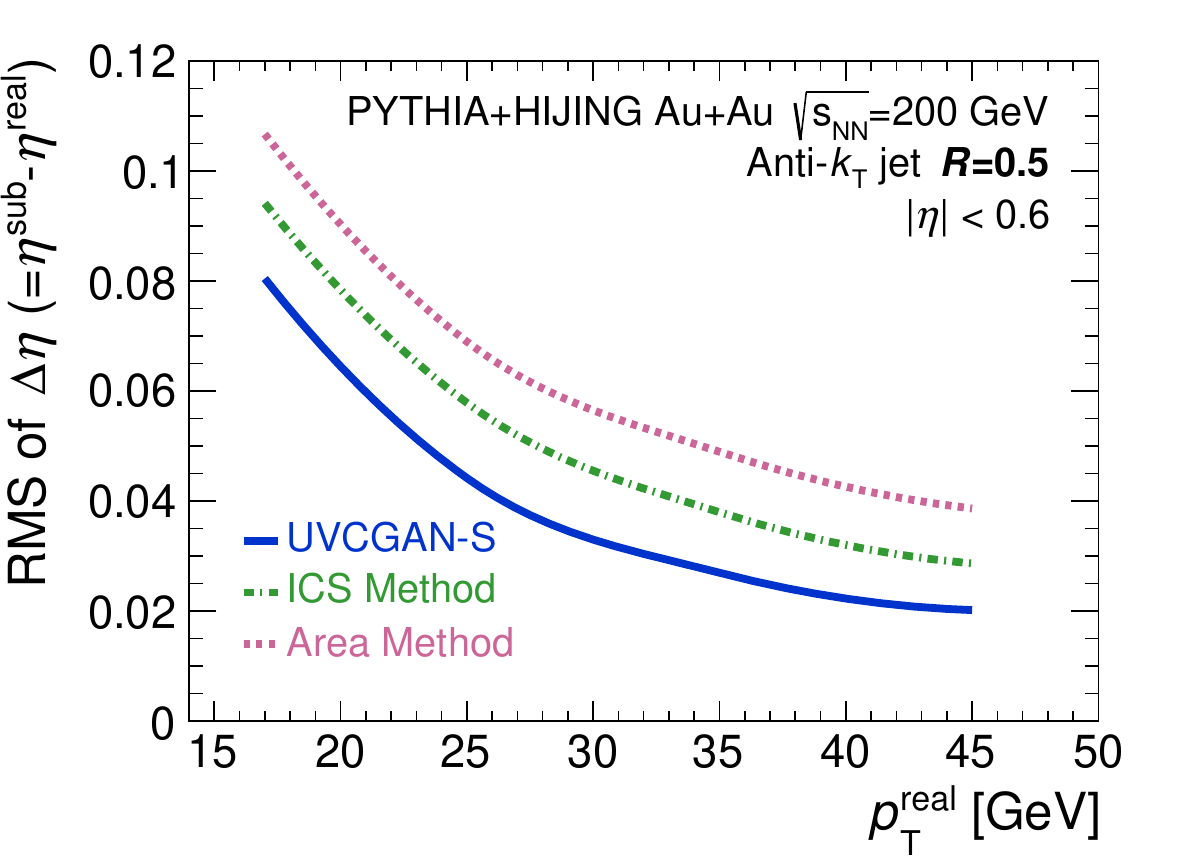}
    \caption{
        Jet $\eta$ position difference ($\Delta\eta = \eta^{\mathrm{sub}} - \eta^{\mathrm{real}}$) distribution (Left) for $R=0.5$ at $30<\ptreal<40$~GeV. The right panel shows the root mean square (RMS) of $\Delta\eta$ (a measure of position resolution) as a function of the ground truth jet \ptreal for $R=0.5$, demonstrating superior performance of the \ourmodel.}
    \label{fig:detaDist}
\end{figure}

The accuracy of the jet angular position response is evaluated by comparing the background-subtracted jet axis to the ground truth jet axis. The difference of $\Delta\eta=\eta^{\rm sub}-\eta^{\rm real}$ is evaluated in bins of ground truth \ptreal\ and jet radius $R$. \autoref{fig:detaDist} shows representative \deta distributions for $R=0.5$ and the root mean square (RMS) of \deta as a function of \ptreal. For a larger jet radius ($R=0.5$), where background contamination is more significant, the \ourmodel model significantly outperforms the conventional methods. While all methods accurately recover the mean position ($\langle \Delta\eta \rangle \approx 0$), our model achieves up to $\sim40\%$ better position resolution (smaller RMS of \deta). The results for the azimuthal angle $\phi$ are consistent with the $\eta$ findings -- please find the details in Supplementary Note~\ref{app:posres}.

\subsection{Momentum Response}
\begin{figure}[ht!]
    \centering
    \includegraphics[width=0.45\textwidth]{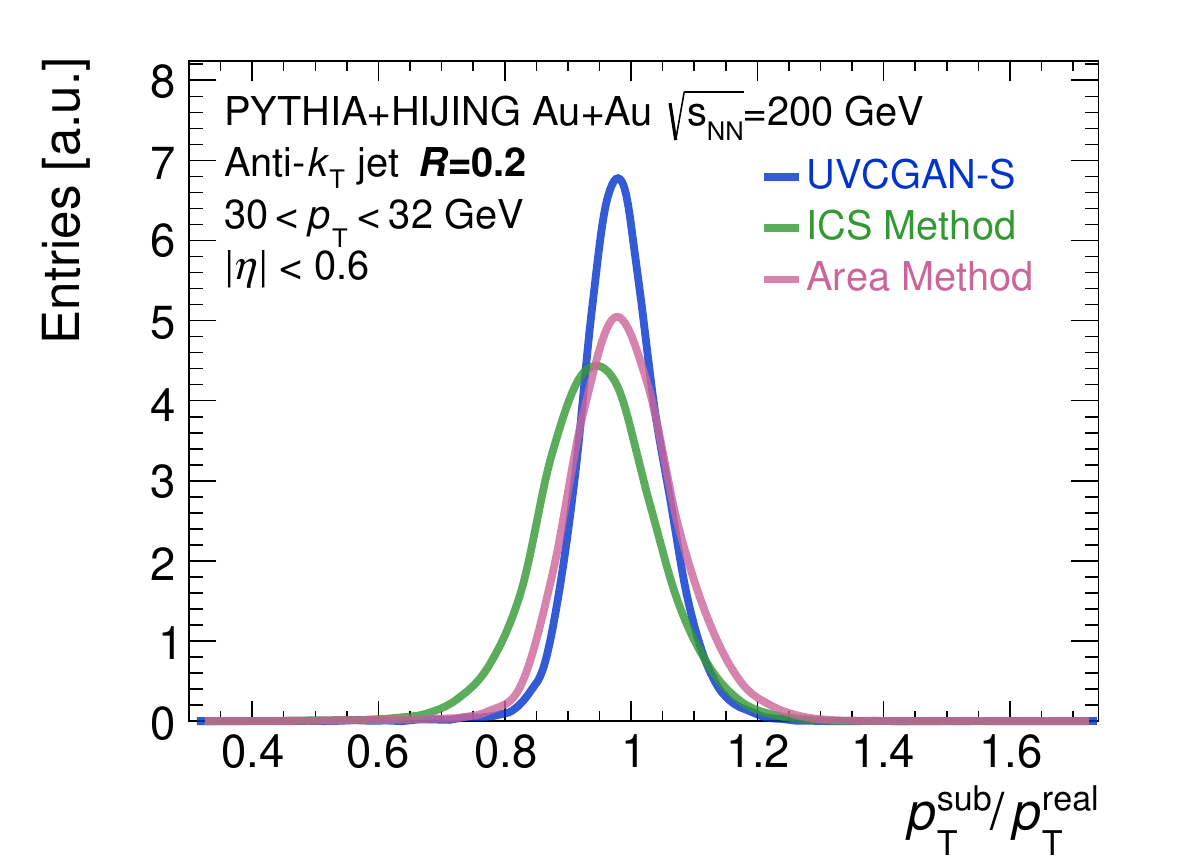}
    \includegraphics[width=0.45\textwidth]{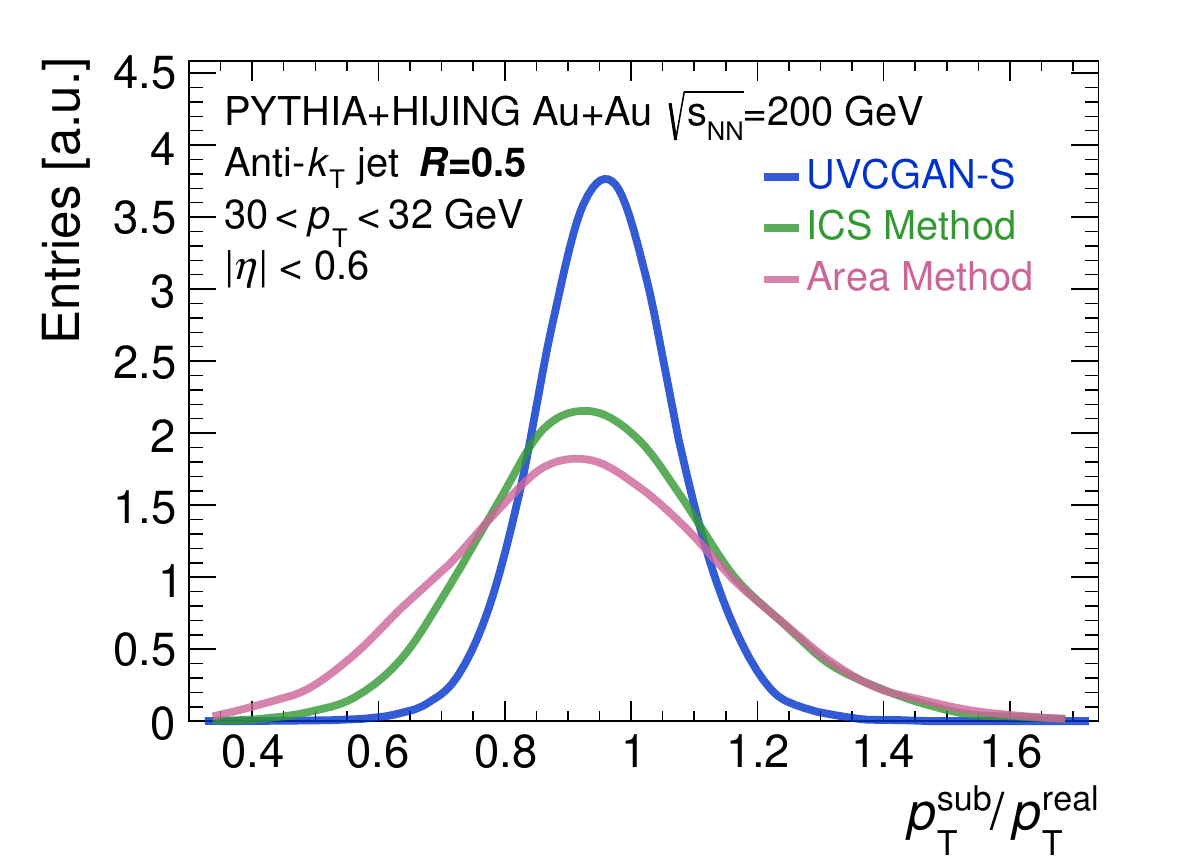}
    \includegraphics[width=0.45\textwidth]{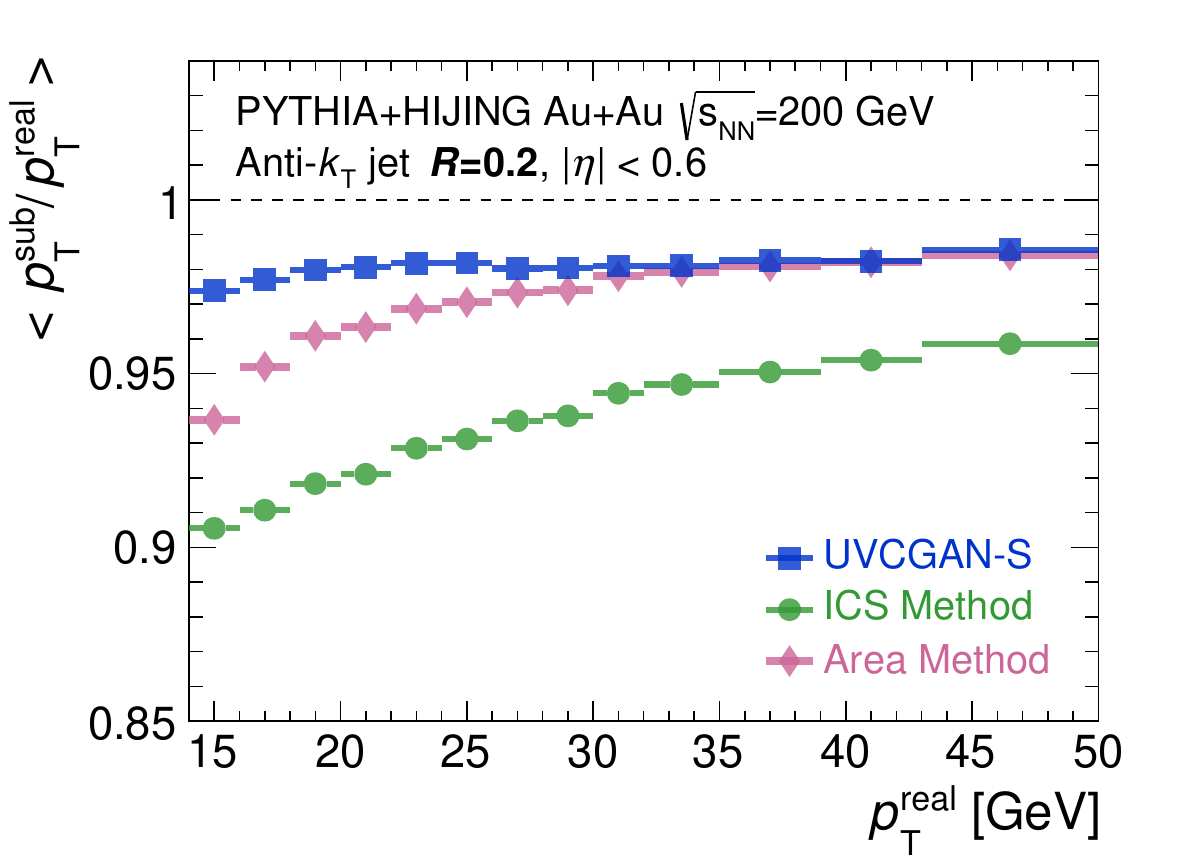}
    \includegraphics[width=0.45\textwidth]{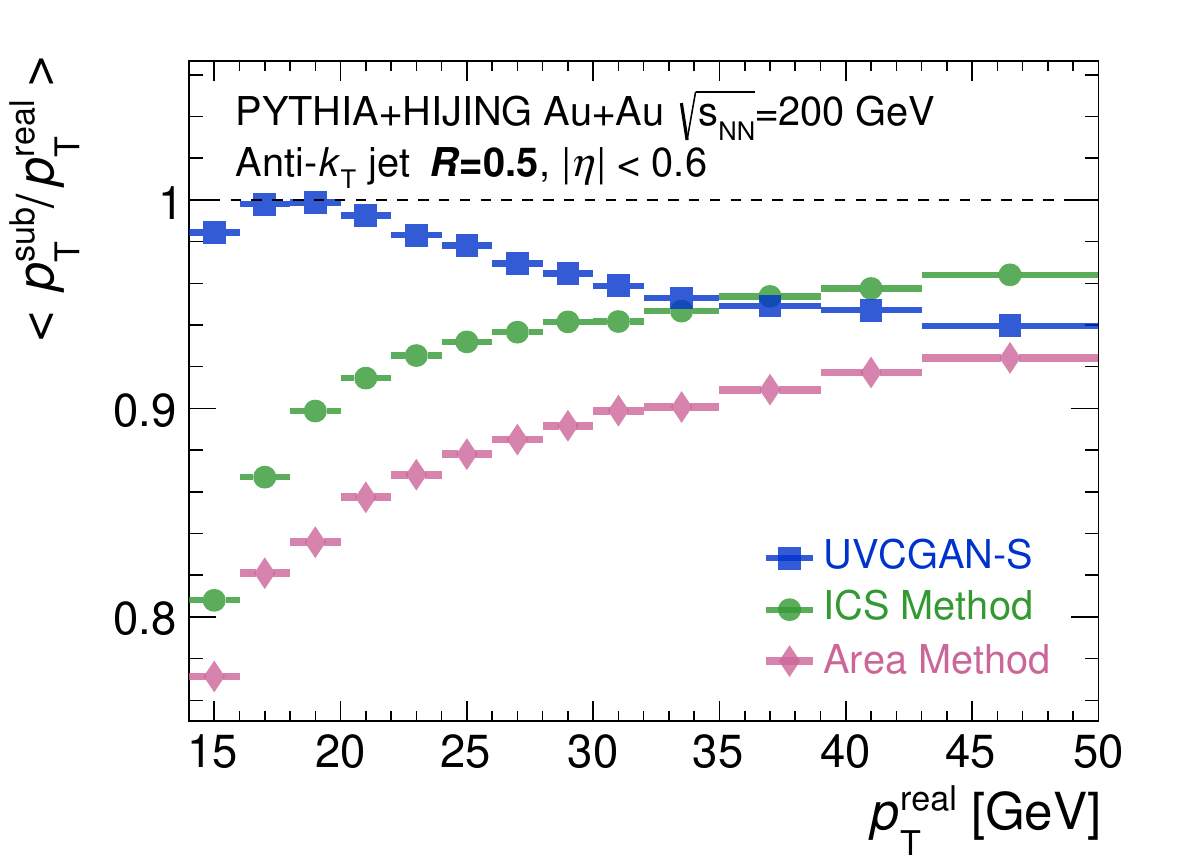}
    \includegraphics[width=0.45\textwidth]{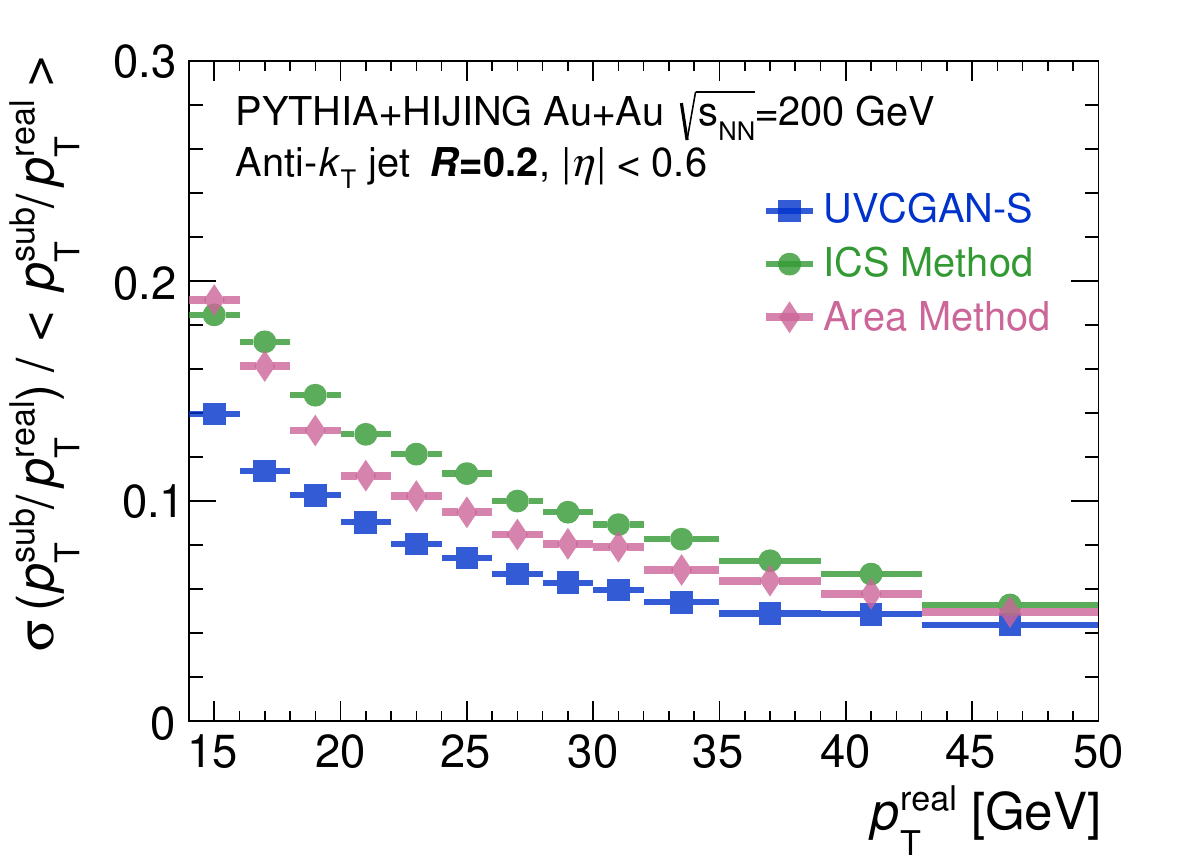}
    \includegraphics[width=0.45\textwidth]{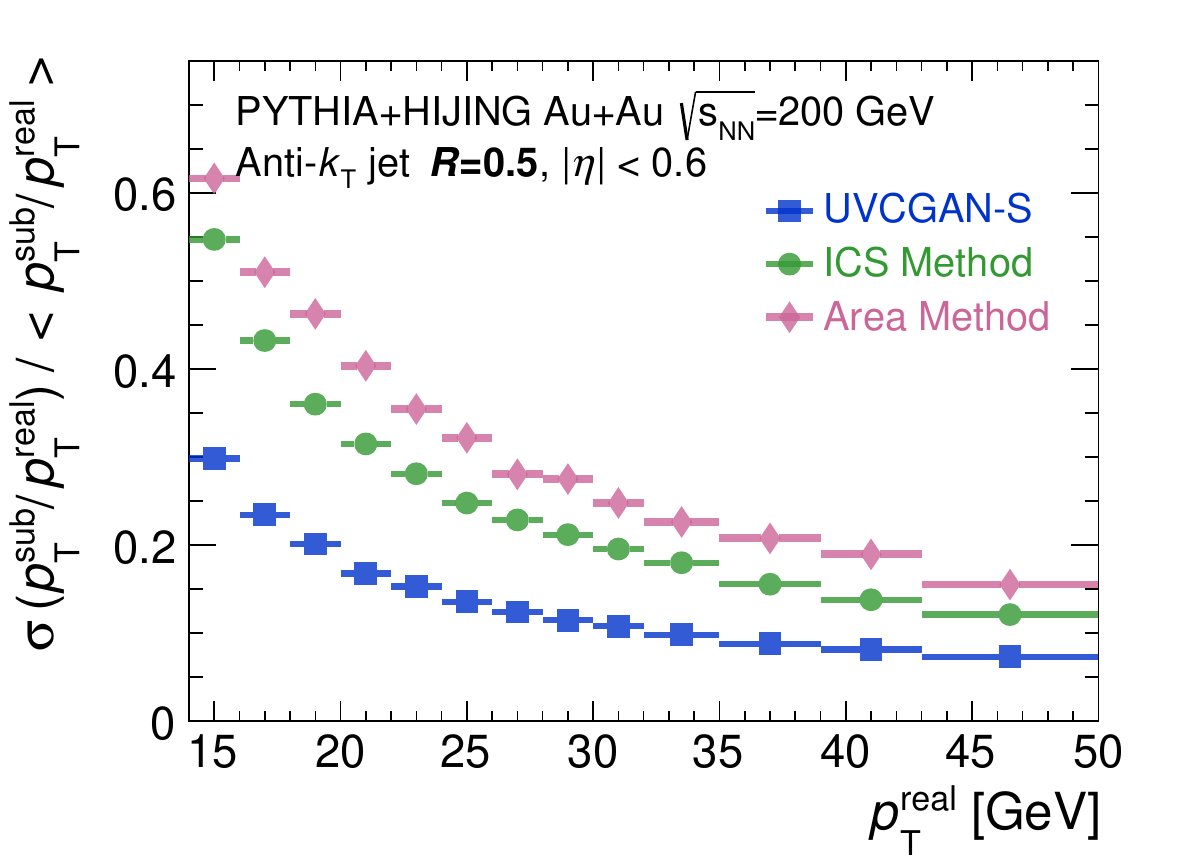}
    \caption{Jet transverse momentum response. Top panels show example distributions of \ptratio for $30<\ptreal<32$~GeV bin for $R=0.2$ (Left) and $R=0.5$ (Right). Panels in the middle row are mean of \ptratio as a function of \ptreal for $R=0.2$ (Left) and $R=0.5$ (Right). Bottom panels are resolution (standard deviation of \ptratio) as a function of \ptreal for $R=0.2$ (Left) and $R=0.5$ (Right). }
    \label{fig:JESJER}
\end{figure}

The jet momentum response, quantified by the ratio \ptratio (where \ptreal is the ground truth jet \pt), is presented in \autoref{fig:JESJER} as a function of \ptreal. For small radius ($R=0.2$), \ourmodel matches or slightly improves the mean response and the resolution relative to conventional methods. However, for large jet radius ($R=0.5$), the \ourmodel model dramatically outperforms the baselines. Specifically, the jet momentum resolution shows an improvement of up to \textbf{$100\%$} relative to the conventional subtraction methods, reflecting the network's superior ability to manage large, fluctuating background contributions.

\subsection{Reconstruction Efficiency and Fake Rates}

\begin{figure}[htpt!]
    \centering
    \includegraphics[width=0.45\textwidth]{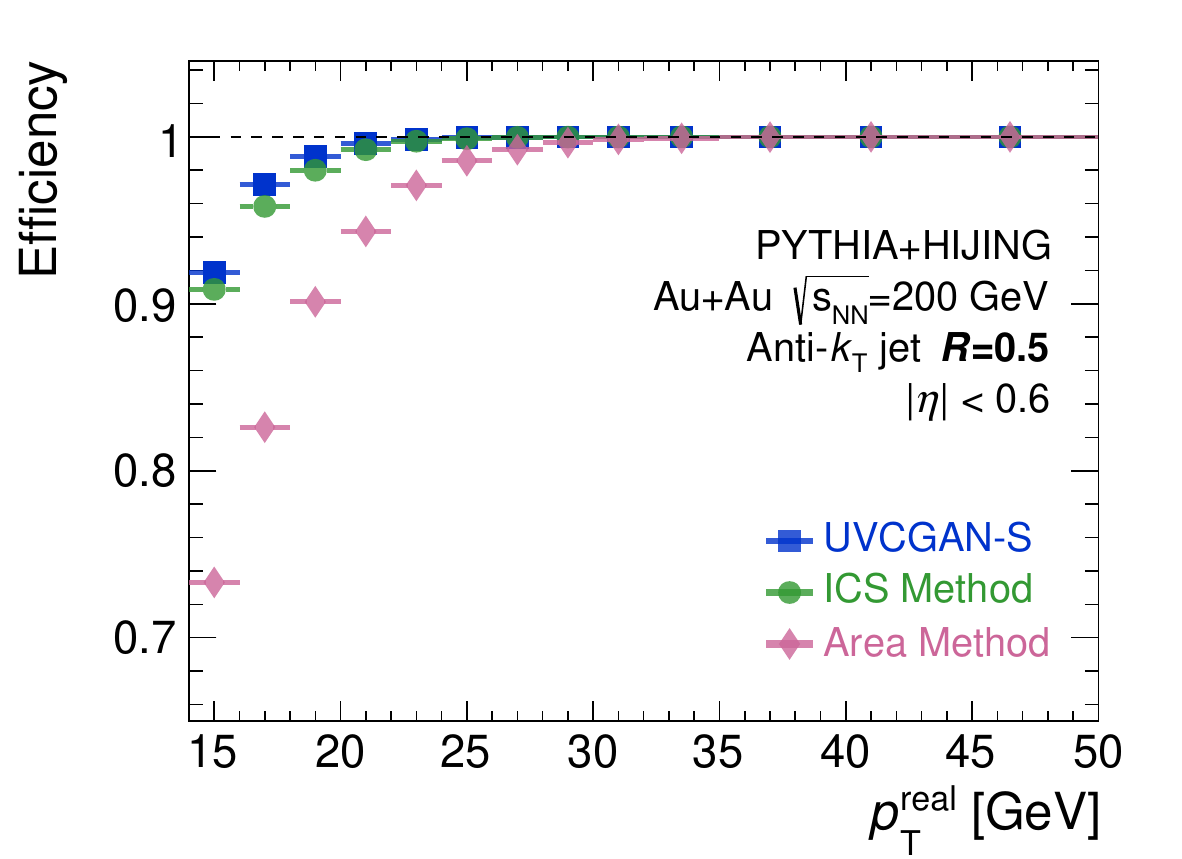}%
    \includegraphics[width=0.45\textwidth]{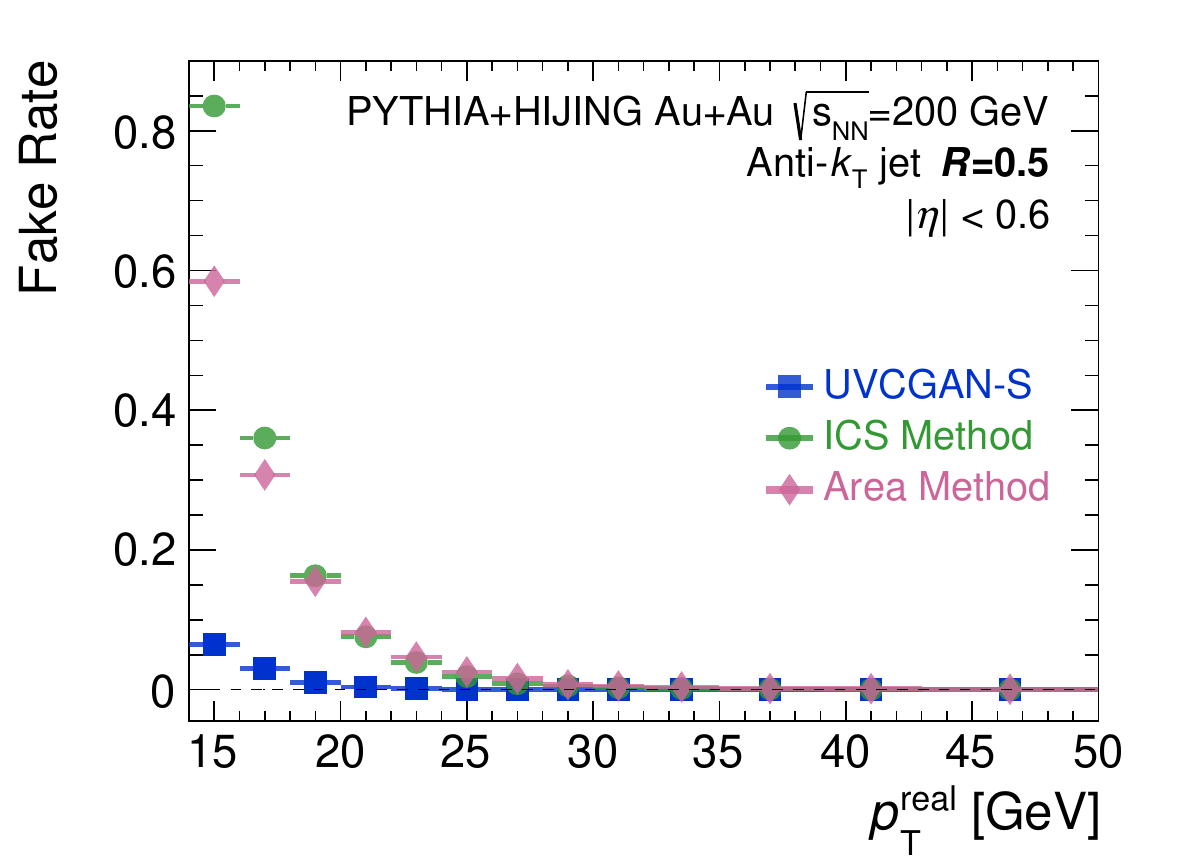}
    \caption{Jet reconstruction efficiency (Left) and fake rate (Right) as a function of the ground truth jet \ptreal for $R=0.5$. The statistical uncertainties are smaller than the marker sizes. }
    \label{fig:eff_fake}
\end{figure}

Background subtraction may remove too much energy and cause ground truth jets to fail to reconstruct. Conversely, local background fluctuations can produce jets that do not originate from a hard scattering, referred to as \textit{fake jets}. 
Therefore, the performance of the subtraction method can be characterized by two metrics: reconstruction efficiency and the fake rate. Efficiency is defined as the fraction of ground truth jets that have a reconstructed partner within a fixed matching criterion of $\Delta R (=0.75\times R)$. For example, background-subtracted jets with $R=0.4$ are considered matched to a ground truth jet if the two axes satisfy $\Delta R<0.30$. This matching window is much larger than the observed $\Delta\eta$ and $\Delta\phi$ resolutions (both $<0.1$ above $14$~GeV). Thus, ground truth jets without a match are interpreted as losses introduced by the subtraction. Conversely, the fake rate is defined as the fraction of reconstructed jets that do not have a matched ground truth partner, indicating that they likely originated from underlying event fluctuations rather than hard scattering.

\autoref{fig:eff_fake} compares these metrics. Crucially, for large radius ($R=0.5$), the \ourmodel significantly outperforms the conventional methods, particularly at low \ptreal. At $\ptreal \sim 14~\text{GeV}$, the efficiency is improved from under $74\%$ to approximately $92\%$, and the fake rate is dramatically reduced from $\sim 84\%$ to $\sim 6\%$. This indicates the model's ability to maintain high sensitivity to the true jet signal while effectively suppressing spurious background structures. 
For small radius ($R=0.2$), all methods perform well, achieving $\sim 99.5\%$ efficiency across the \ptreal range, with UVCGAN-S showing a slight improvement in fake rate from $\sim 2\%$ to $\sim 0.6\%$ at low \pt; detailed results are provided in Appendix~\ref{app:eff_fake}.

\begin{figure}[ht!]
    \centering
    \begin{tikzpicture}
        \def\width{0.46\textwidth}
        \def\xs{0.01\textwidth}
        \def\ys{0.01\textwidth}
        \node[inner sep=0] (fig1) at (0, 0) {\includegraphics[width=\width]{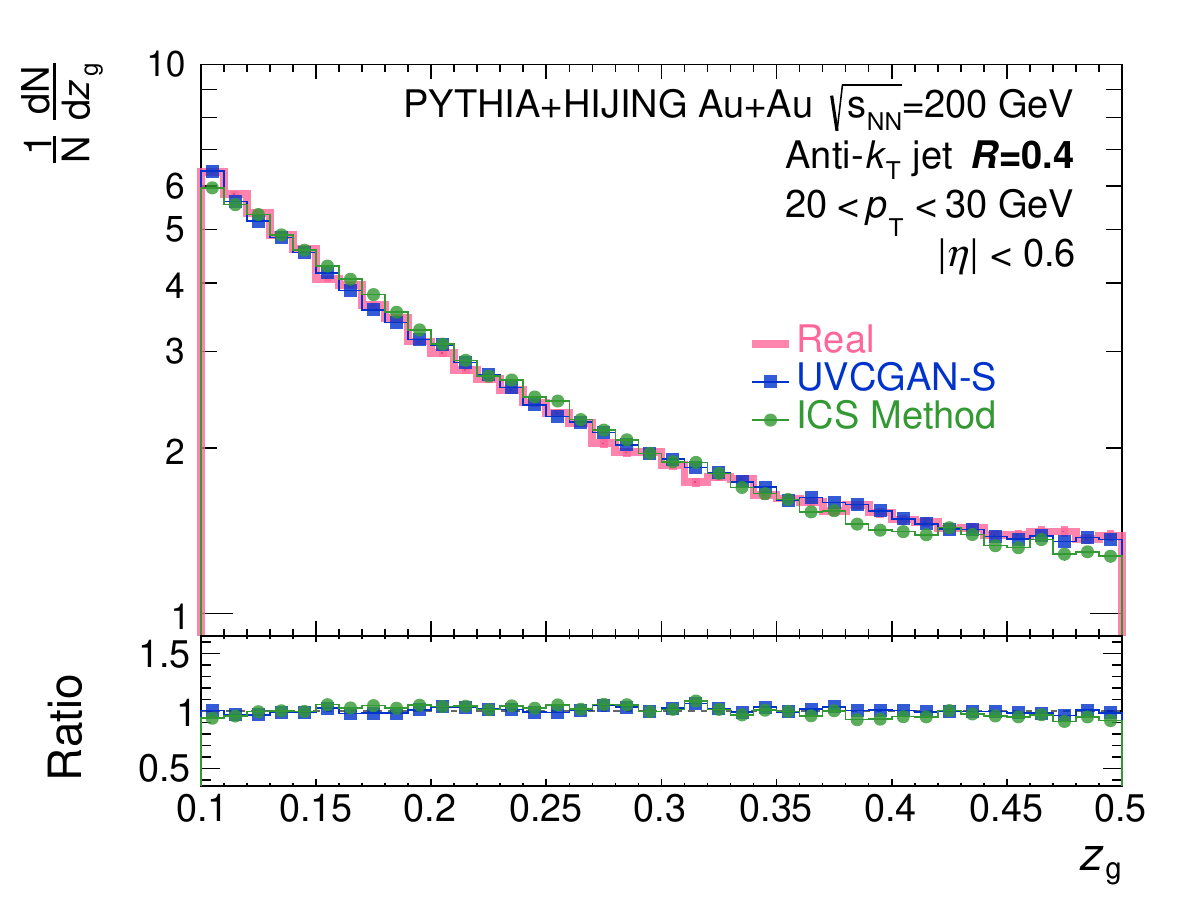}};
        \node[inner sep=0, anchor=west] (fig2) at ([xshift=\xs]fig1.east) {\includegraphics[width=\width]{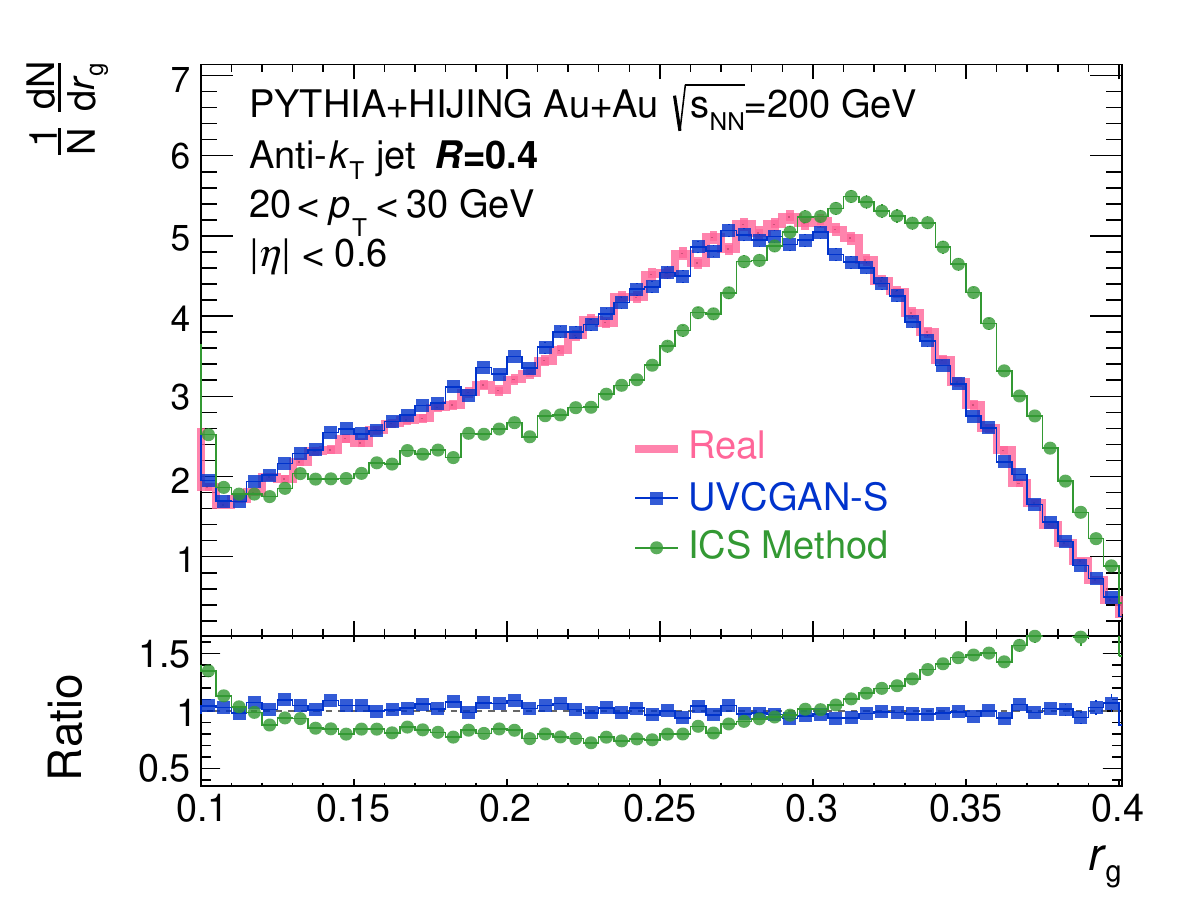}};
        \node[inner sep=0, anchor=north west] (fig3) at ([yshift=-\ys]fig1.south west) {\includegraphics[width=\width]{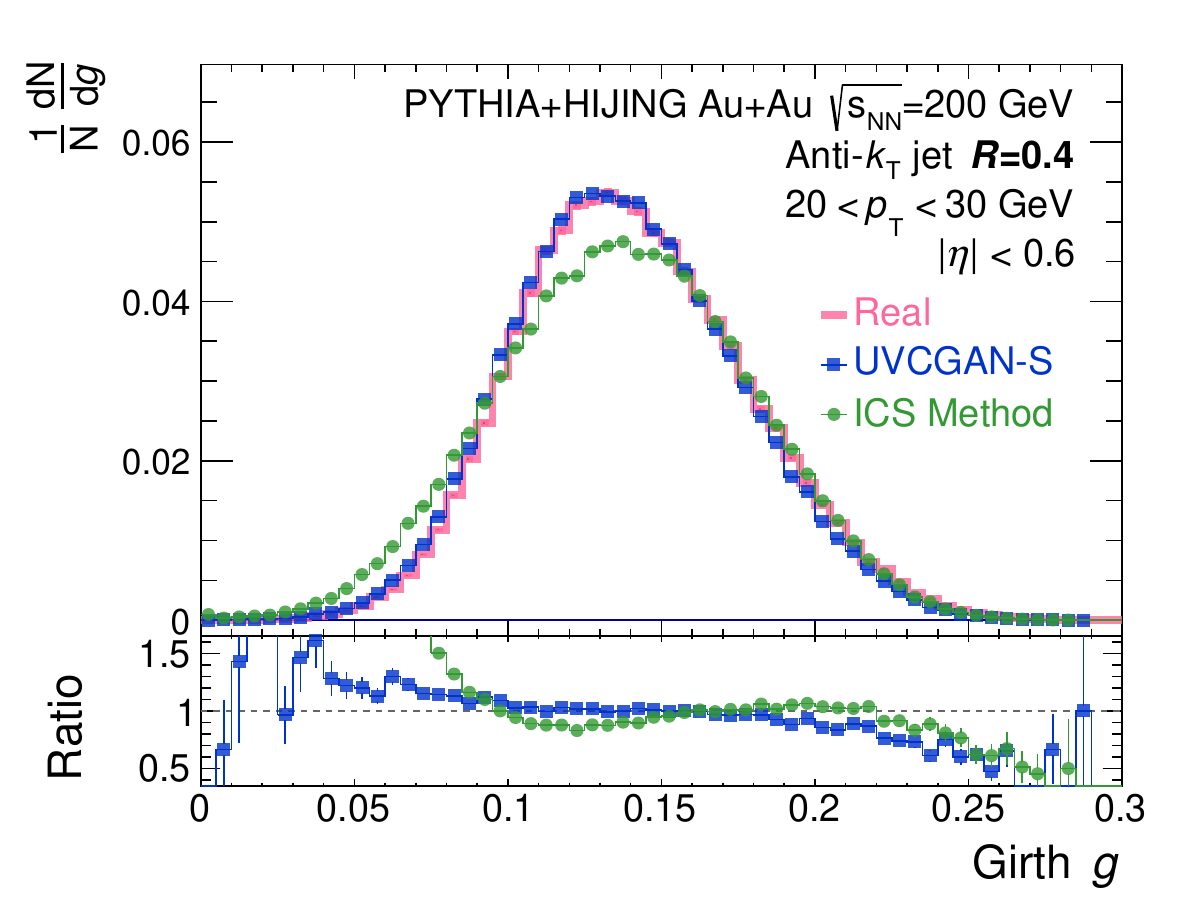}};
        \node[inner sep=0, anchor=west] (fig4) at ([xshift=\xs]fig3.east) {\includegraphics[width=\width]{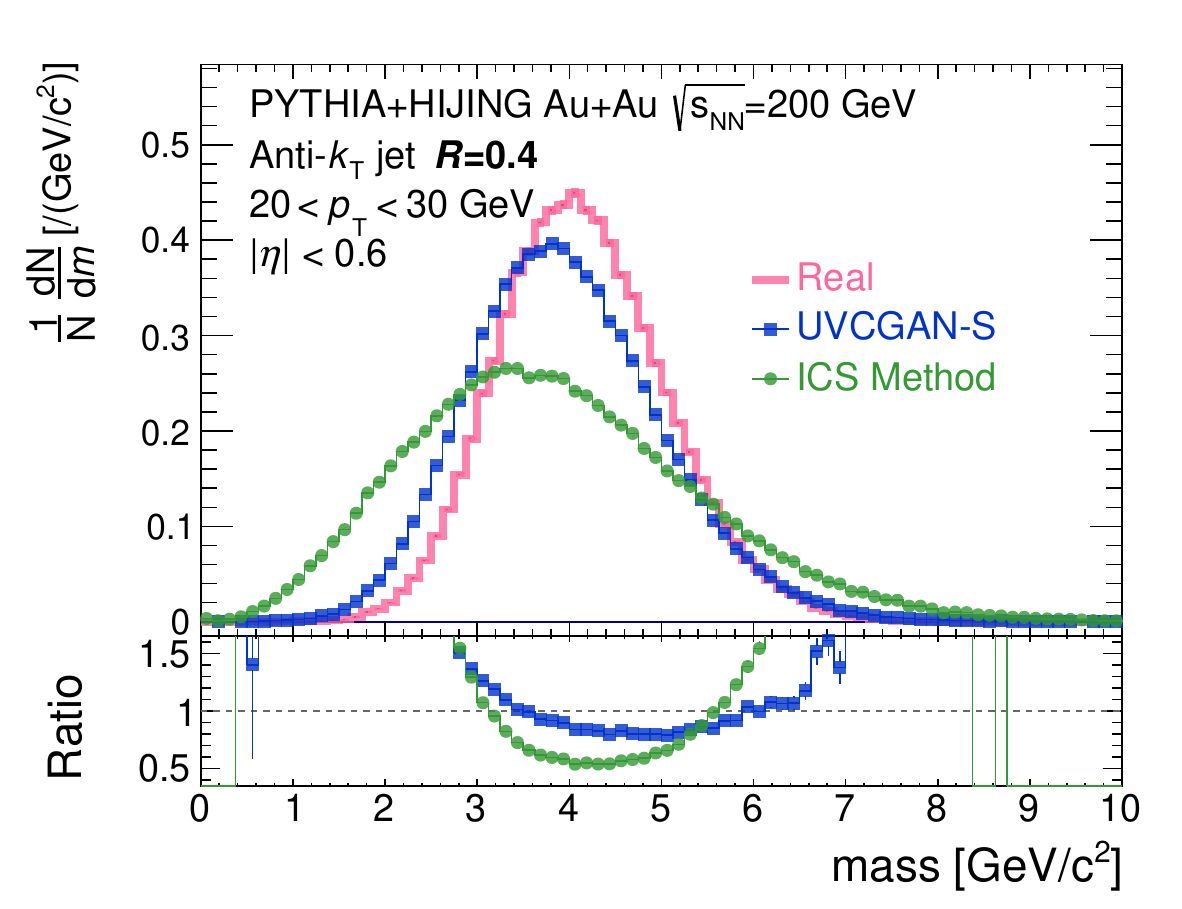}};
        \node[inner sep=0, anchor=north west] (fig5) at ([yshift=-\ys]fig3.south west) {\includegraphics[width=\width]{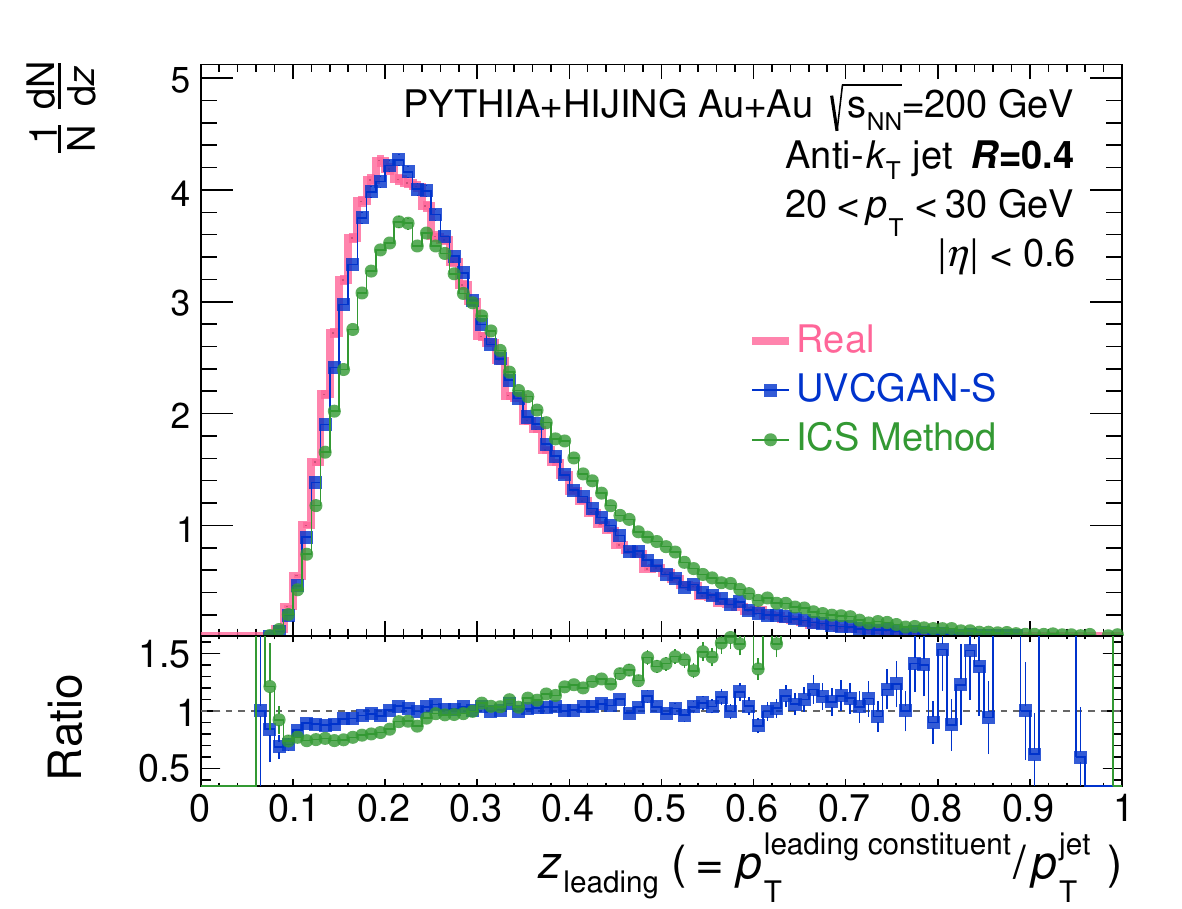}};
        \node[inner sep=0, anchor=west] (fig6) at ([xshift=\xs]fig5.east) {\includegraphics[width=\width]{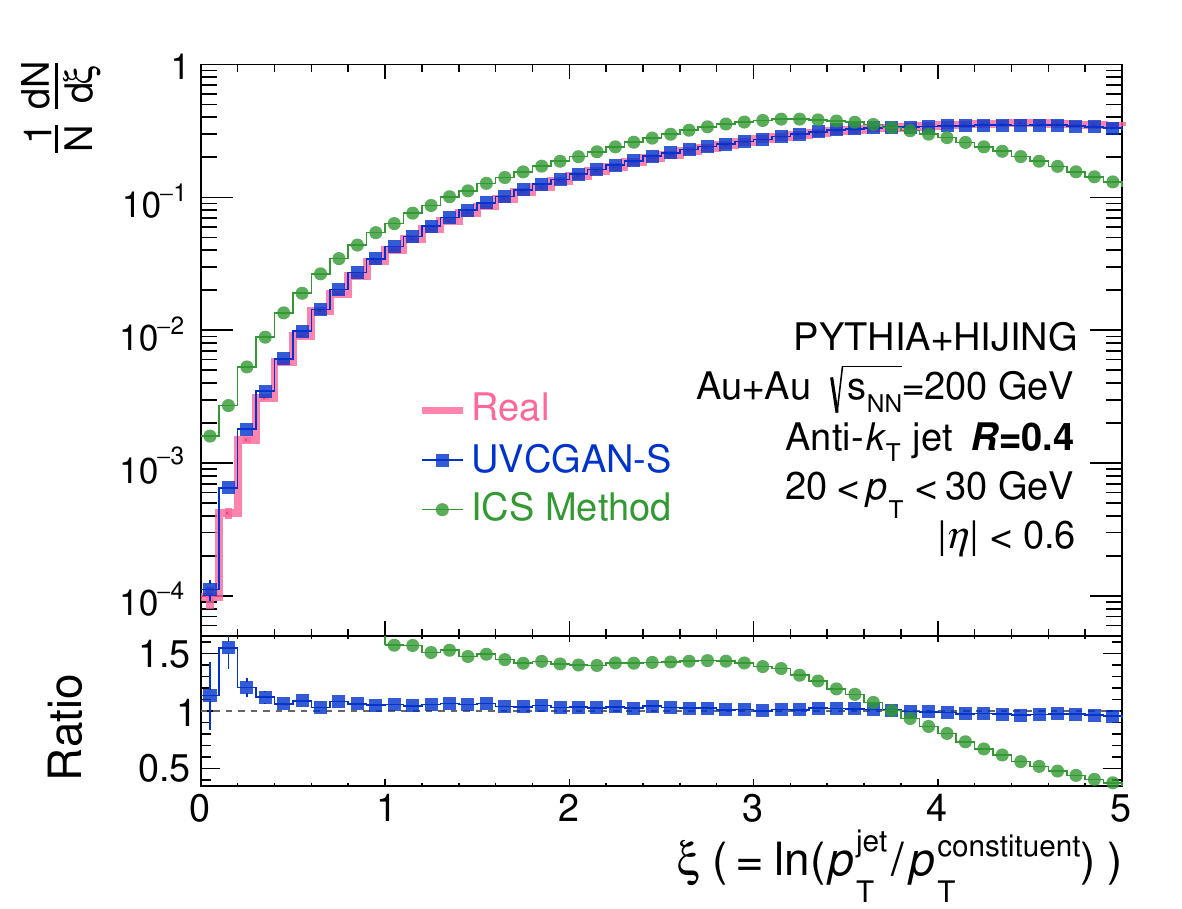}};
        \node[anchor=north west, font=\bfseries] at ([yshift=+10pt]fig1.north west) {(a)};
        \node[anchor=north west, font=\bfseries] at ([yshift=+10pt]fig2.north west) {(b)};
        \node[anchor=north west, font=\bfseries] at ([yshift=+10pt]fig3.north west) {(c)};
        \node[anchor=north west, font=\bfseries] at ([yshift=+10pt]fig4.north west) {(d)};
        \node[anchor=north west, font=\bfseries] at ([yshift=+10pt]fig5.north west) {(e)};
        \node[anchor=north west, font=\bfseries] at ([yshift=+10pt]fig6.north west) {(f)};
    \end{tikzpicture}

    \caption{
    Various jet substructure observable distributions (top panels) and their ratios to the ground truth (``real'') distributions (bottom panels): (a) Groomed momentum sharing fraction \zg, (b) Groomed jet radius \rg, (c) jet girth $g$, (d) jet mass, (e) jet momentum fraction of leading constituents \zleading, (f) jet fragmentation function $\xi$. All distributions are for jet with $R=0.4$ at $20<\ptreal<30$~GeV. The \ourmodel reconstruction is significantly closer to the ``real'' distributions than the Iterative Constituent Subtraction (ICS) Method, demonstrating better preservation of jet substructure.
    }
    \label{fig:substructure}
\end{figure}

\subsection{Jet Substructure}
Jet substructure observables provide a critical test of background subtraction methods, as they probe the fine-grained distribution of energy and momentum within the jet cone, making them highly sensitive to local background fluctuations (see Ref.~\cite{Cunqueiro:2021wls, Kogler:2021kkw} for reviews). We selected six representative observables to assess the ability of our framework to preserve the complex internal energy flow.

We begin with the groomed momentum sharing fraction, \zg, and the groomed jet radius, \rg, derived using the Soft Drop (SD) algorithm~\cite{Larkoski:2014wba}. The variable \zg quantifies the momentum balance between two sub-branches ($p_{\mathrm{T},1}$ and $p_{\mathrm{T},2}$), defined as:
\begin{equation}
z_g = \frac{\min(p_{\mathrm{T},1},\, p_{\mathrm{T},2})}{p_{\mathrm{T},1} + p_{\mathrm{T},2}}
\end{equation}
The SD procedure recursively declusters a jet and identifies the first splitting that satisfies the condition:
\begin{equation}
\frac{\min(p_{\mathrm{T},1},\, p_{\mathrm{T},2})}{p_{\mathrm{T},1} + p_{\mathrm{T},2}} 
> z_{\mathrm{cut}} \cdot \left(\frac{\Delta R_{12}}{R}\right)^{\beta},
\end{equation}
where $\Delta R_{12}$ is the angular separation between the two sub-branches and $R$ is the jet radius. In this study, we use $z_{\mathrm{cut}} = 0.1$ and $\beta = 0$, for which the SD condition simplifies to $z_g > z_{\mathrm{cut}} = 0.1$, removing soft splittings with momentum sharing fraction below $10\%$. The groomed radius \rg is the angular separation ($\Delta R_{12}$) between these two primary sub-branches. These groomed observables are essential for probing hard-splitting dynamics, but their calculation is easily biased if background removal distorts the sub-branch kinematics.

We also examine fundamental ungroomed jet shape variables, including the jet mass ($m$) and jet girth ($g$). Jet mass is a measure of the jet's invariant mass derived from the four-vector sum of its constituents. Girth quantifies the momentum-weighted radial distribution of the constituents, providing insight into the jet's collimation:
\begin{equation}
g = \frac{1}{p_{\mathrm{T}, \text{jet}}} \sum_{i \in \text{jet}} p_{\rm{T},i} \Delta R_{\mathrm{i}, \text{axis}}.
\end{equation}
Finally, we study jet fragmentation function through the momentum fraction of the leading constituent, $z_{\text{leading}} = p_{\mathrm{T}, \text{leading constituent}} / p_{\mathrm{T}, \text{jet}}$, and the fragmentation variable $\xi = \ln(p_{\mathrm{T}, \text{jet}} / p_{\mathrm{T}, \text{constituent}})$. These metrics probe the hard core of the jet as well as the energy spectrum of all jet constituents.

\autoref{fig:substructure} shows the reconstructed distributions for all six substructure variables for $R=0.4$ jets in the $20<\ptreal<30$~GeV bin. In every instance, from the hard-splitting properties (\zg and \rg) to the bulk energy spread ($m$ and $g$), the \ourmodel reconstruction consistently demonstrates a significantly higher fidelity to the ground truth distribution (``real'') compared to the conventional Iterative Constituent Subtraction method. The ICS Method introduces systematic biases, visible as a clear deviation from the ground truth, particularly in the suppression of low \zg values and the excessive broadening of the jet mass and girth distributions. The uniform performance gain achieved by \ourmodel across these diverse and sensitive metrics conclusively demonstrates that our unsupervised, image-based subtraction approach preserves the complex, full-dimensional internal jet structure far more effectively than the conventional method.

%% file: figures/cyclegan/cyclegan.tex



    
    \begin{tikzpicture}
        \def\xs{120pt}
        \def\ys{120pt}
        \def\lw{1pt}
    
        \definecolor{myblue}{HTML}{0FA5C7} 
        \definecolor{myorange}{HTML}{F8A300} 
        \definecolor{yellowgreen}{HTML}{A1CB2E} 
        \definecolor{darkgreen}{HTML}{5AAA4F} 
        \definecolor{mypurple}{HTML}{7000AD} 
    
        \tikzset{
            shadowed path/.style={
                postaction={
                    copy shadow={shadow xshift=0.2em, 
                                 shadow yshift=-0.2em, 
                                 opacity=0.2, 
                                 fill opacity=0}
                }
            }
        }
        
        \tikzset{
            block/.style={
                rectangle, 
                rounded corners=10, 
                blur shadow={
                    shadow blur steps=5, 
                    shadow xshift=1pt, 
                    shadow yshift=-1pt
                },
                align=center,
            },
            flow/.style={
                -{Latex[length=4mm, width=4mm]}, 
                rounded corners=10pt, 
                line width=4 * \lw,  
                shadowed path,
                shorten >= 2pt,
                shorten <= 2pt
            },
            flowinfer/.style={
                -{Latex[length=4mm, width=4mm]}, 
                rounded corners=10pt, 
                line width=4 * \lw,
                densely dashed,
                draw=cyan,
                opacity=.8,
                shorten >= 2pt,
                shorten <= 2pt
            },
        }
    
        \tikzstyle{bs} = [blur shadow={shadow blur steps=5,
                                       shadow blur extra rounding=1pt, 
                                       shadow xshift=1pt,
                                       shadow yshift=-1pt}]
        \tikzstyle{trapezoid} = [inner sep=2,
                                 shape=trapezium, 
                                 trapezium angle=60,
                                 minimum width=25,
                                 anchor=south, 
                                 rounded corners=3pt, 
                                 bs]
        \newcommand{\neuralTranslator}[1]{
            \begin{tikzpicture}
                \node[trapezoid, 
                      draw=#1, 
                      fill=#1!75,
                      rotate=-90] (enc) at (0, 0) {};
                \node[inner sep=0,
                      draw=#1, 
                      fill=#1!75,
                      minimum width=4, 
                      minimum height=13, 
                      rounded corners=1pt, 
                      bs, 
                      anchor=west] (mid) at ([xshift=2pt]enc.north) {};
                \node[trapezoid, 
                      draw=#1, 
                      fill=#1!75,
                      rotate=90, 
                      anchor=north] (dec) at ([xshift=2pt]mid.east) {};
            \end{tikzpicture}
        }
        
        \def\cyclexshift{1.5*\xs}
        \def\myscale{.51}
        \def\folder{figures/cyclegan}

        \node[inner sep=5] {
            \begin{tikzpicture}
                \node[inner sep=0, scale=\myscale, transform shape] (training) at (0, 0) {
                    \begin{tikzpicture}
                        \node[inner sep=5, block, fill=myblue!50] (mix) at (0, 0) {\includegraphics[scale=.2]{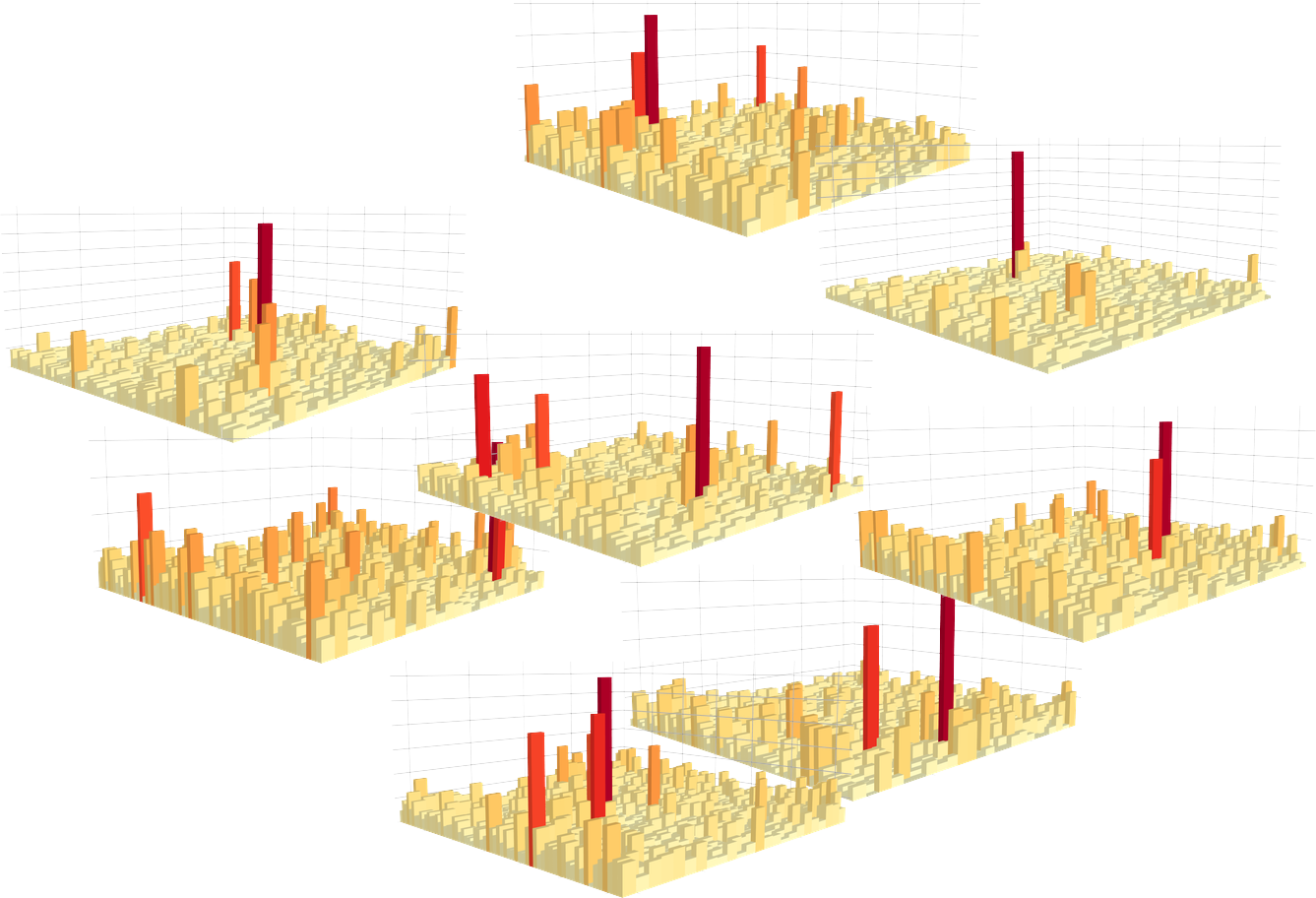}};
                        \tikzmath{coordinate \C;\C=(mix.north east)-(mix.south west);}
                        \node[inner sep=8, anchor=north west, align=left, font=\fontsize{12}{12}\selectfont] (mix_title) at (mix.north west) {\textbf{Domain B}:\\background+signal\\combined data};
                        \node[inner sep=5, block, fill=myblue!70!black, anchor=west, minimum height=\Cy] (bkg_sgn) at ([xshift=.3*\Cx]mix.east) {\includegraphics[scale=.2]{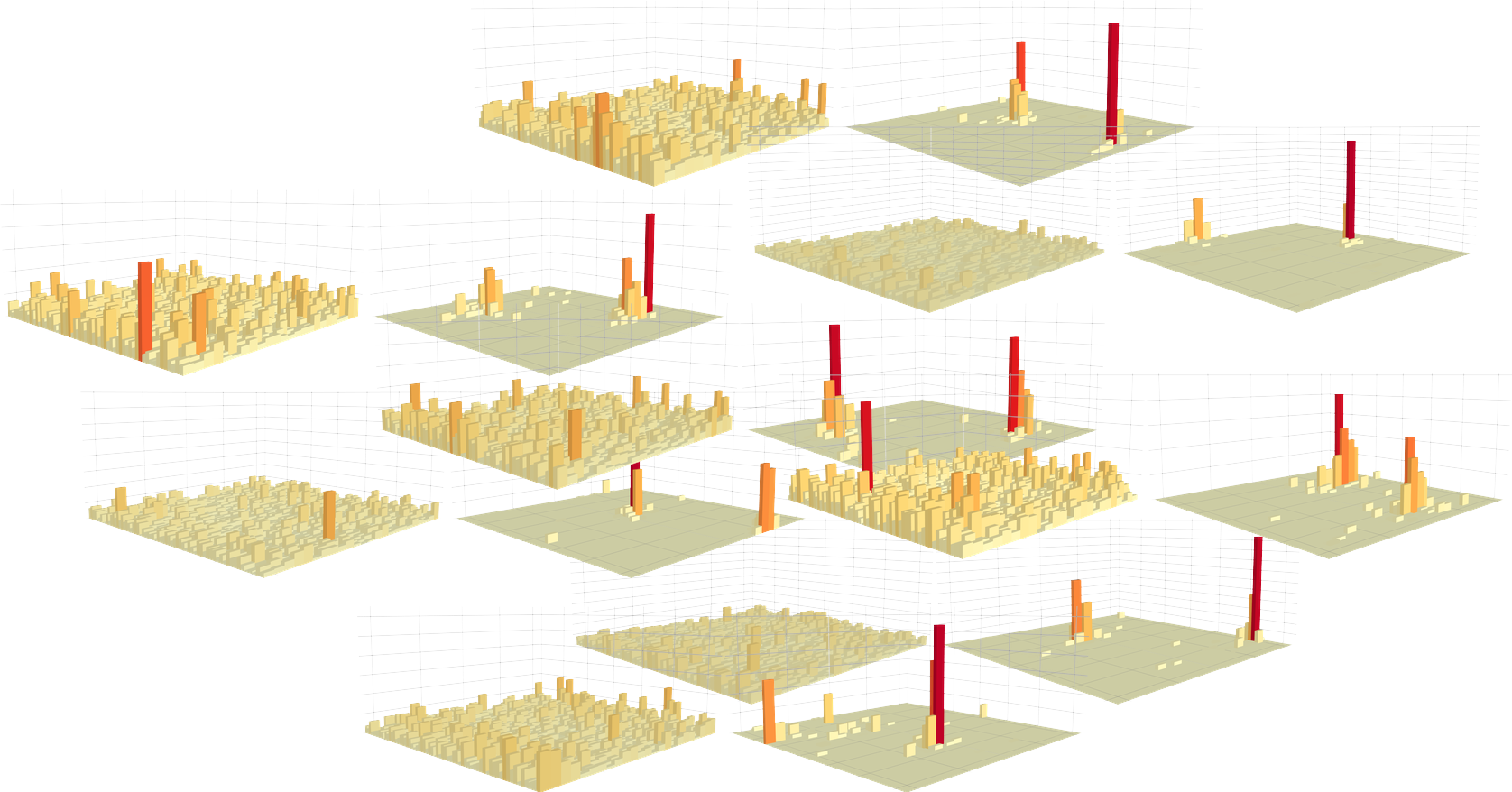}};
                        \node[inner sep=8, anchor=north west, align=left, text=white, font=\fontsize{12}{12}\selectfont] (bkg_sgn_title) at (bkg_sgn.north west) {\textbf{Domain A}:\\background and signal\\data, separately};
                        \def\offset{.25 * \Cy}
                        \coordinate (A1) at ([yshift=\offset]mix.east);
                        \coordinate (A2) at ([yshift=-\offset]mix.east);
                        \coordinate (B1) at ([yshift=\offset]bkg_sgn.west);
                        \coordinate (B2) at ([yshift=-\offset]bkg_sgn.west);
                        \node[inner sep=0] (Gab) at ($(A1)!.5!(B1)$) {\neuralTranslator{yellowgreen}};
                        \node[inner sep=0] (Gba) at ($(A2)!.5!(B2)$) {\neuralTranslator{darkgreen}};
                        \node[inner sep, anchor=center, align=center, font=\fontsize{12}{12}\selectfont] (translator) at ($(Gab)!.5!(Gba)$) {\ourmodel\\Neural\\translators};
                        \draw[flow, draw=black!75] (A1) -- (Gab);
                        \draw[flow, draw=black!75] (Gab) -- (B1);
                        \draw[flow, draw=black!75] (B2) -- (Gba);
                        \draw[flow, draw=black!75] (Gba) -- (A2);   
                    \end{tikzpicture}
                };
                \node[inner xsep=0, inner ysep=2, anchor=south west] (training_title) at (training.north west) {\textbf{(a)} Training Neural Translators on Unpaired Domains};

                \node[inner sep=0, anchor=north west, scale=\myscale, transform shape] (inference) at ([yshift=-1cm]training.south west) {
                    \begin{tikzpicture}
                        \node[inner sep=0, anchor=north] (mix_one) at (0, 0) {\includegraphics[scale=.19]{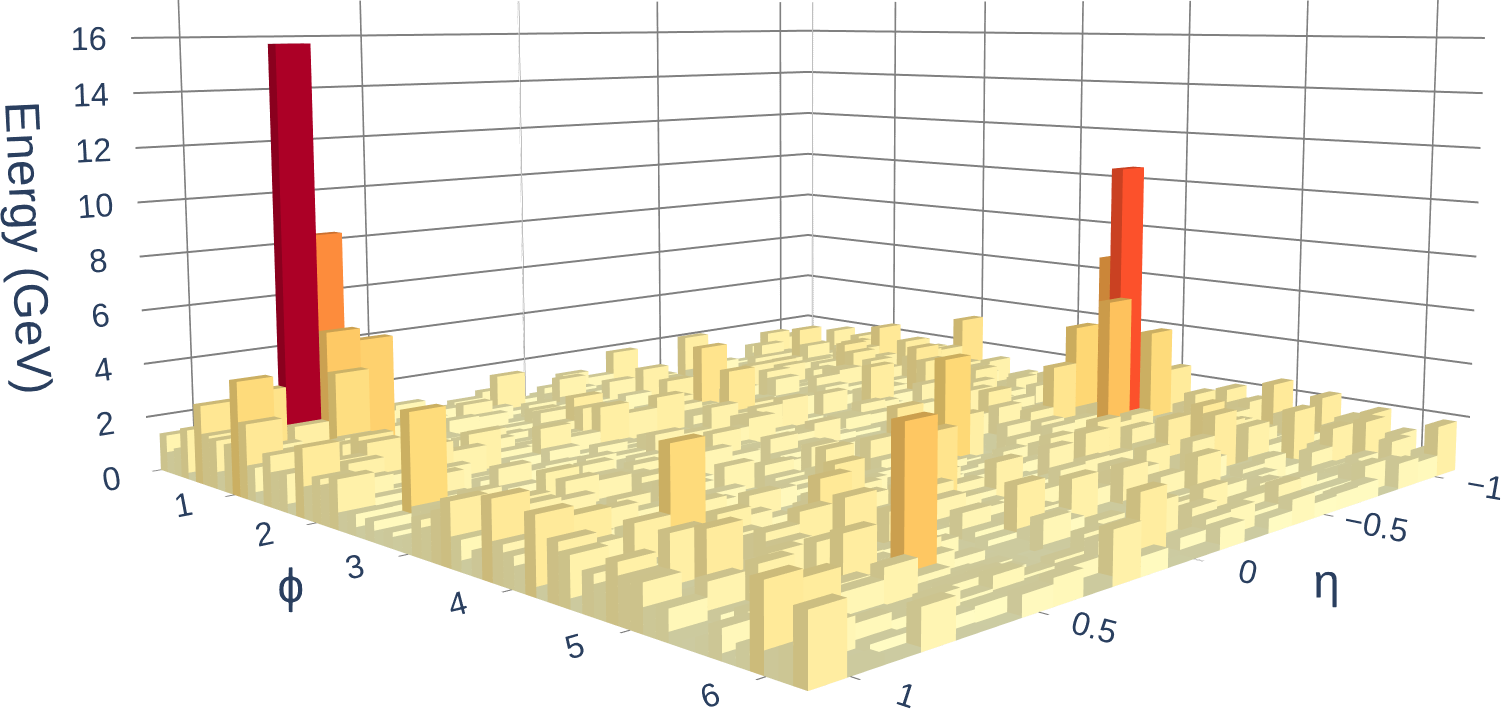}};
                        \node[inner sep=0, anchor=east] (sgn_one) at (bkg_sgn.east |- mix_one.east) {\includegraphics[scale=.19]{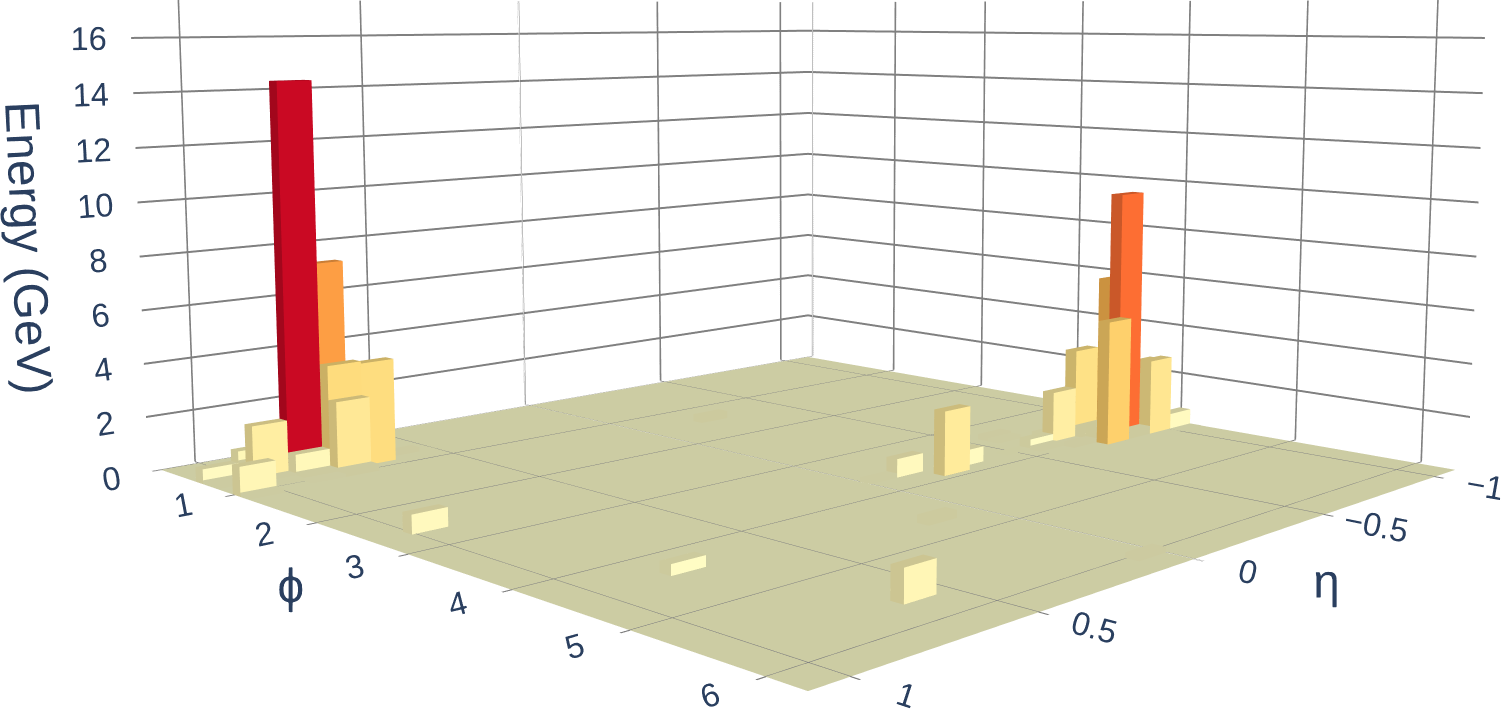}};
                        \node[inner sep=0, anchor=west] (Gab_infer) at ($(mix_one.east)!.23!(sgn_one.west)$) {\neuralTranslator{yellowgreen}};
                        \node[circle, draw=black, inner sep=1, anchor=east] (split) at ($(Gab_infer.east)!.62!(sgn_one.west)$) {$\boldsymbol{\ominus}$};
                        \node[inner sep=0, anchor=south, align=center] at ([yshift=2pt]split.north) {projection\\to one\\component};
                        \draw[flow, draw=black!75] (mix_one) -- (Gab_infer);
                        \draw[flowinfer] (mix_one) -- (Gab_infer);
                        \draw[flow, draw=black!75] (Gab_infer) -- (split);
                        \draw[flowinfer] (Gab_infer) -- (split);
                        \draw[flow, draw=black!75] (split) -- (sgn_one);
                        \draw[flowinfer] (split) -- (sgn_one);  
                    \end{tikzpicture}
                };
                \node[inner xsep=0, inner ysep=2, anchor=south west] (inference_title) at (inference.north west) {\textbf{(b)} Use the Neural Translator to Parse the Signal from the Combined data};
            \end{tikzpicture}
        };
    \end{tikzpicture}

%% file: src/discussion.tex
\section{Discussion}\label{sec:discussion}

This work demonstrates that an unsupervised, unpaired image-to-image translation framework can perform background subtraction directly at the level of detector images. The performance of \ourmodel, particularly its ability to effectively suppress fluctuating background components while preserving the kinematics and substructure of jets, provides a critical advance for high-energy jet measurements. Our results demonstrate superior accuracy compared to conventional subtraction methods, especially for large jet radii ($R=0.5$) and lower jet \pt, a regime historically difficult to access experimentally. This enables the precision measurements highly sought after in high-energy heavy-ion physics. Detailed performance results are provided in~\autoref{app:perf}.

\begin{figure}[ht!]
\centering
\includegraphics[width=0.45\textwidth]{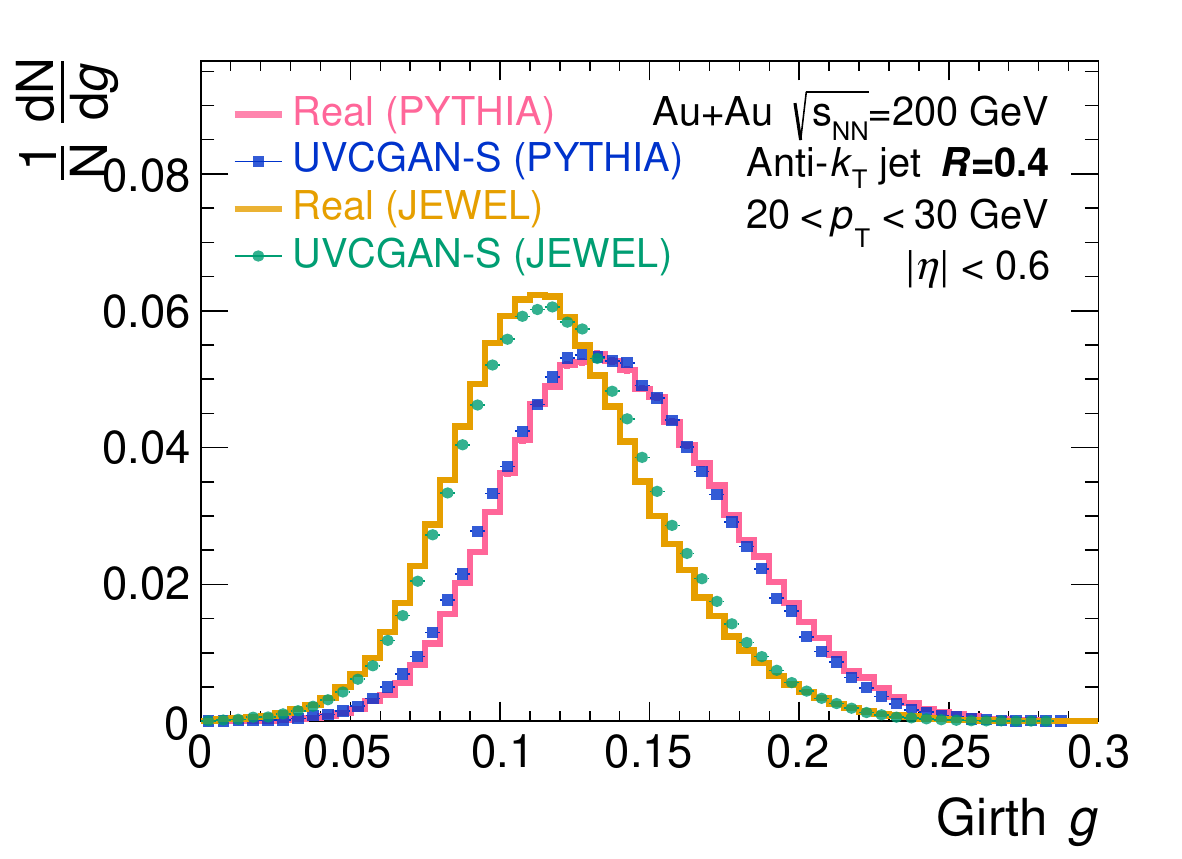}%
\includegraphics[width=0.45\textwidth]{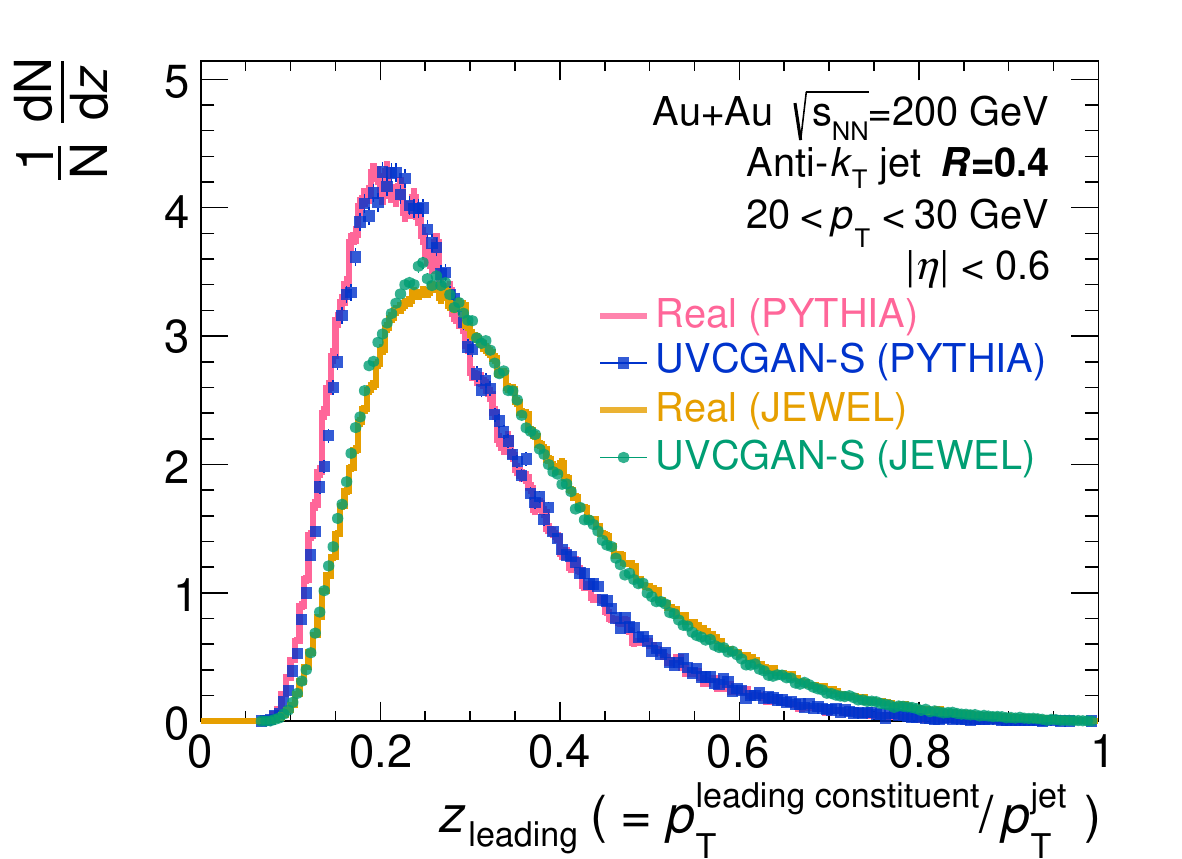}%
\caption{
Jet girth $g$ (Left) and \zleading (Right). Ground truth distributions (Real) and \ourmodel distributions for \pythia (unquenched) and \jewel (quenched) jets, respectively. The \ourmodel trained only on \pythia successfully reconstructs the distinct characteristics of both the quenched \jewel jets and unquenched \pythia jets, demonstrating strong generalization beyond the training domain.
}
\label{fig:jewel_pythia_comp}
\end{figure}

A central question for practical use is \textit{generalization} beyond the training domain. The \ourmodel model was trained exclusively on unquenched vacuum jets (\pythia) embedded in background (\hijing). To test the robustness of the learned features, we applied the trained model to quenched jet inputs, which represent a significant shift from the training domain. Quenched jets were generated using \jewel2.3.0~\cite{Zapp:2012ak}, a MC event generator that simulates the interaction of \pythia vacuum jets with the QGP, thereby modifying the jet's energy and substructure properties. \autoref{fig:jewel_pythia_comp} shows that the substructure distributions for unquenched \pythia jets and quenched \jewel jets differ markedly. Crucially, even though the model was trained exclusively on unquenched \pythia jets, when \jewel jets are used as input for inference, the model successfully reconstructs the distinct quenched jet properties and maintains the medium-induced modifications. This success confirms that background subtraction is an \textit{emergent property} of our method: the network does not simply learn the average statistical properties of the training sample but rather successfully separates the signal from the background regardless of the jet's properties. This robust separation demonstrates that the model learns the emergent properties of the background and removes them effectively while preserving the underlying jet modification. 
The resulting performance in the \jewelhijing environment demonstrates comparable fidelity to the \pythiahijing case, significantly outperforming conventional subtraction methods on the quenched jets. Detailed results are provided in Appendix~\ref{app:jewel}.

The network achieves near closure under this domain shift, which indicates that it successfully learned the minimal necessary features to separate background from signal, rather than relying on a narrow mapping tied to a specific generator. This is especially relevant when subtraction implicitly requires \textit{in-painting} of missing substructure after large energy loss. Unlike supervised AI methods, in which in-painting can be heavily biased by the training data, our method's emergent subtraction property allows the reconstruction to rely on correlations learned from the surrounding image features, thereby reducing bias in the reconstructed jet core. In contrast, supervised predictors trained on a single domain can inherit biases when the jet internal structure differs between training and application. The unsupervised formulation mitigates this risk because it does not require labeled targets and enforces consistency through the reconstruction of the input image from the separated components.

A key advantage of this unpaired, unsupervised framework is that it does not require MC simulations for training. Since the framework requires only unpaired, statistically representative samples from each domain --- ``combined,'' ``pure background,'' and ``pure signal'' --- with no event-by-event correspondence between them, it can in principle be trained on experimental data, removing the dependence on specific MC generators. For example, proton-proton jet events can serve as the signal domain, minimum-bias heavy-ion events as the background domain, and proton-proton jets embedded into minimum-bias heavy-ion events as the combined domain. This data-driven approach would allow the model to learn directly from real detector data, naturally capturing effects that are difficult to simulate perfectly, such as detector response, acceptance, noise, and event-by-event fluctuations in the underlying event.

While the \ourmodel achieves impressive performance gains, any remaining systematic non-closure between the reconstructed jet and the ground truth can be quantified and addressed through established calibration and correction procedures. For instance, the trained model can be tested on dedicated labeled simulation samples (both quenched and unquenched) to quantify and correct residual effects. The systematic uncertainty of the method is most meaningfully characterized by this residual non-closure itself, evaluated on an observable-by-observable basis, as presented in Figs.~\ref{fig:JESJER}, \ref{fig:substructure}, and Appendix~\ref{app:substructure}. Given that the network already significantly outperforms conventional methods across all key observables, the magnitude of these final, analysis-level corrections will be substantially smaller.

Finally, there are natural extensions to this work. The same formulation applies whenever signal and background superpose in an image-like representation. This suggests applications beyond jets in heavy-ion collisions, including proton-proton collision environments, neutrino detectors, and other imaging systems in nuclear and particle physics.

In summary, we have presented \ourmodel, an unsupervised, unpaired image-to-image translation framework for jet background subtraction in heavy-ion collisions, and evaluated its performance on sPHENIX simulated Au+Au events at $\sqrt{s_\mathrm{_{NN}}}=200~\text{GeV}$. The key innovation is a stratified CycleGAN architecture that decomposes combined heavy-ion calorimeter images into signal and background components without requiring any labeled or paired training data. Compared to widely-used conventional methods (Area-based and Iterative Constituent Subtraction), \ourmodel achieves up to 40\% better jet position resolution and up to 100\% improvement in momentum resolution for large-radius (R = 0.5) jets, while simultaneously improving reconstruction efficiency and reducing fake rates significantly. Crucially, the model preserves jet substructure observables — including jet mass, groomed splitting radius, momentum sharing fraction, and the jet fragmentation function — significantly more faithfully than conventional approaches. The robustness of the framework was further demonstrated by applying the model, trained exclusively on \pythia jets, to \jewel-generated quenched jets without retraining, showing that the method is not rigidly dependent on the physics model used in training. These results establish unsupervised image-to-image translation as a promising new direction for background subtraction in heavy-ion jet measurements, with the potential to extend the kinematic reach of jet analyses to lower momenta and larger cone radii where conventional methods face fundamental limitations.

%% file: src/methods.tex
\section{Methods}\label{sec:method}
In the preceding sections, we demonstrated that \ourmodel achieves significant improvements in jet momentum resolution, reconstruction efficiency, and substructure preservation compared to conventional methods. In this section, we describe the unsupervised, unpaired image-to-image translation framework that enables these gains, including the CycleGAN foundation, the motivation for our stratified domain decomposition, and the \ourmodel architecture and training procedure.

\subsection{CycleGAN Setup}
\label{sec:cyclegan_setup}

CycleGAN~\cite{8237506} combines two Generative Adversarial Networks (GANs)~\cite{Goodfellow:2014upx} to learn a bidirectional mapping between two domains A and B without requiring paired training examples.
The architecture employs two generators $\mathcal{G}_{A \to B}$ and $\mathcal{G}_{B \to A}$ that translate samples between domains, along with two discriminators $\mathcal{D}_{A}$ and $\mathcal{D}_{B}$ that distinguish real samples from generated ones in their respective domains.

The CycleGAN generators $\mathcal{G}_{A \to B}$ and $\mathcal{G}_{B \to A}$ are trained adversarially against the corresponding discriminators $\mathcal{D}_{A}$ and $\mathcal{D}_{B}$ in a minimax game common to GAN training.
The adversarial losses for both translation directions are:
\begin{align}
    \label{eq:cyclegan_loss_adv}
    \mathcal{L}_{\text{GAN}}(\mathcal{G}_{A \to B}, \mathcal{D}_B)
        &= \mathbb{E}_b[f(\mathcal{D}_B(b))] + \mathbb{E}_a[g(\mathcal{D}_B(\mathcal{G}_{A \to B}(a)))] \\
    \mathcal{L}_{\text{GAN}}(\mathcal{G}_{B \to A}, \mathcal{D}_A)
        &= \mathbb{E}_a[f(\mathcal{D}_A(a))] + \mathbb{E}_b[g(\mathcal{D}_A(\mathcal{G}_{B \to A}(b)))]
\end{align}
where $a \in A$ and $b \in B$ denote samples from the domain A and B respectively, and $f$ and $g$ define the adversarial objective (e.g., log loss, least squares, Wasserstein distance~\cite{DBLP:journals/corr/ArjovskyCB17}).

However, adversarial losses alone are insufficient to ensure meaningful domain translation, as generators may learn to ignore input structure and produce arbitrary outputs from the target domain.
CycleGAN addresses this by introducing the concept of cycle consistency, requiring that sequential application of both generators returns to the original input $\mathcal{G}_{B \to A}(\mathcal{G}_{A \to B}(a)) \approx a$.
In practice, this requirement is enforced by introducing a cycle consistency loss to the CycleGAN training:
\begin{equation}
    \mathcal{L}_{\text{cycle}}^{A\to B \to A} =
        \mathbb{E}_a[||\mathcal{G}_{B \to A}(\mathcal{G}_{A \to B}(a)) - a||_1]
\end{equation}
and a similar loss in the opposite direction.

To further stabilize training, CycleGAN introduces identity losses that encourage generators to preserve samples when applied to their target domain:
\begin{equation}
    \label{eq:cyclegan_idt}
    \mathcal{L}_{\text{idt}}^{B \to A} = \mathbb{E}_a[||\mathcal{G}_{B \to A}(a) - a||_1]
\end{equation}
with a similar loss applied in the $A \to B$ direction.
This constraint prevents unnecessary changes when the input already belongs to the target domain and helps maintain color consistency in image translation tasks.

The complete CycleGAN training objective combines all terms:
\begin{equation}
    \label{eq:cyclegan_loss_total}
    \mathcal{L}_{total}
        = \mathcal{L}_{\text{GAN}}
        + \lambda_\text{cycle} \mathcal{L}_\text{cycle}
        + \lambda_\text{idt} \mathcal{L}_{\text{itd}}
\end{equation}
where $\lambda_\text{cycle}$ and $\lambda_\text{idt}$ are the hyperparameters, controlling the magnitude of identity and cycle-consistency losses.

In this work, we use the UVCGANv2 implementation of CycleGAN~\cite{torbunov2023}, which achieves better performance than the original CycleGAN and has been successfully applied to scientific data analysis~\cite{huang2024unpaired}.

\subsection{Naive CycleGAN fails in Signal Extraction}

The CycleGAN framework appears naturally suited for unsupervised signal extraction in heavy-ion collisions.
We can define two domains: Domain $A$ containing isolated jet images and Domain $B$ containing jet images with QGP backgrounds as observed in heavy-ion collision events (``embedded jets'').
Applying the CycleGAN framework to such domains, the generator $\mathcal{G}_{B \to A}$ should learn to extract clean jet signals from embedded jet images, while $\mathcal{G}_{A \to B}$ should learn to generate realistic heavy-ion collision images from isolated jets.

We attempted to train such a CycleGAN architecture on sPHENIX simulation data. 
However, we observed that the resulting performance is highly sensitive to the hyperparameter configurations, with a large fraction of models converging to simple identity transformations, where generators produced outputs identical to inputs without performing meaningful signal extraction (\autoref{fig:naive_failures}).
These identity mappings represent natural local minima of CycleGAN that trivially minimize both cycle consistency and identity losses.

\begin{figure}[htbp]
    \centering
            \input{figures/naive_failure_v2/naive_failure_v3}
    \caption{
   Failure mode of naive CycleGAN for signal extraction. (a) Ideal signal extraction by $\mathcal{G}_{B \to A}$ burdens $\mathcal{G}_{A \to B}$ to recover the full original input, leading to a large cycle-consistency loss. (b) This cycle-consistency challenge overwhelms training, causing both generators to converge to simple identity functions instead of learning meaningful extraction.
    }
    \label{fig:naive_failures}
\end{figure}

Overall, the CycleGAN framework appears to work for some hyperparameter configurations, but finding stable regimes requires finding careful balance of discriminator architecture, gradient penalty magnitude~\cite{gulrajani2017improved}, and cycle consistency weights.
This hyperparameter sensitivity poses a significant barrier to practical deployment: each new dataset, collision energy, or detector configuration would require substantial tuning effort to find the narrow range of stable training configurations.

What may cause this extreme hyperparameter sensitivity?
The issue seems to stem from a fundamental tension in the cycle-consistency requirement. The mapping from embedded jet data ($b \in B$) to isolated signals ($a^\prime = \mathcal{G}_{B \to A}(b) \in A$) necessarily destroys the background information needed for cycle reconstruction.
We observed that in successful training configurations, generators encode residual background information as subtle perturbations in the extracted signal images -- small-amplitude artifacts that carry just enough information for approximate reconstruction (cf.~\cite{chu2017cyclegan}).
However, for this artifact-based encoding to work, a delicate balance is required: discriminator $\mathcal{D}_A$ must be powerful enough to distinguish generated from real jet signals, yet not so sensitive that it detects these encoding artifacts and penalizes them.
This likely explains the narrow hyperparameter regime where training succeeds -- discriminator capacity must fall within a certain range that permits information smuggling while maintaining adversarial pressure.

\subsection{Stratified CycleGAN Framework}
\label{sec:stratifiedCycleGAN}
We attempt to resolve the fundamental incompatibility between the background subtraction problem and the cycle-consistency constraint by rethinking the CycleGAN setup.
At the core, the failure stems from the fact that the translation $B \to A$ (embedded signal to isolated signal) necessarily leads to irrecoverable information loss making the cycle-consistency impossible to maintain.
Therefore, to enable cycle-consistency, we need to provide additional dimensions for the $B \to A$ translation to preserve this information.

Motivated by this observation, we modify the CycleGAN setup by constructing the domain $A$ out of two components: $A = (A_0, A_1) = (A_\text{bkg}, A_\text{sgn})$, where $A_0$ represents the background component and $A_1$ represents the signal component.
In this setup, the $B \to A$ translation decomposes embedded signal into explicit background and signal components $b \to (a_\text{bkg}, a_\text{sgn})$, preserving all information needed for reconstruction.
The reverse $A \to B$ translation learns to combine independently sampled signal and background components into realistic embedded observations.
Crucially, the cycle $\mathcal{G}_{A \to B}(\mathcal{G}_{B \to A}(b)) \approx b$ can now be satisfied because background information can be preserved in the $B \to A$ translation.

This approach requires three independent data sources: isolated jets, pure background components, and embedded jets, compared to the two domains needed for standard CycleGAN. However, it maintains the key advantage that no ground-truth correspondences between isolated signals and embedded signals are required - the mapping is learned from statistical distributions alone.

In practice, the approach is well-suited to be used for sPHENIX data, where background samples can be obtained from minimum-bias events (where high-$p_T$ jets are absent), signal samples from simulation or proton-proton collisions, and embedded samples from heavy-ion collision data.

We call this approach ``Stratified CycleGAN'' to reflect the decomposition of Domain $A$ into explicit signal and background strata.

\subsection{\thename Implementation}

In this section, we describe the implementation of the Stratified CycleGAN framework on top of UVCGANv2 that we call \thename.

The \thename architecture employs two generators and three discriminators to handle the stratified domain structure.
Generator $\mathcal{G}_{B \to A}$ decomposes embedded signal images from Domain $B$ ($C_b$ channels) into stratified components in Domain $A$, outputting $(C_{a_0} + C_{a_1})$ channel images which are then split channelwise into background ($C_{a_0}$) and signal ($C_{a_1}$) components.
The reverse generator $\mathcal{G}_{A \to B}$ takes concatenated stratified inputs and produces embedded signal in Domain $B$.
Three discriminators enable adversarial training: $\mathcal{D}_{A_0}$ distinguishes real from generated background components, $\mathcal{D}_{A_1}$ handles signal components, and $\mathcal{D}_B$ operates on embedded signal in Domain $B$.

\begin{figure}[htbp]
    \centering
    \ifdefined\isarxiv
        \includegraphics[width=0.90\textwidth]{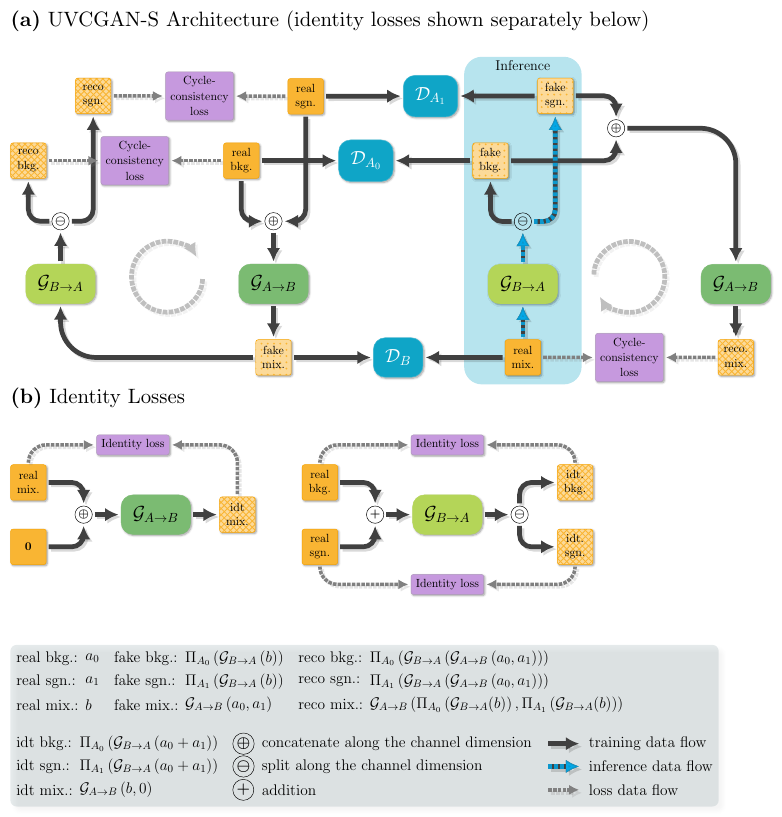}
    \else
        \tikzsetnextfilename{stratified_arch}
        \input{figures/stratified_arch/stratified_arch}
    \fi
    \caption{Schematic Diagram of \ourmodel Architecture}
    \label{fig:stratified_arch}
\end{figure}

The training objective extends the standard CycleGAN loss formulation~\eqref{eq:cyclegan_loss_total} to accommodate the stratified domain structure.
The adversarial losses now include separate terms for each discriminator:
\begin{align}
    \label{eq:stratified_loss_adv}
    \mathcal{L}_{\text{GAN}}(\mathcal{G}_{B \to A}, \mathcal{D}_{A_0}) 
        &= \mathbb{E}_{a_0}[f(\mathcal{D}_{A_0}(a_0))] + \mathbb{E}_b[g(\mathcal{D}_{A_0}(\Pi_{A_0}(\mathcal{G}_{B \to A}(b))))] \\
    \mathcal{L}_{\text{GAN}}(\mathcal{G}_{B \to A}, \mathcal{D}_{A_1}) 
        &= \mathbb{E}_{a_1}[f(\mathcal{D}_{A_1}(a_1))] + \mathbb{E}_b[g(\mathcal{D}_{A_1}(\Pi_{A_1}(\mathcal{G}_{B \to A}(b))))] \\
    \mathcal{L}_{\text{GAN}}(\mathcal{G}_{A \to B}, \mathcal{D}_B) 
        &= \mathbb{E}_b[f(\mathcal{D}_B(b))] + \mathbb{E}_{a_0,a_1}[g(\mathcal{D}_B(\mathcal{G}_{A \to B}(a_0, a_1)))]
\end{align}
where $\Pi_{A_0}: A \to A_0$ and $\Pi_{A_1}: A \to A_1$ denote projection operators extracting the background and signal components from domain $A$ respectively, and $f$ and $g$ define the adversarial objective as described in~\autoref{sec:cyclegan_setup}.

The cycle consistency losses ensure that sequential translations preserve the original input. For the stratified framework, these become:
\begin{align}
    \mathcal{L}_{\text{cycle}}^{B \to A \to B}
        &= \mathbb{E}_b[||\mathcal{G}_{A \to B}(\mathcal{G}_{B \to A}(b)) - b||_1] \\
    \mathcal{L}_{\text{cycle}}^{A \to B \to A}
        &= \mathbb{E}_{a_0,a_1}[||\Pi_{A_0}(\mathcal{G}_{B \to A}(\mathcal{G}_{A \to B}(a_0, a_1))) - a_0||_1] \nonumber \\
    &\quad + \mathbb{E}_{a_0,a_1}[||\Pi_{A_1}(\mathcal{G}_{B \to A}(\mathcal{G}_{A \to B}(a_0, a_1))) - a_1||_1]
\end{align}

To stabilize training, the original CycleGAN work introduces identity losses~\eqref{eq:cyclegan_idt}.
In the general stratified framework, identity losses can be designed in various ways depending on the relationship between domains and the available domain knowledge.
For sPHENIX data, we found that the specific form of the identity losses has minimal impact on final performance, allowing us flexibility in design choices.
Therefore, in this work, we settle on a simple form of identity losses, exploiting the fact that the channel dimensions are equal ($C_{a_0} = C_{a_1} = C_b = 1$) and the background component $A_0$ shares similar statistical properties with the mixed data in $B$.
\begin{align}
    \mathcal{L}_{\text{idt}}^{B \to A}
        &= \mathbb{E}_{a_0,a_1}[||\Pi_{A_0}(\mathcal{G}_{B \to A}(a_0 + a_1)) - a_0||_1] \nonumber \\
    &\quad + \mathbb{E}_{a_0,a_1}[||\Pi_{A_1}(\mathcal{G}_{B \to A}(a_0 + a_1)) - a_1||_1] \\
    \mathcal{L}_{\text{idt}}^{A \to B}
        &= \mathbb{E}_b[||\mathcal{G}_{A \to B}(b, \mathbf{0}) - b||_1]
\end{align}

The first identity loss tests the generator's ability to perform consistent decomposition on controlled inputs, ensuring stable training behavior.
The second identity loss ensures that when no signal is present (signal component is zero), the generator preserves the background data without artificial modifications.
These constraints guide the training toward physically meaningful decompositions while maintaining the unsupervised nature of the approach.

The complete \thename training objective combines all loss terms:
\begin{align}
    \mathcal{L}_{\text{total}}
        &= \frac{1}{2}\left[
              \mathcal{L}_{\text{GAN}}(\mathcal{G}_{B \to A}, \mathcal{D}_{A_0})
            + \mathcal{L}_{\text{GAN}}(\mathcal{G}_{B \to A}, \mathcal{D}_{A_1})
            \right]
    + \mathcal{L}_{\text{GAN}}(\mathcal{G}_{A \to B}, \mathcal{D}_B) \nonumber \\
    &\quad + \frac{1}{2}\lambda_{\text{cycle}}^{A \to B \to A} \mathcal{L}_{\text{cycle}}^{A \to B \to A} + \lambda_{\text{cycle}}^{B \to A \to B} \mathcal{L}_{\text{cycle}}^{B \to A \to B} \nonumber \\
    &\quad + \frac{1}{2}\lambda_{\text{idt}}^{B \to A} \mathcal{L}_{\text{idt}}^{B \to A} + \lambda_{\text{idt}}^{A \to B} \mathcal{L}_{\text{idt}}^{A \to B}
\end{align}
where the $\lambda$ parameters control the relative importance of cycle consistency and identity constraints for each direction.

Our implementation employs UVCGANv2's hybrid Vision Transformer and Style-Modulated UNet generators, as they have proven to work well for our problem.
We modified the discriminator architecture, however, from a shallow ResNet-based design to a deeper configuration, which provided better signal-background separation performance.
We also increased batch sizes to improve training stability and convergence.
Complete training procedures, hyperparameter settings, and computational requirements are detailed in~\autoref{sec:training_details}.

%% file: figures/naive_failure_v2/naive_failure_v3.tex



    
    \begin{tikzpicture}
        \def\xs{130pt}
        \def\ys{130pt}
        \def\lw{1pt}
    
        \definecolor{myblue}{HTML}{0FA5C7} 
        \definecolor{myorange}{HTML}{F8A300} 
        \definecolor{yellowgreen}{HTML}{A1CB2E} 
        \definecolor{darkgreen}{HTML}{5AAA4F} 
        \definecolor{mypurple}{HTML}{7000AD} 
        
        \definecolor{failure}{HTML}{D9534F} 
        \definecolor{success}{HTML}{5CB85C} 
    
        \tikzset{
            shadowed path/.style={
                postaction={
                    copy shadow={shadow xshift=0.2em, 
                                 shadow yshift=-0.2em, 
                                 opacity=0.2, 
                                 fill opacity=0}
                }
            }
        }
        
        \tikzset{
            block/.style={
                rectangle, 
                rounded corners=5, 
                blur shadow={
                    shadow blur steps=5, 
                    shadow xshift=1pt, 
                    shadow yshift=-1pt
                },
                align=center,
            },
            flow/.style={
                -{Latex[length=4mm, width=4mm]}, 
                rounded corners=20pt, 
                line width=4 * \lw,  
                shadowed path,
                shorten >= 2pt,
                shorten <= 2pt
            },
            flowinfer/.style={
                -{Latex[length=4mm, width=4mm]}, 
                rounded corners=20pt, 
                line width=4 * \lw,
                densely dashed,
                draw=cyan,
                opacity=.8,
                shorten >= 2pt,
                shorten <= 2pt
            },
            flowloss/.style={
                -{Latex[length=3mm, width=3mm]},
                rounded corners=5pt, 
                line width=3 * \lw, 
                densely dotted, 
                draw=black!50,
                shadowed path,
                shorten >= 2pt,
                shorten <= 2pt
            },
        }
    
        \tikzstyle{bs} = [blur shadow={shadow blur steps=5,
                                       shadow blur extra rounding=1pt, 
                                       shadow xshift=1pt,
                                       shadow yshift=-1pt}]
        \tikzstyle{trapezoid} = [inner sep=2,
                                 shape=trapezium, 
                                 trapezium angle=60,
                                 minimum width=30,
                                 anchor=south, 
                                 rounded corners=3pt, 
                                 bs]
        \newcommand{\neuralTranslator}[2]{
            \begin{tikzpicture}
                \node[trapezoid, 
                      draw=#1, 
                      fill=#1!75,
                      rotate=-90, 
                      opacity=#2] (enc) at (0, 0) {};
                \node[inner sep=0,
                      draw=#1, 
                      fill=#1!75,
                      minimum width=4, 
                      minimum height=15, 
                      rounded corners=1pt, 
                      bs, 
                      anchor=west, 
                      opacity=#2] (mid) at ([xshift=2pt]enc.north) {};
                \node[trapezoid, 
                      draw=#1, 
                      fill=#1!75,
                      rotate=90, 
                      anchor=north, 
                      opacity=#2] (dec) at ([xshift=2pt]mid.east) {};
            \end{tikzpicture}
        }
        \newcommand{\mycross}[2]{
            \begin{tikzpicture}
                \node[scale=#1, transform shape] {
                    \begin{tikzpicture}
                        \draw[line width=6pt, draw=#2, shadowed path] (-0.5,-0.5) -- (0.5,0.5);
                        \draw[line width=6pt, draw=#2, shadowed path] (-0.5,0.5) -- (0.5,-0.5);
                    \end{tikzpicture}
                };
            \end{tikzpicture}
        }
        
        \newcommand{\gab}{\mathcal{G}_{A \to B}}
        \newcommand{\gba}{\mathcal{G}_{B \to A}}
        
        \def\cyclexshift{1.5*\xs}
        \def\myscale{.45}
        \def\folder{figures/cyclegan}
        \def\figscale{.14}

        \node[inner sep=2] {
            \begin{tikzpicture}
                \node[scale=\myscale, transform shape, anchor=west] (A) at (0, 0) {
                    \begin{tikzpicture}
                        \node[inner sep=0, anchor=north] (embedded) at (0, 0) {\includegraphics[scale=\figscale]{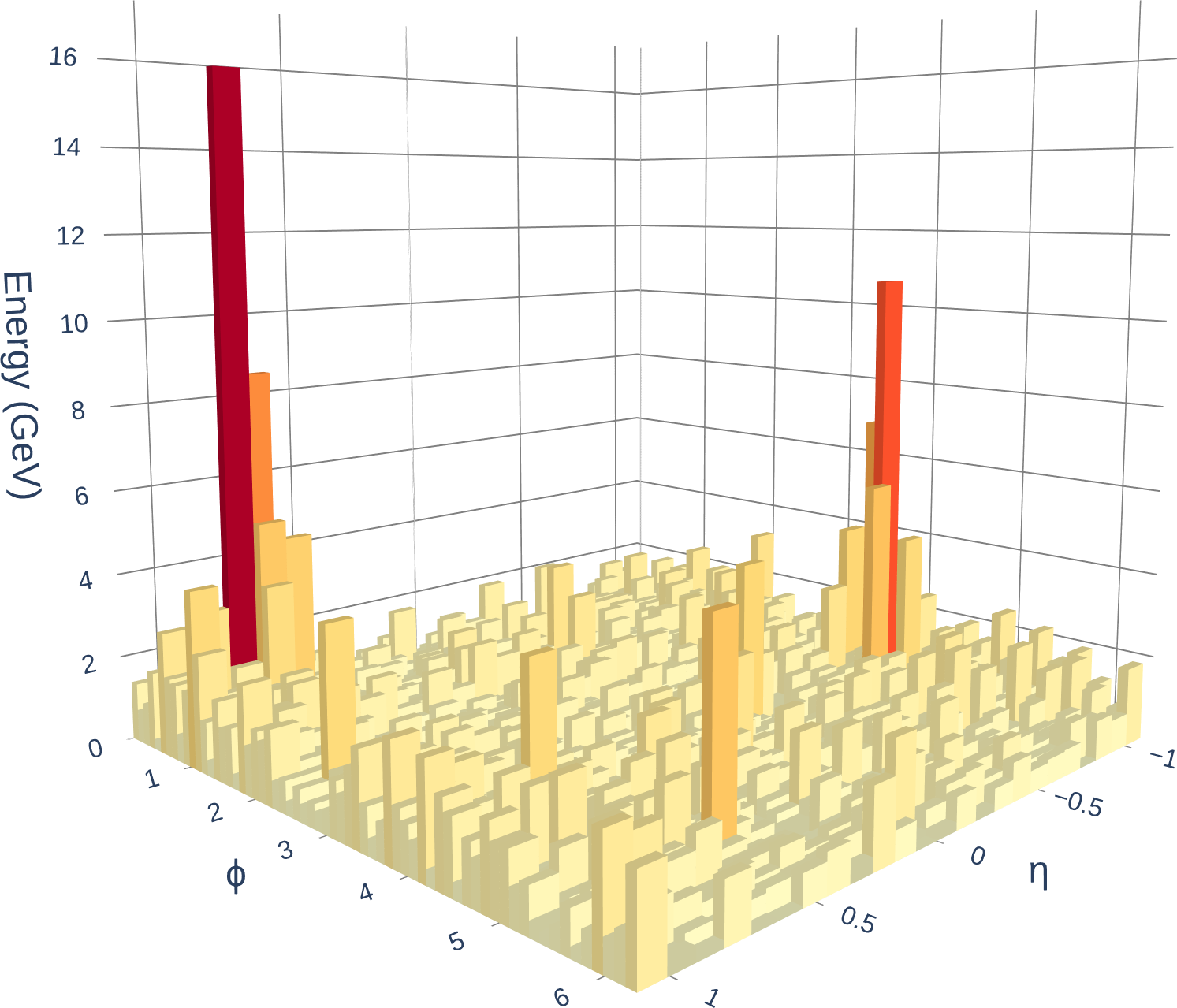}};
                        \node[block, inner sep=5, anchor=west, fill=success!20] (sgn) at ([xshift=.6*\xs]embedded.east) {\includegraphics[scale=\figscale]{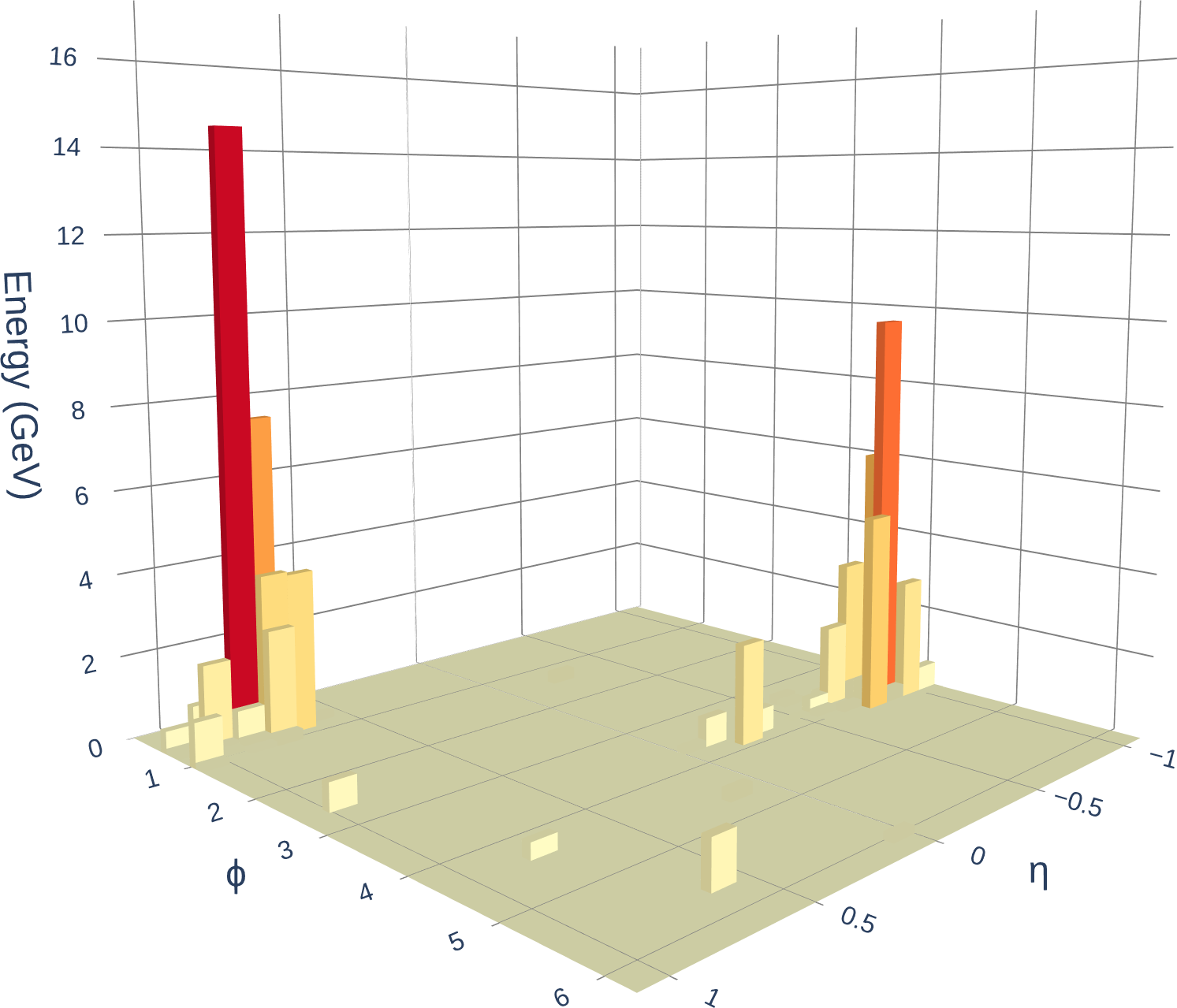}};
                        \node[block, inner sep=5, anchor=south west, fill=success!70!black, text=white, align=center] (comment) at ([yshift=-.08 * \ys]sgn.north west) {\Large \textbf{Effective signal extraction}\\\large and background removal};
                        \node[block, anchor=west, inner xsep=.1 * \xs, inner ysep=.6 * \ys, fill=failure!70!black, rounded corners=1] (embedded_recon) at ([xshift=.6*\xs]sgn.east) {};
                        
                        \node[inner sep=0, anchor=center] (gba_icon) at ($(embedded.east)!.5!(sgn.west)$) {\neuralTranslator{yellowgreen}{1}};
                        \node[inner sep=2, anchor=south] (gba_label) at (gba_icon.north) {\Large$\gba$};
                        \node[inner sep=0, anchor=center] (gab_icon) at ($(sgn.east)!.5!(embedded_recon.west)$) {\neuralTranslator{darkgreen}{.5}};
                        \node[inner sep=2, anchor=south] (gab_label) at (gab_icon.north) {\Large$\gab$};
                        \node[inner sep=2, anchor=center] at ([yshift=-.08*\ys]gab_icon) {\mycross{1}{red}};

                        \draw[flow, draw=black!75] (embedded) -- (gba_icon);
                        \draw[flow, draw=black!75] (gba_icon) -- (sgn);
                        \draw[flow, draw=black!30] (sgn) -- (gab_icon);
                        \draw[flow, draw=black!30] (gab_icon) -- (embedded_recon);

                        \node[inner sep=0, rotate=90, anchor=center, text=white] at (embedded_recon) {\Large Cycle-consistency};
                    \end{tikzpicture}
                };
                \node[inner xsep=0, inner ysep=5, anchor=south west] at (A.north west) {\textbf{(a)} Difficulty of naive CycleGAN to achieve effective signal extraction};

                \node[scale=\myscale, transform shape, anchor=north west] (B) at ([yshift=-.18*\ys]A.south west) {
                    \begin{tikzpicture}
                        \node[inner sep=0, anchor=north] (embedded) at (0, 0) {\includegraphics[scale=\figscale]{figures/naive_failure_v2/event-309_mix_narrow.png}};
                        \node[block, inner sep=5, anchor=west, fill=failure!20] (sgn) at ([xshift=.6*\xs]embedded.east) {\includegraphics[scale=\figscale]{figures/naive_failure_v2/event-309_mix_narrow.png}};
                        \node[block, inner sep=5, anchor=south west, fill=failure!70!black, text=white] (comment) at ([yshift=-.05 * \ys]sgn.north west) {\Large \textbf{No signal extraction!}};
                        \node[inner sep=0, anchor=west] (embedded_recon) at ([xshift=.6*\xs]sgn.east) {\includegraphics[scale=\figscale]{figures/naive_failure_v2/event-309_mix_narrow.png}};
                        
                        \node[inner sep=0, anchor=center] (gba_icon) at ($(embedded.east)!.5!(sgn.west)$) {\neuralTranslator{yellowgreen}{1}};
                        \node[inner sep=2, anchor=south] at (gba_icon.north) {\Large$\gba$};
                        \node[inner sep=2, anchor=north] at (gba_icon.south) {\Large$\approx\text{Identity}$};
                        \node[inner sep=0, anchor=center] (gab_icon) at ($(sgn.east)!.5!(embedded_recon.west)$) {\neuralTranslator{darkgreen}{1}};
                        \node[inner sep=2, anchor=south] at (gab_icon.north) {\Large$\gab$};
                        \node[inner sep=2, anchor=north] at (gab_icon.south) {\Large$\approx\text{Identity}$};

                        \draw[flow, draw=black!75] (embedded) -- (gba_icon);
                        \draw[flow, draw=black!75] (gba_icon) -- (sgn);
                        \draw[flow, draw=black!75] (sgn) -- (gab_icon);
                        \draw[flow, draw=black!75] (gab_icon) -- (embedded_recon);
                    \end{tikzpicture}
                };
                \node[inner xsep=0, inner ysep=5, anchor=south west] at (B.north west) {\textbf{(b)} Failure mode of naive CycleGAN};
            \end{tikzpicture}
        };
    \end{tikzpicture}

%% file: figures/stratified_arch/stratified_arch.tex



        
    \begin{tikzpicture}
    
        \newcommand{\fake}[1]{{#1}^{\textrm{f}}}
        \newcommand{\real}[1]{{#1}^{\textrm{r}}}
        \newcommand{\iden}[1]{#1_{\textrm{i}}}
        \newcommand{\gab}{\mathcal{G}_{A \to B}}
        \newcommand{\gba}{\mathcal{G}_{B \to A}}
        \newcommand{\dabkg}{\mathcal{D}_{A_0}}
        \newcommand{\dasgn}{\mathcal{D}_{A_1}}
        \newcommand{\db}{\mathcal{D}_{B}}
        \newcommand{\paren}[1]{\left(#1\right)}
        
        \def\xs{120pt}
        \def\ys{120pt}
        \def\lw{1pt}
    
        \definecolor{myblue}{HTML}{0FA5C7} 
        \definecolor{myorange}{HTML}{F8A300} 
        \definecolor{yellowgreen}{HTML}{A1CB2E} 
        \definecolor{darkgreen}{HTML}{5AAA4F} 
        \definecolor{mypurple}{HTML}{7000AD} 
    
        \tikzset{
            shadowed path/.style={
                postaction={
                    copy shadow={shadow xshift=0.2em, 
                                 shadow yshift=-0.2em, 
                                 opacity=0.2, 
                                 fill opacity=0}
                }
            }
        }
        
        \tikzset{
            block/.style={
                rectangle, 
                rounded corners=2, 
                blur shadow={
                    shadow blur steps=5, 
                    shadow xshift=1pt, 
                    shadow yshift=-1pt
                },
                align=center,
            },
            data/.style={
                block,
                minimum size=.25 * \xs
            },
            realdata/.style={
                data,
                fill=myorange!80,
                draw=myorange,
            },
            fakedata/.style={
                rectangle, 
                rounded corners=2,
                align=center,
                minimum size = .25 * \xs,
                pattern=dots,
                pattern color=myorange!80,
                draw=myorange,
            },
            recodata/.style={
                rectangle, 
                rounded corners=2,
                align=center,
                minimum size = .25 * \xs,
                pattern=crosshatch,
                pattern color=myorange!80,
                draw=myorange,
            },
            gen/.style={
                block, 
                rounded corners=10,
                minimum width=.3 * \xs,
                minimum height=.2 * \ys,
                text=black, 
                inner sep=10pt,
                font=\Large,
            },
            genab/.style={
                gen,
                draw=darkgreen!100!black!80, 
                fill=darkgreen!80, 
            },
            genba/.style={
                gen,
                draw=yellowgreen!100!black!80, 
                fill=yellowgreen!80, 
            },
            dis/.style={
                block, 
                rounded corners=10,
                minimum width=.3 * \xs,
                minimum height=.2 * \ys, 
                text=white, 
                draw=myblue!100!black!80, 
                fill=myblue, 
                inner sep=10pt,
                font=\Large,
            },
            loss/.style={
                block, 
                inner sep=4pt, 
                execute at begin node=\setlength{\baselineskip}{3ex}
            },
            genloss/.style={
                loss, 
                draw=myorange!50!black!40, 
                fill=myorange!40, 
            },
            consistloss/.style={
                loss, 
                draw=mypurple!50!black!40, 
                fill=mypurple!40, 
                text opacity=1.,
            },
            disloss/.style={
                loss, 
                draw=myblue!50!black!40, 
                fill=myblue!40, 
            },
            flow/.style={
                -{Latex[length=4mm, width=4mm]}, 
                rounded corners=10pt, 
                line width=4 * \lw, 
                draw=black!75, 
                shadowed path,
                shorten >= 2pt,
                shorten <= 2pt
            },
            flowinfer/.style={
                -{Latex[length=4mm, width=4mm]}, 
                rounded corners=10pt, 
                line width=4 * \lw,
                densely dashed,
                draw=cyan,
                opacity=.8,
                shorten >= 2pt,
                shorten <= 2pt
            },
            flowloss/.style={
                -{Latex[length=3mm, width=3mm]},
                rounded corners=5pt, 
                line width=3 * \lw, 
                densely dotted, 
                draw=black!50,
                shadowed path,
                shorten >= 2pt,
                shorten <= 2pt
            },
            textnode/.style={
                inner sep=1pt,
            }
        }
        \newcommand{\makefakenode}[1]{
            \begin{tikzpicture}
                \node[data, fill=myorange!40] at (0, 0) {};
                \node[fakedata] at (0, 0) {#1};
            \end{tikzpicture}
        }
        \newcommand{\makereconode}[1]{
            \begin{tikzpicture}
                \node[data, fill=myorange!40] at (0, 0) {};
                \node[recodata] at (0, 0) {#1};
            \end{tikzpicture}
        }
        
        \newcommand{\makeidtnode}[1]{
            \begin{tikzpicture}
                \node[data, fill=myorange!40] at (0, 0) {};
                \node[recodata] at (0, 0) {#1};
            \end{tikzpicture}
        }
        
        \def\cyclexshift{1.5*\xs}
        \node[inner sep=5] {
        \begin{tikzpicture}
            \def\myscale{.57}
            \node[inner sep=0, scale=\myscale, transform shape] (architecture) at (0, 0) {
                \begin{tikzpicture}
                    \node[realdata] (realbkg) at (0, 0) {real\\bkg.};
                    \node[realdata, anchor=south west] (realsgn) at ([xshift=.2 * \xs,yshift=.2*\ys]realbkg.north east) {real\\sgn.};
                    \node[inner sep=0, anchor=west] (fakebkg) at ([xshift=\cyclexshift]realbkg.east) {\makefakenode{fake\\bkg.}};
                    \node[inner sep=0, anchor=south west] (fakesgn) at ([xshift=.2 * \xs,yshift=.2*\ys]fakebkg.north east) {\makefakenode{fake\\sgn.}};
        
                    \node[inner sep=0] (recobkg) at ([xshift=-\cyclexshift]realbkg) {\makereconode{reco\\bkg.}};
                    \node[inner sep=0, anchor=south west] (recosgn) at ([xshift=.2 * \xs,yshift=.2*\ys]recobkg.north east) {\makereconode{reco\\sgn.}};
                    
                    \coordinate (D) at ($(realbkg)!.5!(realsgn)$);
                    \node[genab, anchor=north] (gab) at ([yshift=-.6*\ys]realbkg.south -| D) {$\gab$};
                    \coordinate (D) at ($(fakebkg)!.5!(fakesgn)$);
                    \node[genba, anchor=north] (gba) at ([yshift=-.6*\ys]fakebkg.south -| D) {$\gba$};
                    \node[genba] (gba_cyc) at ([xshift=-\cyclexshift]gab) {$\gba$};
                    \node[genab] (gab_cyc) at ([xshift=\cyclexshift]gba) {$\gab$};
        
                    \coordinate (D) at ($(realbkg.south)!.5!(gab.north)$);
                    \node[circle, draw=black, inner sep=1] (concat) at (D -| gab) {$\boldsymbol{\oplus}$};
                    \draw[flow] (realbkg.south) |- (concat);
                    \draw[flow] (realsgn.south) |- (concat);
                    \draw[flow] (concat) -- (gab);
        
                    \tikzmath{coordinate \L; \L=(fakebkg.south)-(gba.north);}
                    \coordinate (D) at ($(fakebkg.north)!.5!(fakesgn.south)$);
                    \node[circle, draw=black, inner sep=1] (concat_cyc) at ([xshift=\Ly/2]D -| fakesgn.east) {$\boldsymbol{\oplus}$};
                    \draw[flow] (fakesgn.east) -| (concat_cyc.north);
                    \draw[flow] (fakebkg.east) -| (concat_cyc.south);
                    \draw[flow, rounded corners=30] (concat_cyc) -| (gab_cyc);
        
                    \coordinate (D) at ($(fakebkg.south)!.5!(gba.north)$);
                    \node[circle, draw=black, inner sep=1] (split) at (D -| gba) {$\boldsymbol{\ominus}$};
                    \draw[flow] (split) -| (fakebkg.south);
                    \draw[flow] (split) -| (fakesgn.south);
                    \draw[flowinfer] (split) -| (fakesgn.south);
                    \draw[flow] (gba) -- (split);
                    \draw[flowinfer] (gba) -- (split);
        
                    \node[circle, draw=black, inner sep=1] (split_cyc) at (concat -| gba_cyc) {$\boldsymbol{\ominus}$};
                    \draw[flow] (split_cyc) -| (recobkg.south);
                    \draw[flow] (split_cyc) -| (recosgn.south);
                    \draw[flow] (gba_cyc) -- (split_cyc);
        
                    \node[inner sep=0, anchor=north] (fakemix) at ([yshift=-.25*\ys]gab.south) {\makefakenode{fake\\mix.}};
                    \node[realdata] (realmix) at (gba |- fakemix) {real\\mix.};
                    \node[inner sep=0] (recomix) at (realmix -| gab_cyc) {\makereconode{reco.\\mix.}};
                    \draw[flow] (gab) -- (fakemix);
                    \draw[flow] (realmix.north) -- (gba.south);
                    \draw[flowinfer] (realmix.north) -- (gba.south);
                    \draw[flow, rounded corners=30] (fakemix) -| (gba_cyc);
                    \draw[flow] (gab_cyc) |- (recomix.north);
        
                    \node[consistloss, align=center] (cyclelossmix) at ($(realmix)!.5!(recomix)$) {Cycle-\\consistency\\loss};
                    \node[consistloss, align=center] (cyclelossbkg) at ($(realbkg)!.5!(recobkg)$) {Cycle-\\consistency\\loss};
                    \node[consistloss, align=center] (cyclelosssgn) at ($(realsgn)!.5!(recosgn)$) {Cycle-\\consistency\\loss};
                    \draw[flowloss] (recomix) -- (cyclelossmix);
                    \draw[flowloss] (realmix) -- (cyclelossmix);
                    \draw[flowloss] (recobkg) -- (cyclelossbkg);
                    \draw[flowloss] (realbkg) -- (cyclelossbkg);
                    \draw[flowloss] (recosgn) -- (cyclelosssgn);
                    \draw[flowloss] (realsgn) -- (cyclelosssgn);
                    
                    \node[dis] (disbkg) at ($(realbkg)!.5!(fakebkg)$) {$\dabkg$};
                    \node[dis] (dissgn) at ($(realsgn)!.5!(fakesgn)$) {$\dasgn$};
                    \node[dis] (dismix) at ($(realmix)!.5!(fakemix)$) {$\db$};
                    \draw[flow] (fakemix) -- (dismix);
                    \draw[flow] (realmix) -- (dismix);
                    \draw[flow] (fakebkg) -- (disbkg);
                    \draw[flow] (realbkg) -- (disbkg);
                    \draw[flow] (fakesgn) -- (dissgn);
                    \draw[flow] (realsgn) -- (dissgn);
        
                    \def\radius{.25*\xs}
                    \def\degree{330}
                    \draw[flow, draw=black!25, densely dotted] ([xshift=-\cyclexshift / 2 + \radius,yshift=.05*\ys]gab) arc[start angle=0, end angle=-\degree, radius=\radius];
                    \draw[flow, draw=black!25, densely dotted] ([xshift=\cyclexshift / 2 - \radius,yshift=.05*\ys]gba) arc[start angle=180, end angle=180 - \degree, radius=\radius];

                    \tikzmath{coordinate \C;\C=(fakesgn.north east)-(fakebkg.west |- realmix.south);}
                    \def\inferxsep{.58 * \Cx}
                    \def\inferysep{.55 * \Cy}
                    \begin{pgfonlayer}{back}
                        \def\op{.3}
                        \node[
                            fill=myblue,
                            inner xsep=\inferxsep, 
                            inner ysep=\inferysep, 
                            rounded corners=15, 
                            draw opacity=\op,
                            fill opacity=\op,
                        ] (inference) at ([yshift=.52*\Cy]realmix.south) {};
                        \node[inner sep=3, anchor=north] at (inference.north) {\large Inference};
                    \end{pgfonlayer}
                    
                \end{tikzpicture}
            };
            \node[inner sep=0, anchor=south west] at ([yshift=.1 * \ys]architecture.north west) {\textbf{(a)} \ourmodel Architecture (identity losses shown separately below)};

            \node[inner sep=0, anchor=north west, scale=\myscale, transform shape] (idt) at ([yshift=-.2*\ys]architecture.south west) {
                \begin{tikzpicture}
                    \def\idtxs{.2*\xs}
                    \node[realdata, anchor=north west] (realmix_idt) at (0, 0) {real\\mix.};
                    \node[realdata, anchor=north] (zero) at ([yshift=-.2*\ys]realmix_idt.south) {$\mathbf{0}$};
                    \coordinate (D) at ($(realmix_idt.south east)!.5!(zero.north east)$);
                    \node[circle, draw=black, inner sep=1, anchor=west] (concat_idt) at ([xshift=\idtxs]D) {$\boldsymbol{\oplus}$};
                    \node[genab, anchor=west] (gab_idt) at ([xshift=\idtxs]concat_idt.east) {$\gab$};
                    
                    \node[inner sep=0, anchor=west] (idtmix) at ([xshift=\idtxs]gab_idt.east) {\makeidtnode{idt\\mix.}};
                    \draw[flow] (realmix_idt) -| (concat_idt);
                    \draw[flow] (zero) -| (concat_idt);
                    \draw[flow] (concat_idt) -- (gab_idt);
                    \draw[flow] (gab_idt) -- (idtmix);

                    \coordinate (D) at ($(realmix_idt)!.5!(idtmix)$);
                    \node[consistloss, align=center, anchor=north] (idtlossmix) at ([yshift=.45*\ys]D) {Identity loss};
                    
                    \draw[flowloss, rounded corners=15] (realmix_idt) |- (idtlossmix);
                    \draw[flowloss, rounded corners=15] (idtmix) |- (idtlossmix);

                    \node[realdata, anchor=west] (realbkg_idt) at ([xshift=1.8*\xs]realmix_idt.east) {real\\bkg.};
                    \node[realdata, anchor=west] (realsgn_idt) at ([xshift=1.8*\xs]zero.east) {real\\sgn.};
                    \coordinate (D) at ($(realsgn_idt.south east)!.5!(realbkg_idt.north east)$);
                    \node[circle, draw=black, inner sep=1, anchor=west] (add_idt) at ([xshift=\idtxs]D) {$\boldsymbol{+}$};
                    \node[genba, anchor=west] (gba_idt) at ([xshift=\idtxs]add_idt.east) {$\gba$};
                    \node[circle, draw=black, inner sep=1, anchor=west] (split_idt) at ([xshift=\idtxs]gba_idt.east) {$\boldsymbol{\ominus}$};
                    \coordinate (D) at ([xshift=\idtxs]split_idt.east);
                    \node[inner sep=0, anchor=west] (idt_bkg) at (D |- realmix_idt) {\makeidtnode{idt.\\bkg.}};
                    \node[inner sep=0, anchor=west] (idt_sgn) at (D |- zero) {\makeidtnode{idt.\\sgn.}};
                    
                    \draw[flow] (realbkg_idt) -| (add_idt);
                    \draw[flow] (realsgn_idt) -| (add_idt);
                    \draw[flow] (add_idt) -- (gba_idt);
                    \draw[flow] (gba_idt) -- (split_idt);
                    \draw[flow] (split_idt) |- (idt_bkg);
                    \draw[flow] (split_idt) |- (idt_sgn);

                    \node[consistloss, align=center, anchor=center] (idtlossbkg) at (gba_idt |- idtlossmix) {Identity loss};
                    \tikzmath{coordinate \C;\C=(idtlossbkg)-(gba_idt);}
                    \node[consistloss, align=center, anchor=center] (idtlosssgn) at ([yshift=-\Cy]gba_idt) {Identity loss};
                    \draw[flowloss, rounded corners=15] (realbkg_idt) |- (idtlossbkg);
                    \draw[flowloss, rounded corners=15] (idt_bkg) |- (idtlossbkg);
                    \draw[flowloss, rounded corners=15] (realsgn_idt) |- (idtlosssgn);
                    \draw[flowloss, rounded corners=15] (idt_sgn) |- (idtlosssgn);
                \end{tikzpicture}
            };
            \node[inner sep=0, anchor=south west] at ([yshift=.1 * \ys]idt.north west) {\textbf{(b)}  Identity Losses};

            \node[
                inner sep=5,
                rounded corners=3,
                blur shadow,
                fill=myblue!30!black,
                anchor=north west, 
                scale=\myscale, 
                transform shape,
                opacity=.1,
                text opacity=1,
                font=\fontsize{13}{13}\selectfont
            ] (legend) at ([yshift=-.2*\ys]idt.south west) {
                \begin{tikzpicture}
                    \def\xsp{.05 * \xs}
                    \def\ysp{.15 * \ys}
                    \coordinate (R1) at (0, 0);
                    \coordinate (R2) at (0, -1.4 * \ysp);
                    \coordinate (R3) at (0, -2.5 * \ysp);
                    
                    \node[inner sep=0] (realbkg_t) at (R1) {real bkg.:};
                    \node[inner sep=0, anchor=west] (realbkg_n) at ([xshift=\xsp]realbkg_t.east) {$a_0$};
                    \node[inner sep=0, anchor=west] (realsgn_t) at (realbkg_t.west |- R2) {real sgn.:};
                    \node[inner sep=0, anchor=west] (realsgn_n) at (realbkg_n.west |- R2) {$a_1$};
                    \node[inner sep=0, anchor=west] (realmix_t) at (realsgn_t.west |- R3) {real mix.:};
                    \node[inner sep=0, anchor=west] (realmix_n) at (realbkg_n.west |- R3) {$b$};
                    
                    \node[inner sep=0, anchor=west] (fakebkg_t) at ([xshift=2 * \xsp]realbkg_n.east) {fake bkg.:};
                    \node[inner sep=0, anchor=west] (fakebkg_n) at ([xshift=\xsp]fakebkg_t.east) {$\Pi_{A_0}\paren{\gba\paren{b}}$};
                    \node[inner sep=0, anchor=west] (fakesgn_t) at (fakebkg_t.west |- realsgn_n) {fake sgn.:};
                    \node[inner sep=0, anchor=west] (fakesng_n) at (fakebkg_n.west |- realsgn_t) {$\Pi_{A_1}\paren{\gba\paren{b}}$};
                    \node[inner sep=0, anchor=west] (fakemix_t) at (fakebkg_t.west |- realmix_n) {fake mix.:};
                    \node[inner sep=0, anchor=west] (fakemix_n) at (fakebkg_n.west |- realmix_t) {$\gab\paren{a_0, a_1}$};
                    
                    \node[inner sep=0, anchor=west] (recobkg_t) at ([xshift=2 * \xsp]fakebkg_n.east) {reco bkg.:};
                    \node[inner sep=0, anchor=west] (recobkg_n) at ([xshift=\xsp]recobkg_t.east) {$\Pi_{A_0}\paren{\gba\paren{\gab\paren{a_0, a_1}}}$};
                    \node[inner sep=0, anchor=west] (recosgn_t) at (recobkg_t.west |- R2) {reco sgn.:};
                    \node[inner sep=0, anchor=west] (recosng_n) at (recobkg_n.west |- R2) {$\Pi_{A_1}\paren{\gba\paren{\gab\paren{a_0, a_1}}}$};
                    \node[inner sep=0, anchor=west] (recomix_t) at (recobkg_t.west |- R3) {reco mix.:};
                    \node[inner sep=0, anchor=west] (recomix_n) at (recobkg_n.west |- R3) {$\gab\paren{\Pi_{A_0}\paren{\gba(b)}, \Pi_{A_1}\paren{\gba(b)}}$};

                    \def\ysblock{4 * \ysp}
                    \node[inner sep=0, anchor=west] (idtbkg_t) at ([yshift=-\ysblock]realbkg_t.west) {idt bkg.:};
                    \node[inner sep=0, anchor=west] (idtbkg_n) at ([xshift=\xsp]idtbkg_t.east) {$\Pi_{A_0}\paren{\gba\paren{a_0 + a_1}}$};
                    \node[inner sep=0, anchor=west] (idtsgn_t) at ([yshift=-\ysblock]idtbkg_t.west |- R2) {idt sgn.:};
                    \node[inner sep=0, anchor=west] (idtsng_n) at ([yshift=-\ysblock]idtbkg_n.west |- R2) {$\Pi_{A_1}\paren{\gba\paren{a_0 + a_1}}$};
                    \node[inner sep=0, anchor=west] (idtmix_t) at ([yshift=-\ysblock]idtbkg_t.west |- R3) {idt mix.:};
                    \node[inner sep=0, anchor=west] (idtmix_n) at ([yshift=-\ysblock]idtbkg_n.west |- R3) {$\gab\paren{b, 0}$};

                    \node[circle, inner sep=1, anchor=west, draw=black] (concat_t) at ([xshift=2 * \xsp]idtbkg_n.east) {$\boldsymbol{\oplus}$};
                    \node[inner sep=0, anchor=west] (concat_n) at ([xshift=\xsp]concat_t.east) {concatenate along the channel dimension};
                    \node[circle, inner sep=1, anchor=west, draw=black] (split_t) at ([yshift=-\ysblock]concat_t.west |- R2) {$\boldsymbol{\ominus}$};
                    \node[inner sep=0, anchor=west] (split_n) at ([yshift=-\ysblock]concat_n.west |- R2) {split along the channel dimension};
                    \node[circle, inner sep=1, anchor=west, draw=black] (add_t) at ([yshift=-\ysblock]concat_t.west |- R3) {$\boldsymbol{+}$};
                    \node[inner sep=0, anchor=west] (add_n) at ([yshift=-\ysblock]concat_n.west |- R3) {addition};

                    \coordinate (flow_tail) at ([xshift=2 * \xsp]concat_n.east |- concat_t);
                    \coordinate (flow_head) at ([xshift=5 * \xsp]flow_tail);
                    \draw[flow] (flow_tail) -- (flow_head);
                    \node[inner sep=0, anchor=west] at ([xshift=\xsp]flow_head.east) {training data flow};
                    
                    \coordinate (infer_tail) at (split_t -| flow_tail.east);
                    \coordinate (infer_head) at (infer_tail -| flow_head);
                    \draw[flow] (infer_tail) -- (infer_head);
                    \draw[flowinfer] (infer_tail) -- (infer_head);
                    \node[inner sep=0, anchor=west] at ([xshift=\xsp]infer_head.east) {inference data flow};
                    
                    \coordinate (loss_tail) at (add_t -| flow_tail.east);
                    \coordinate (loss_head) at (loss_tail -| flow_head);
                    \draw[flowloss] (loss_tail) -- (loss_head);
                    \node[inner sep=0, anchor=west] at ([xshift=\xsp]loss_head.east) {loss data flow};
                \end{tikzpicture}
            };
        \end{tikzpicture}
        };
    \end{tikzpicture}

%% file: src/appendix.tex

\section{\ourmodel Performance}\label{app:perf}

\subsection{Position Response}\label{app:posres}
\begin{figure}[ht!]
    \centering
    \includegraphics[width=0.43\textwidth]{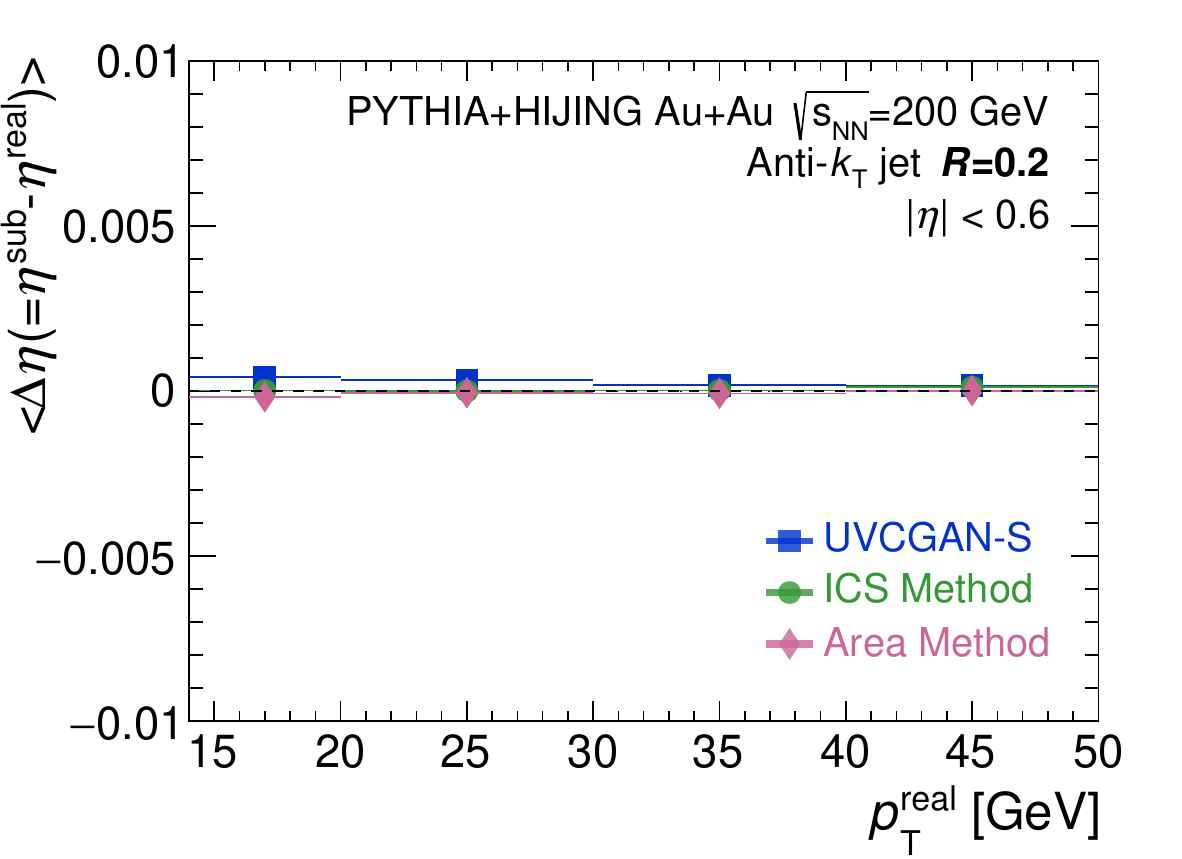}%
    \includegraphics[width=0.43\textwidth]{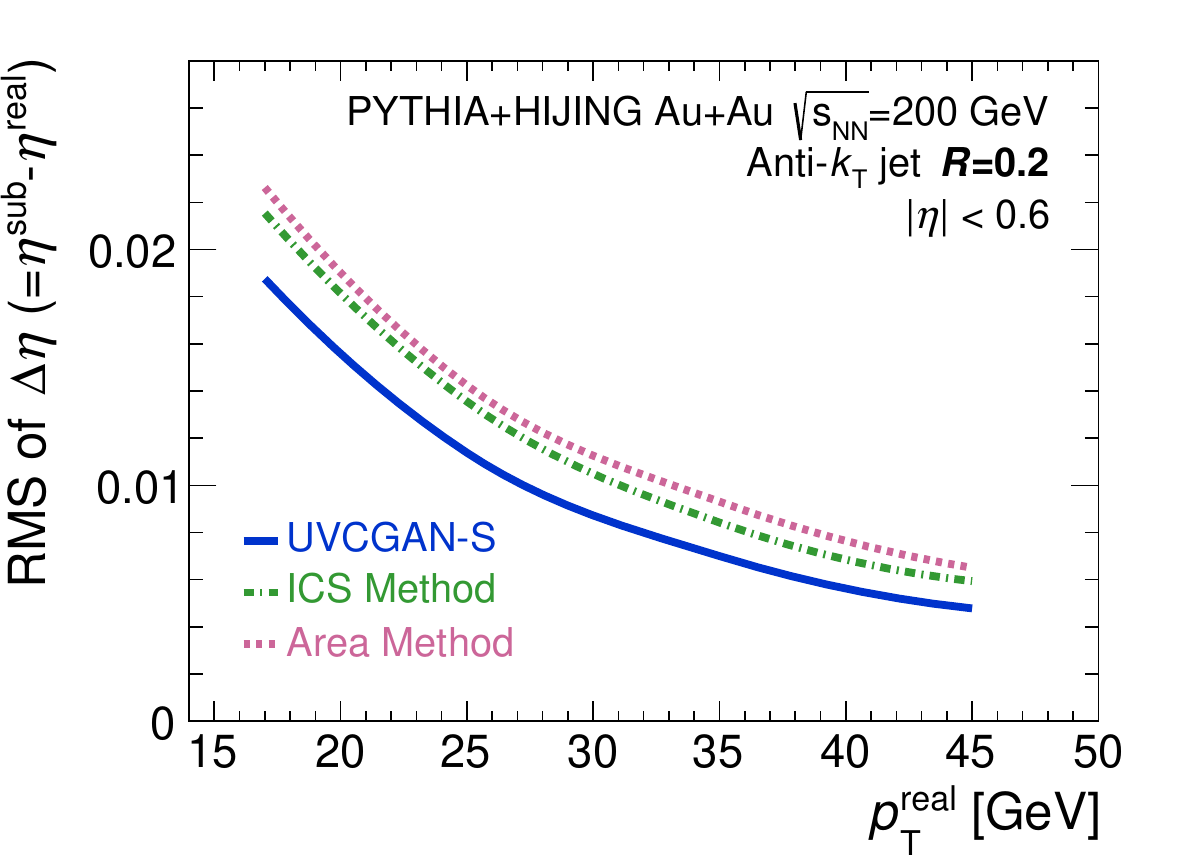}\\
    \includegraphics[width=0.43\textwidth]{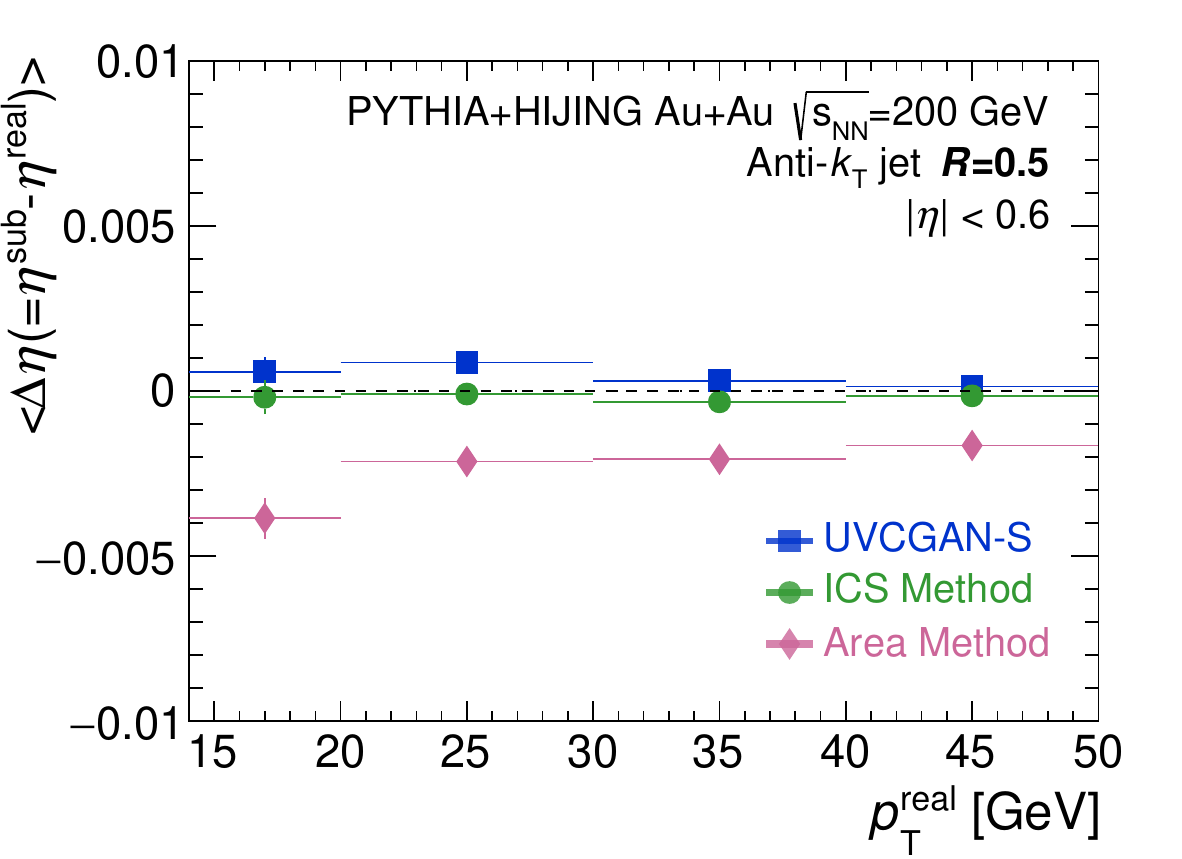}%
    \includegraphics[width=0.43\textwidth]{figures/results/jet_deta_rms_vs_refPt_v8_PythiaPlusHijing_paperFigure0_refpp_Cent0to10_Eta0p00to0p60_refJetPP_R5.pdf}%
    \caption{
        Mean (Left) and RMS (Right) of jet $\eta$ position difference ($\Delta\eta=\eta^{\rm sub}-\eta^{\rm real}$) distributions as a function of jet \pt for $R=0.2$ (Top) and $R=0.5$ (Bottom) in \pythiahijing.}
    \label{fig:detaDist_app}
\end{figure}

\begin{figure}[ht!]
    \centering
    \includegraphics[width=0.43\textwidth]{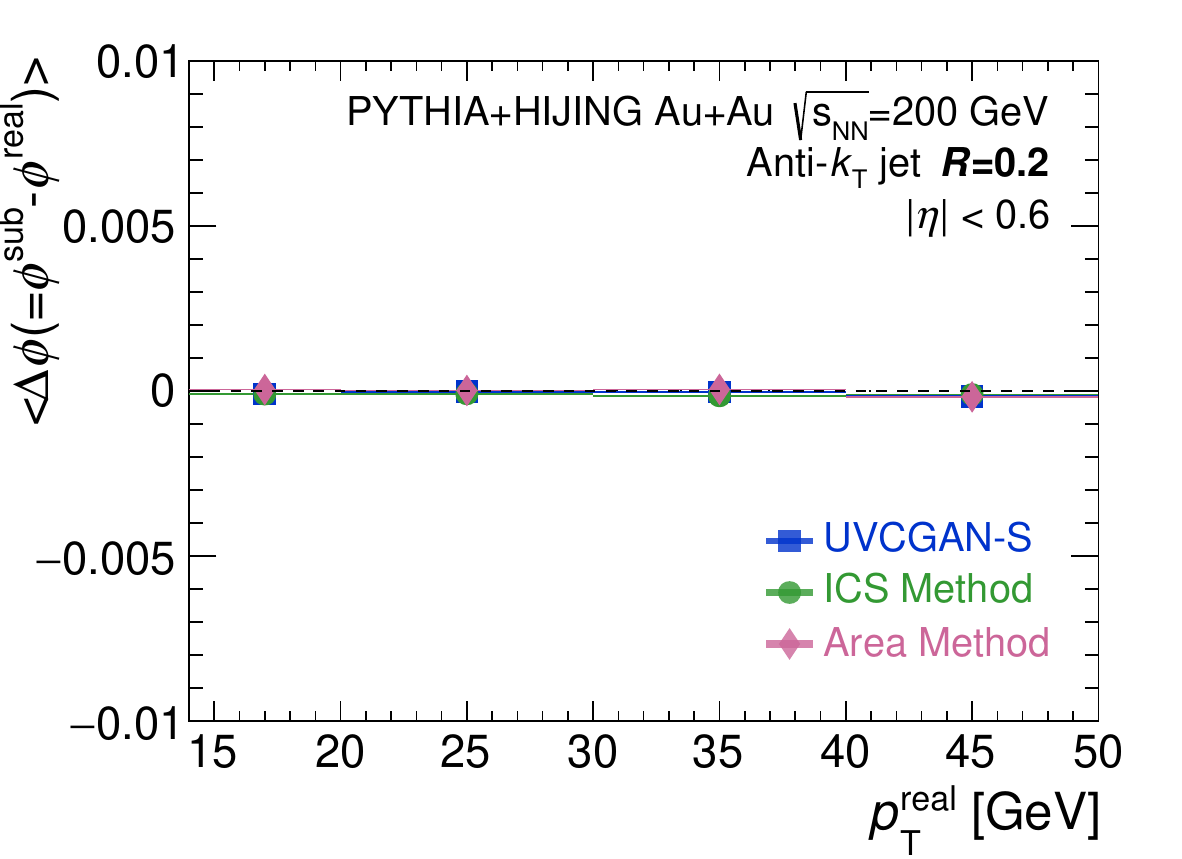}%
    \includegraphics[width=0.43\textwidth]{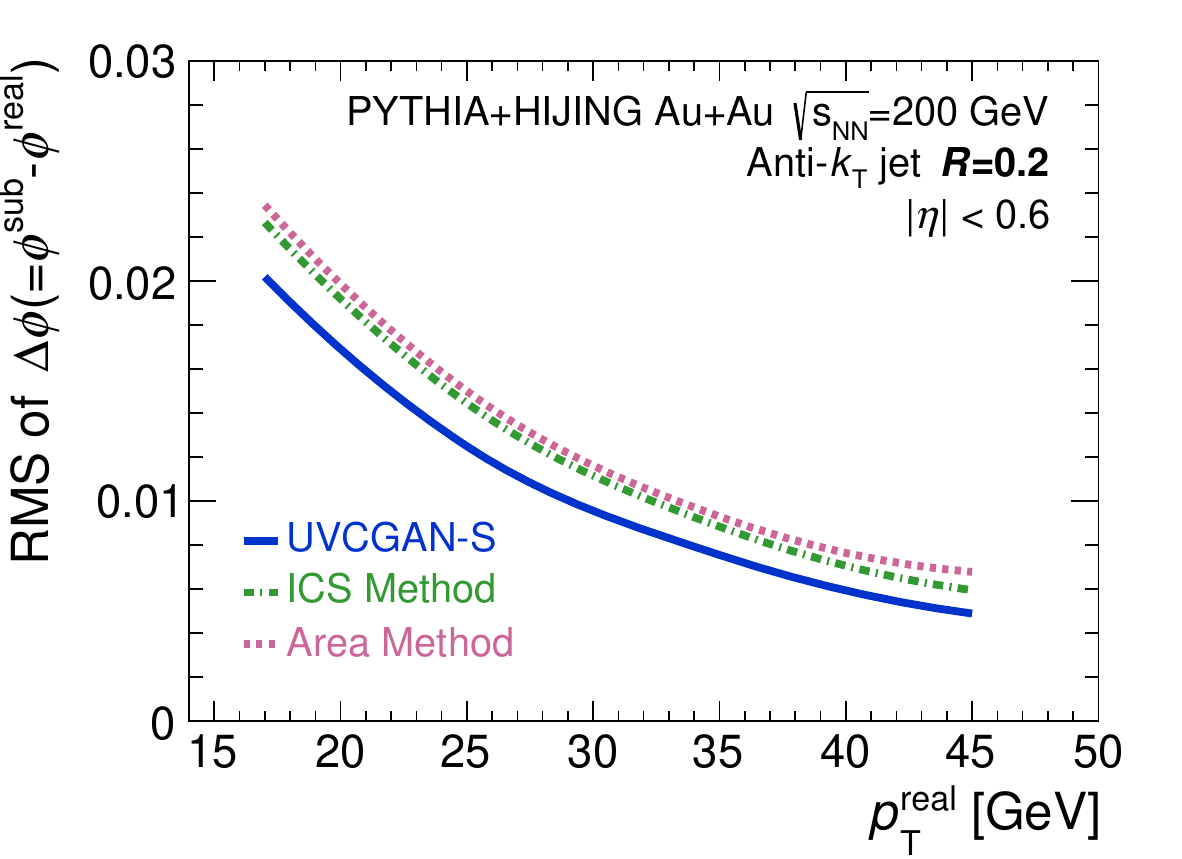}\\
    \includegraphics[width=0.43\textwidth]{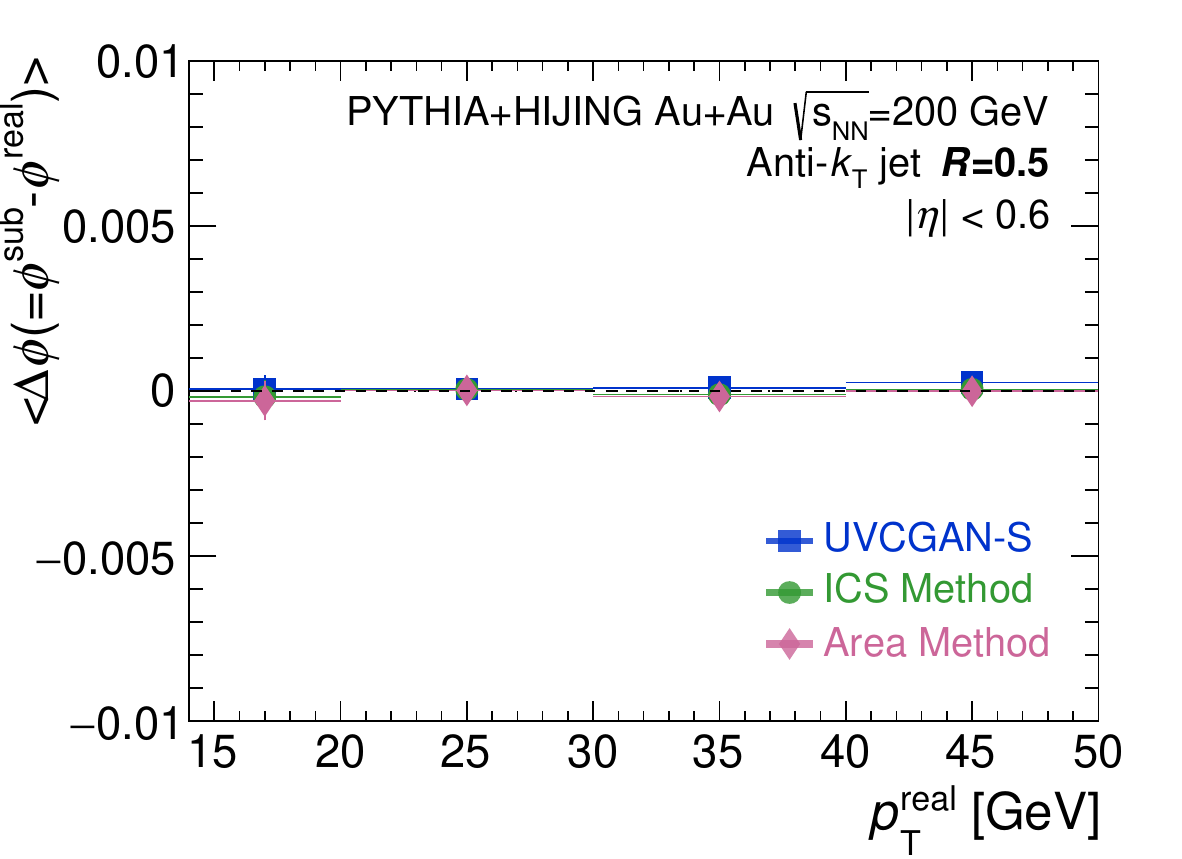}%
    \includegraphics[width=0.43\textwidth]{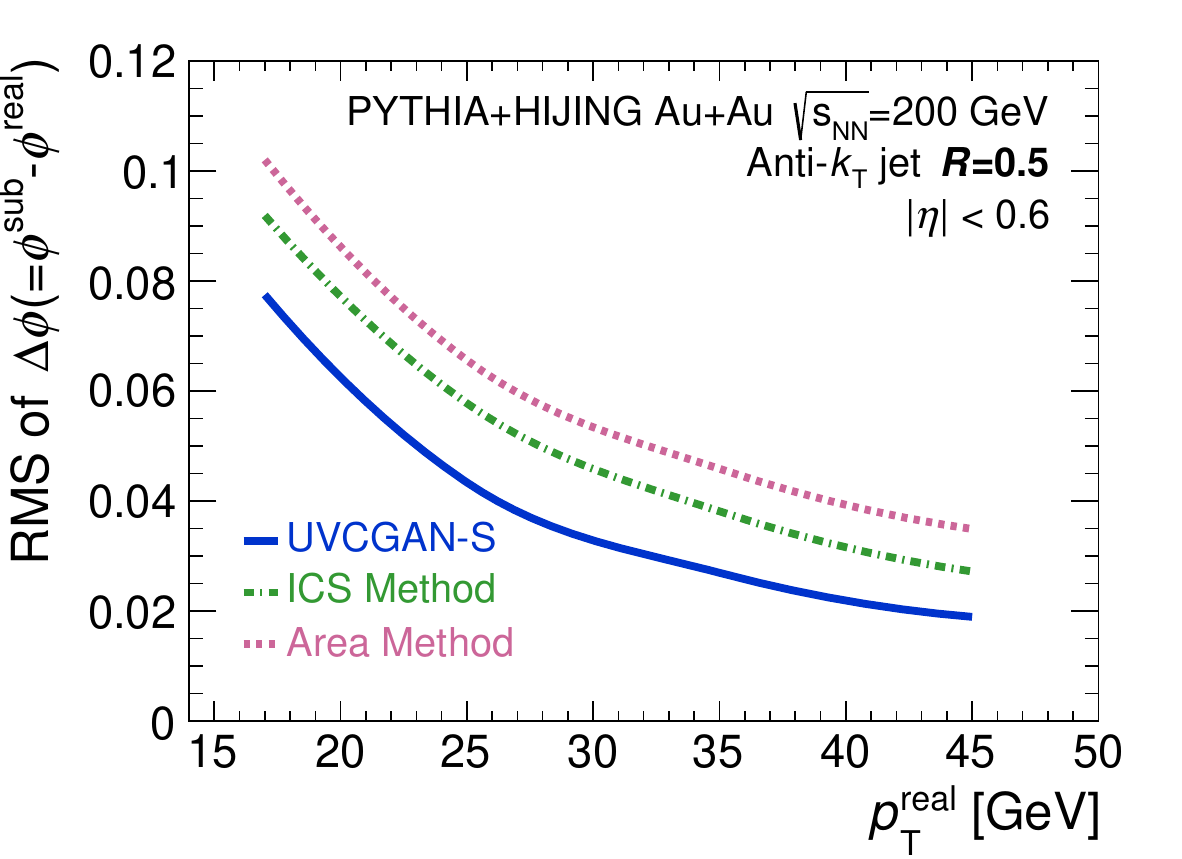}%
    \caption{
        Mean (Left) and RMS (Right) of jet $\phi$ position difference ($\Delta\phi=\phi^{\rm sub}-\phi^{\rm real}$) distributions as a function of jet \pt for $R=0.2$ (Top) and $R=0.5$ (Bottom) in \pythiahijing.}
    \label{fig:dphiDist_app}
\end{figure}

The jet position response is evaluated using the differences $\Delta\eta=\eta^{\rm sub}-\eta^{\rm real}$ and $\Delta\phi=\phi^{\rm sub}-\phi^{\rm real}$ in bins of ground truth jet \pt\ and jet radius $R$, where ``sub'' denotes background-subtracted jets and ``real'' denotes the corresponding detector-level signal jets. \autoref{fig:detaDist_app}-\ref{fig:dphiDist_app} show the mean and the root mean square (RMS) of \deta and \dphi distributions for $R=0.2$ and $R=0.5$. For small jet radius ($R=0.2$), the mean jet positions are similar across the three methods, and \ourmodel provides a modest improvement in the position resolution. For larger jet radius ($R=0.5$), all methods again recover the mean position within a small bias, while our model achieves up to $40\%$ better position resolution (smaller RMS). 

\subsection{Reconstruction Efficiency and Fake Rates}
\label{app:eff_fake}
\begin{figure}[ht!]
    \centering
    \includegraphics[width=0.43\textwidth]{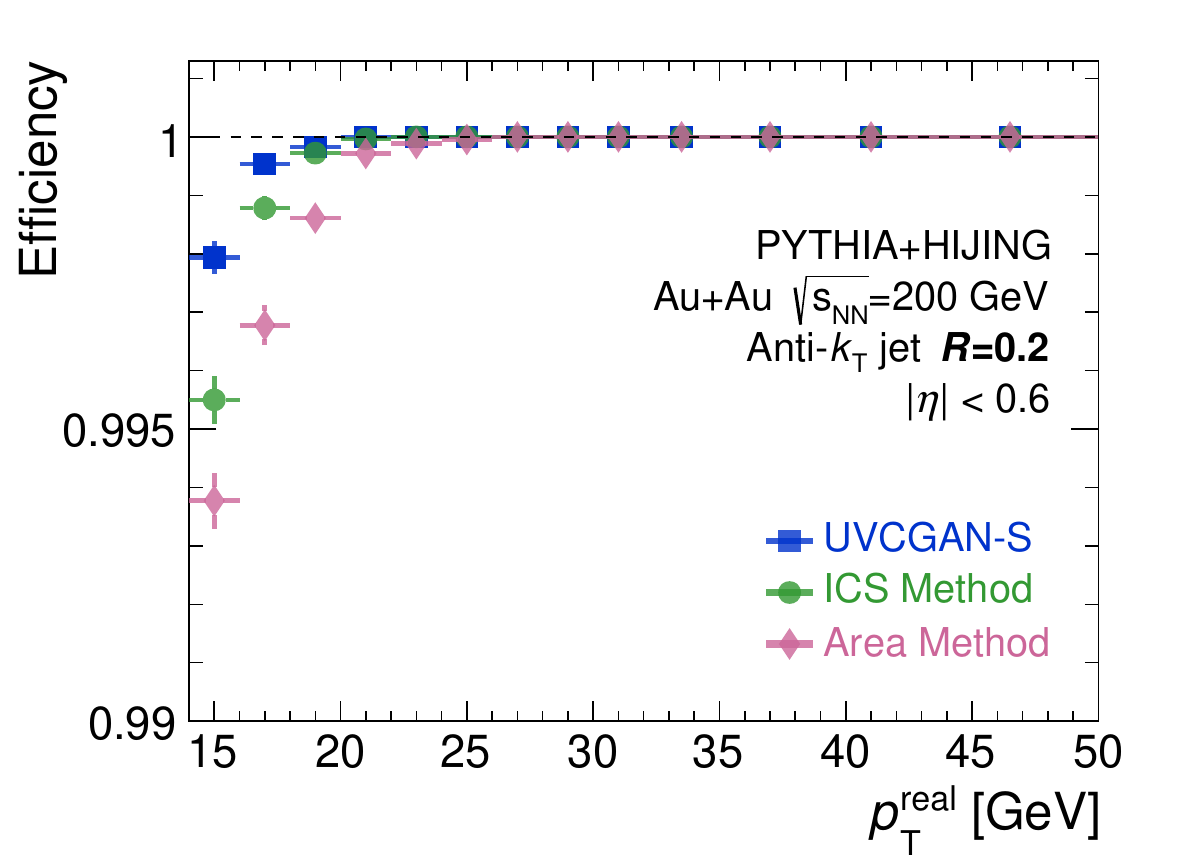}%
    \includegraphics[width=0.43\textwidth]{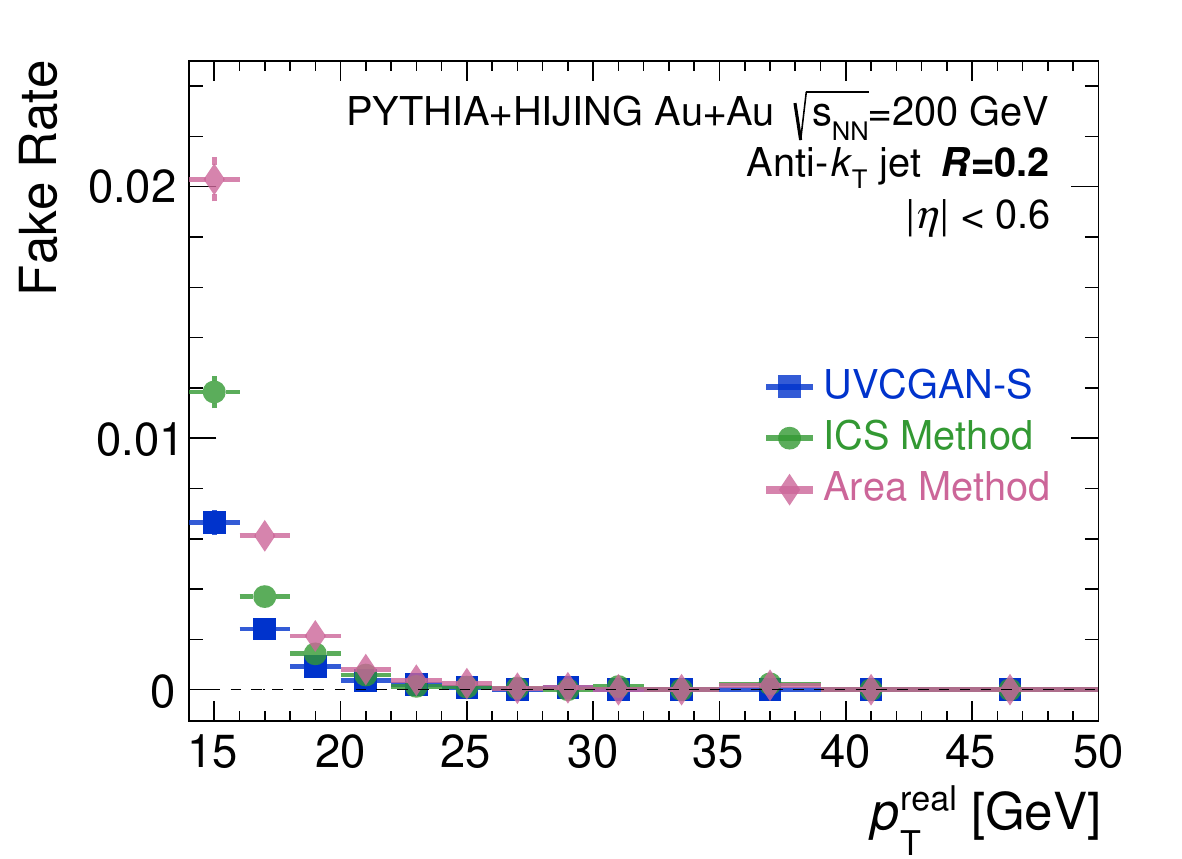}\\
    \caption{Jet reconstruction efficiency (Left) and fake rate (Right) as a function of the ground-truth jet \ptreal for $R=0.2$ in \pythiahijing. The vertical lines associated with the markers are statistical uncertainties.}
    \label{fig:eff_fake_small}
\end{figure}

\autoref{fig:eff_fake_small} shows the reconstruction efficiency and fake rates for jets reconstructed with a small radius of $R=0.2$. For this small jet cone, all background subtraction methods perform well, achieving an efficiency consistently above $99.7\%$ and maintaining a fake rate below $0.4\%$ across the entire \ptreal range. The \ourmodel framework demonstrates a marginal but consistent improvement over the conventional methods in both efficiency and fake rate metrics.

\subsection{Jet Substructures}
\label{app:substructure}
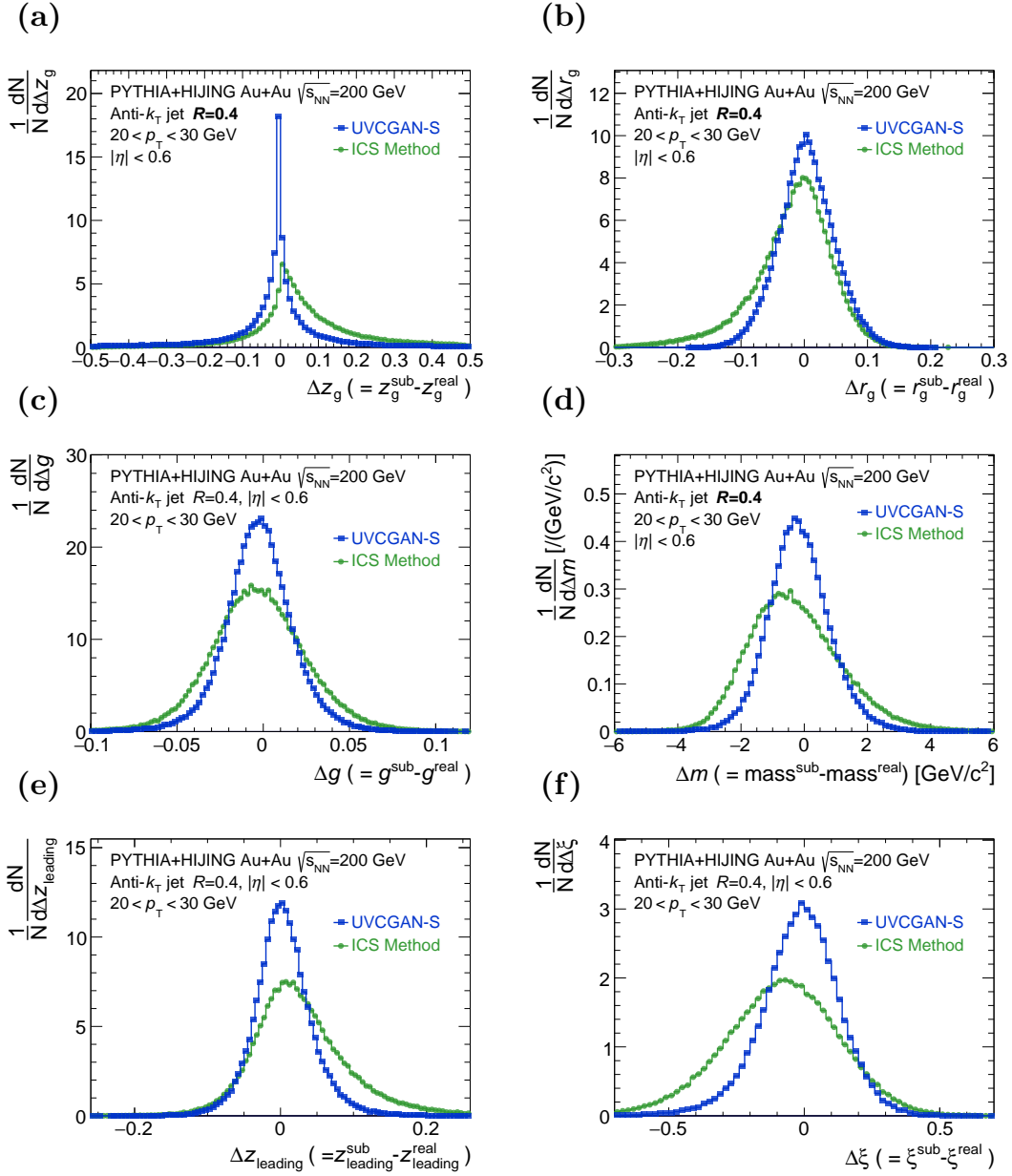
\begin{figure}[ht!]
    \centering
    \ifdefined\isarxiv
    \includegraphics[width=0.90\textwidth]{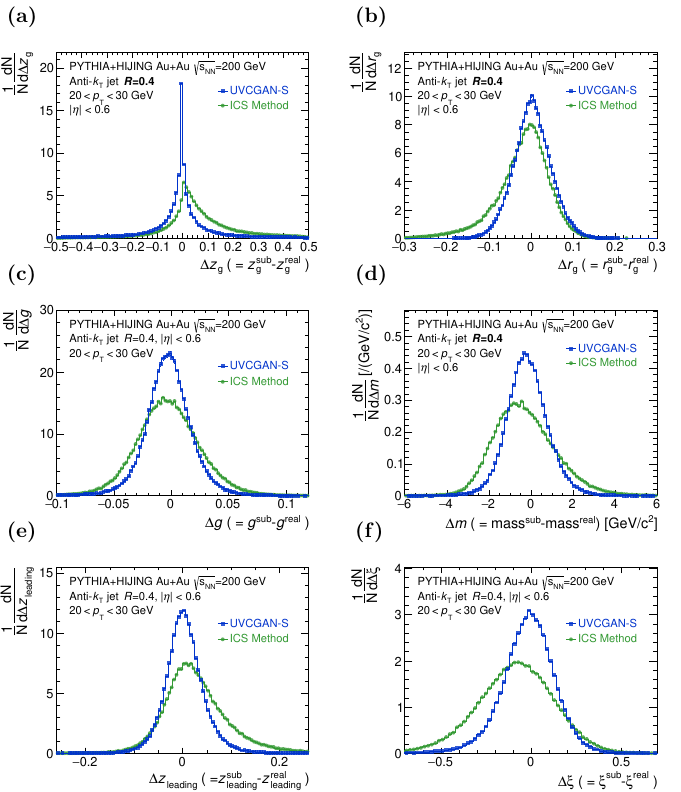}
    \else
    \tikzsetnextfilename{substructure_diff}
        \begin{tikzpicture}
        \def\width{0.43\textwidth}
        \def\xs{0.02\textwidth}
        \def\ys{0.02\textwidth}
        \node[inner sep=0] (fig1) at (0, 0) {\includegraphics[width=\width]{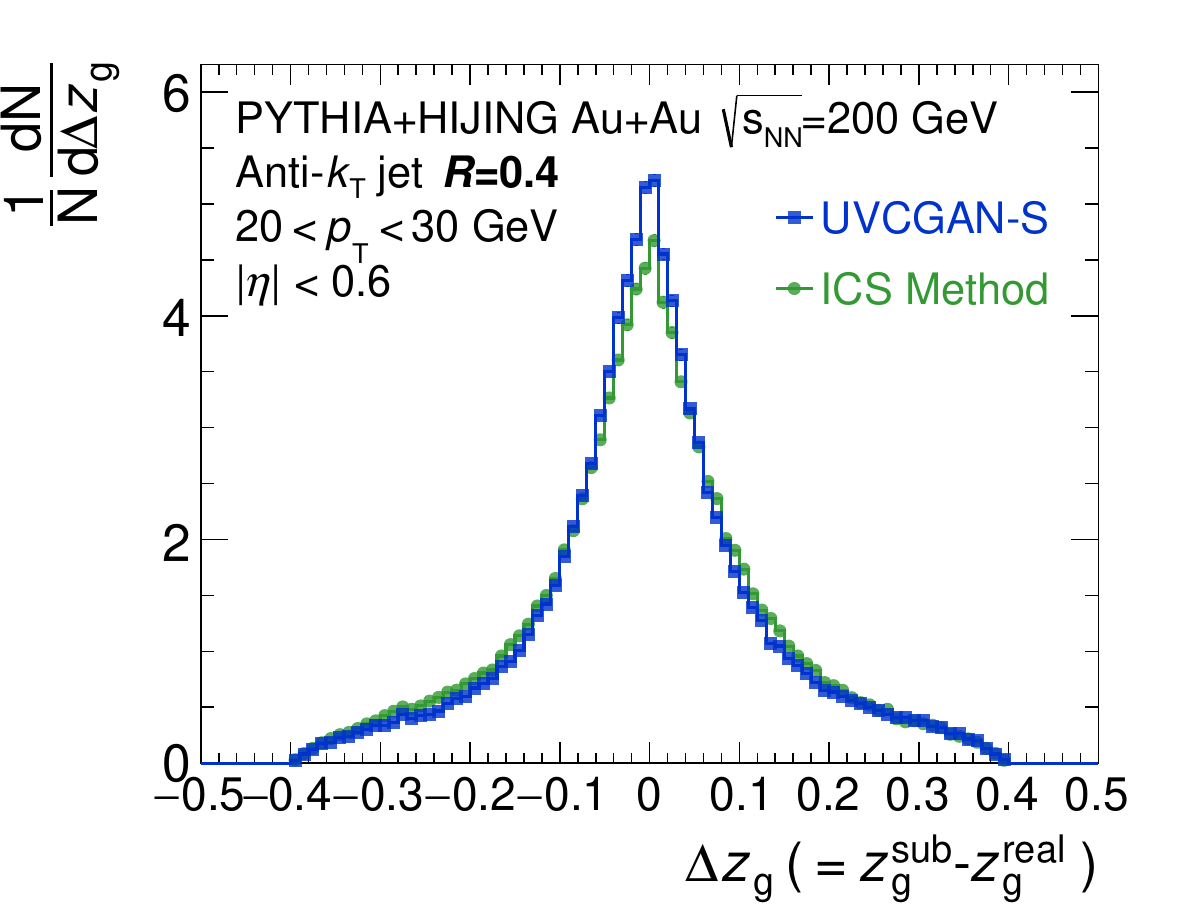}};
        \node[inner sep=0, anchor=west] (fig2) at ([xshift=\xs]fig1.east) {\includegraphics[width=\width]{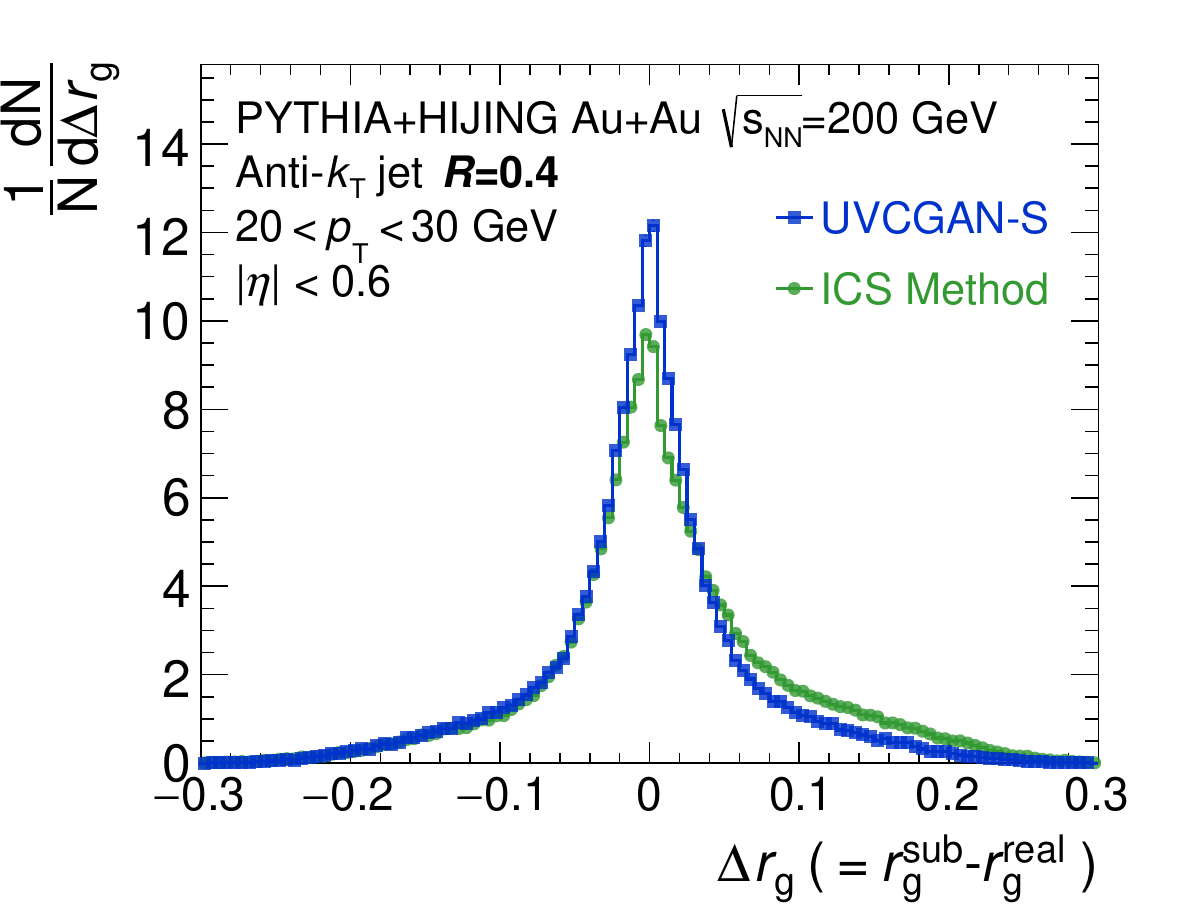}};
        \node[inner sep=0, anchor=north west] (fig3) at ([yshift=-\ys]fig1.south west) {\includegraphics[width=\width]{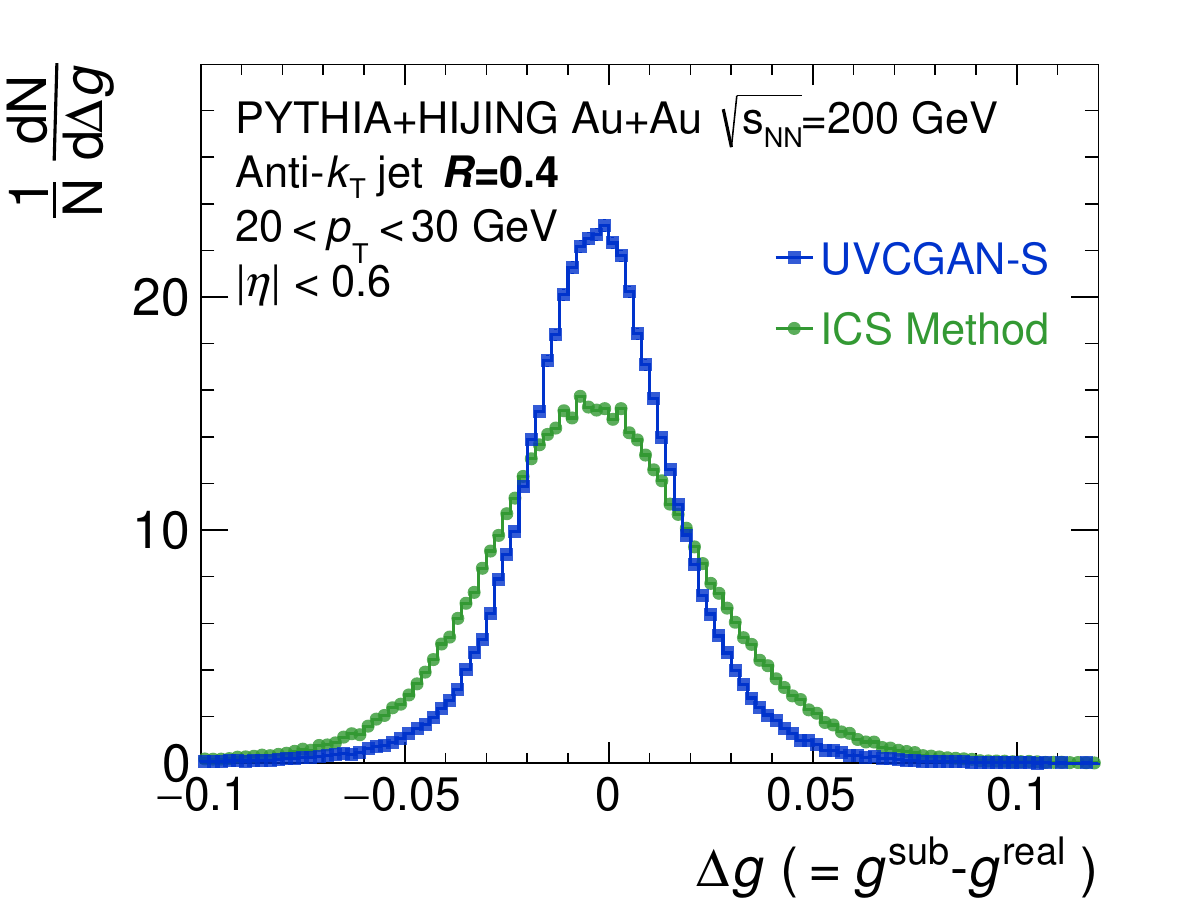}};
        \node[inner sep=0, anchor=west] (fig4) at ([xshift=\xs]fig3.east) {\includegraphics[width=\width]{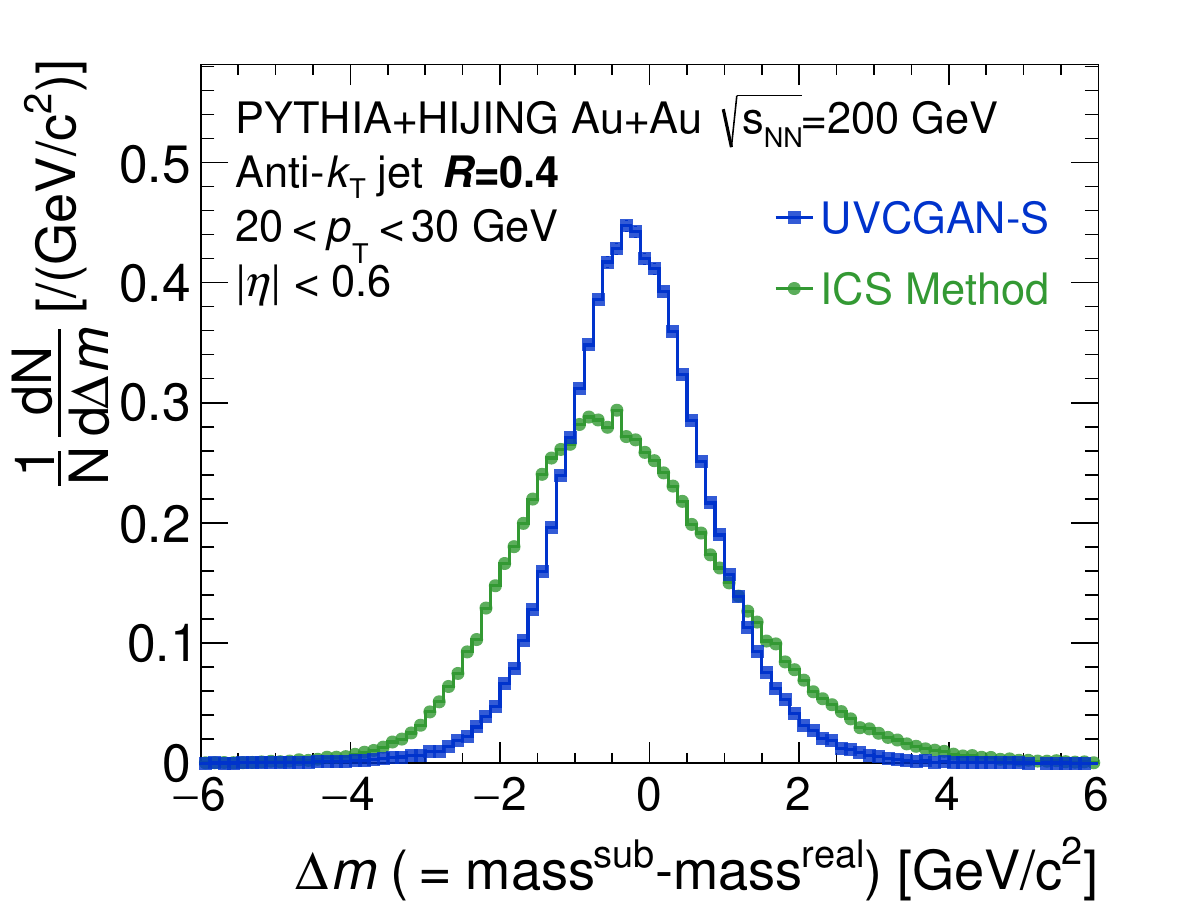}};
        \node[inner sep=0, anchor=north west] (fig5) at ([yshift=-\ys]fig3.south west) {\includegraphics[width=\width]{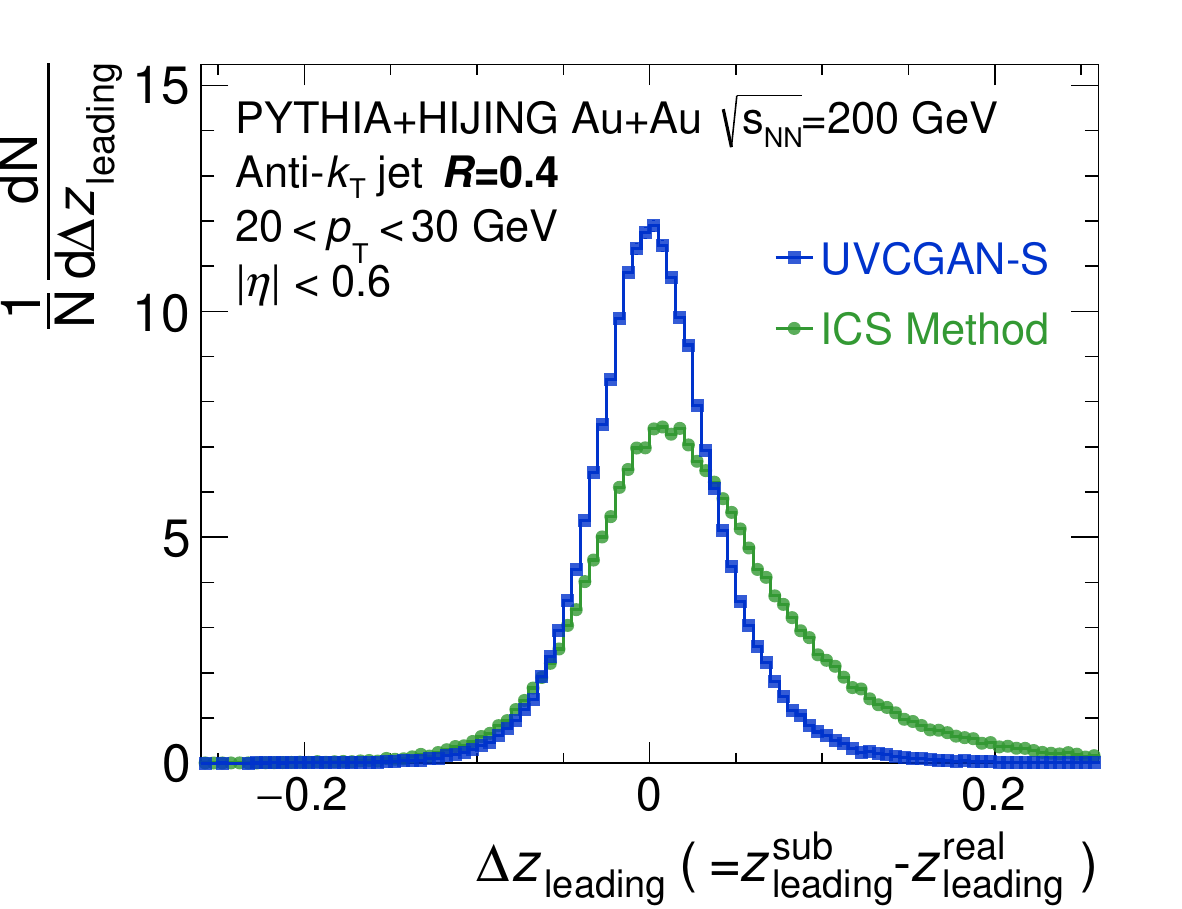}};
        \node[inner sep=0, anchor=west] (fig6) at ([xshift=\xs]fig5.east) {\includegraphics[width=\width]{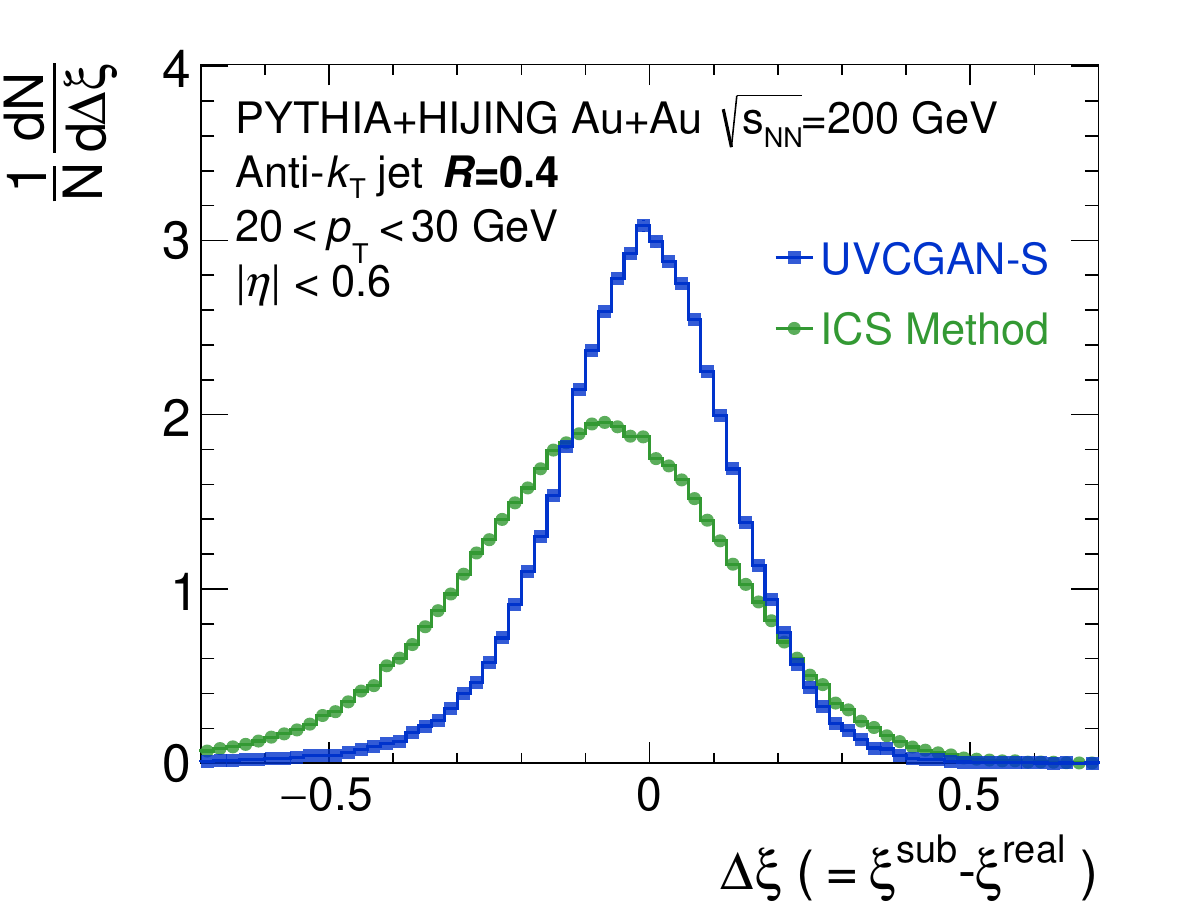}};
        \node[anchor=north west, font=\bfseries] at ([yshift=+10pt]fig1.north west) {(a)};
        \node[anchor=north west, font=\bfseries] at ([yshift=+10pt]fig2.north west) {(b)};
        \node[anchor=north west, font=\bfseries] at ([yshift=+10pt]fig3.north west) {(c)};
        \node[anchor=north west, font=\bfseries] at ([yshift=+10pt]fig4.north west) {(d)};
        \node[anchor=north west, font=\bfseries] at ([yshift=+10pt]fig5.north west) {(e)};
        \node[anchor=north west, font=\bfseries] at ([yshift=+10pt]fig6.north west) {(f)};
    \end{tikzpicture}
    \fi 

    \caption{
    ``sub''-``real'' distributions of various jet substructure observables in \pythiahijing. (a) Groomed momentum sharing fraction \zg, (b) Groomed jet radius \rg, (c) jet girth $g$, (d) jet mass, (e) momentum fraction of leading constituents \zleading, (f) jet fragmentation function $\xi$. All distributions are for jet with $R=0.4$ at $20<\pt<30$~GeV.
    }
    \label{fig:substructure_diff}
\end{figure}

\autoref{fig:substructure_diff} presents the distributions for six distinct jet substructure observables, quantifying the residual bias, resolution, and uncertainty after background subtraction. Each panel displays the difference between the reconstructed (background-subtracted) value and the ground truth value ($\text{Observable}^{\text{sub}} - \text{Observable}^{\text{real}}$), which allows for a direct assessment of subtraction fidelity. The variables include the groomed momentum sharing fraction ($\Delta \zg$), groomed jet radius ($\Delta \rg$), jet girth ($\Delta g$), jet mass ($\Delta m$), the momentum fraction of the leading constituent ($\Delta \zleading$), and the fragmentation variable ($\Delta \xi$). All distributions are shown for jets with $R=0.4$ in the range of $20<\pt<30~\text{GeV}$. The \ourmodel reconstruction consistently shows distributions that are tighter and more precisely centered at zero compared to the conventional Iterative Constituent Subtraction Method, indicating a superior preservation of the intrinsic jet structure across all tested metrics.

\subsection{Generalization and Tests on Quenched Jets}\label{app:jewel}

To assess the model's robustness and generalization capabilities beyond the training domain, we applied the \ourmodel (trained on unquenched \pythia jets embedded in \hijing background) to quenched jet inputs. These inputs represent a significant out-of-distribution shift, as the jet properties are modified by the medium.

Quenched jet events were generated using the \jewel Monte Carlo event generator, utilizing the $0$--$10\%$ most central events with default parameters (no jet recoil) to simulate jet modification in the highest background density. These generated quenched jets were subsequently propagated through a full sPHENIX detector simulation. The resulting calorimeter images containing \jewel jets embedded in \hijing background events (\jewelhijing) were used as input for \ourmodel to perform background subtraction. The ability of the \ourmodel to accurately reconstruct these quenched jet properties, despite never having explicitly trained on quenched jets, provides a stringent test of its unsupervised learning mechanism.

\begin{figure}[ht!]
    \centering
    \includegraphics[width=0.43\textwidth]{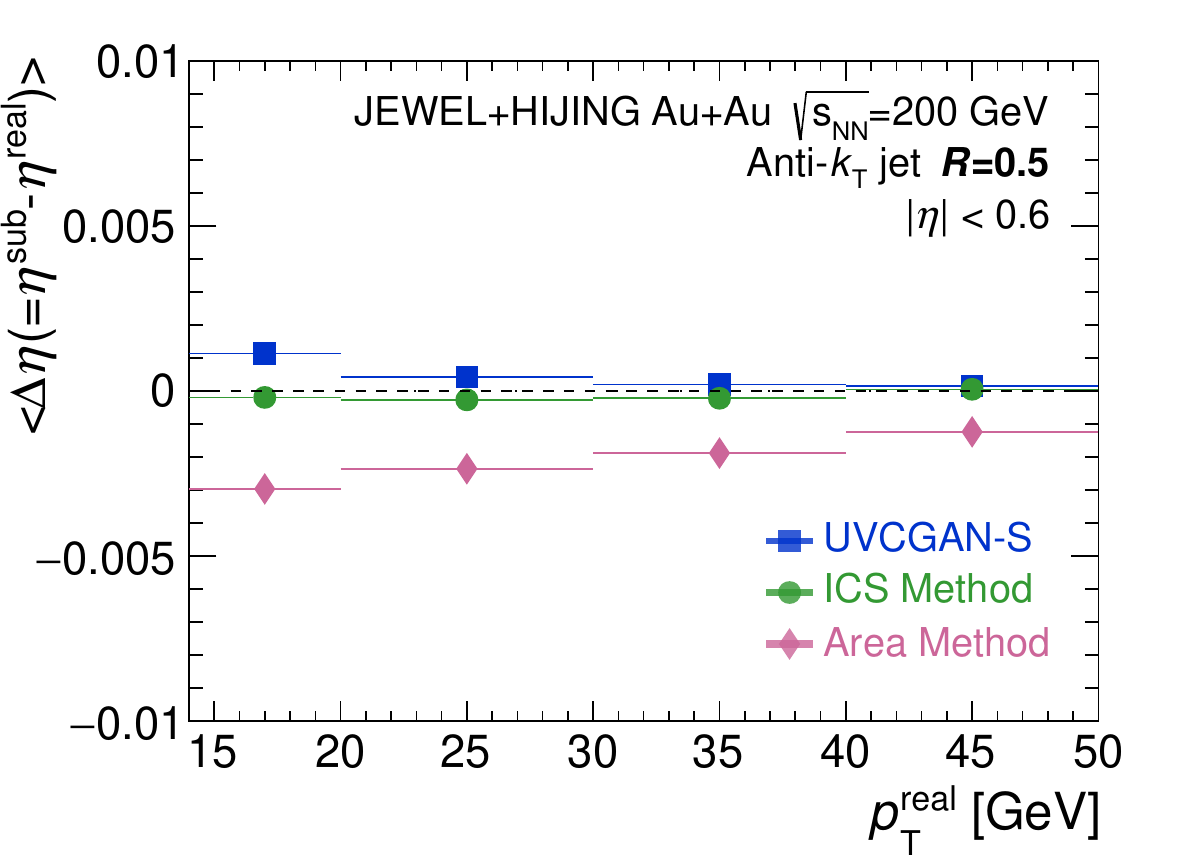}%
    \includegraphics[width=0.43\textwidth]{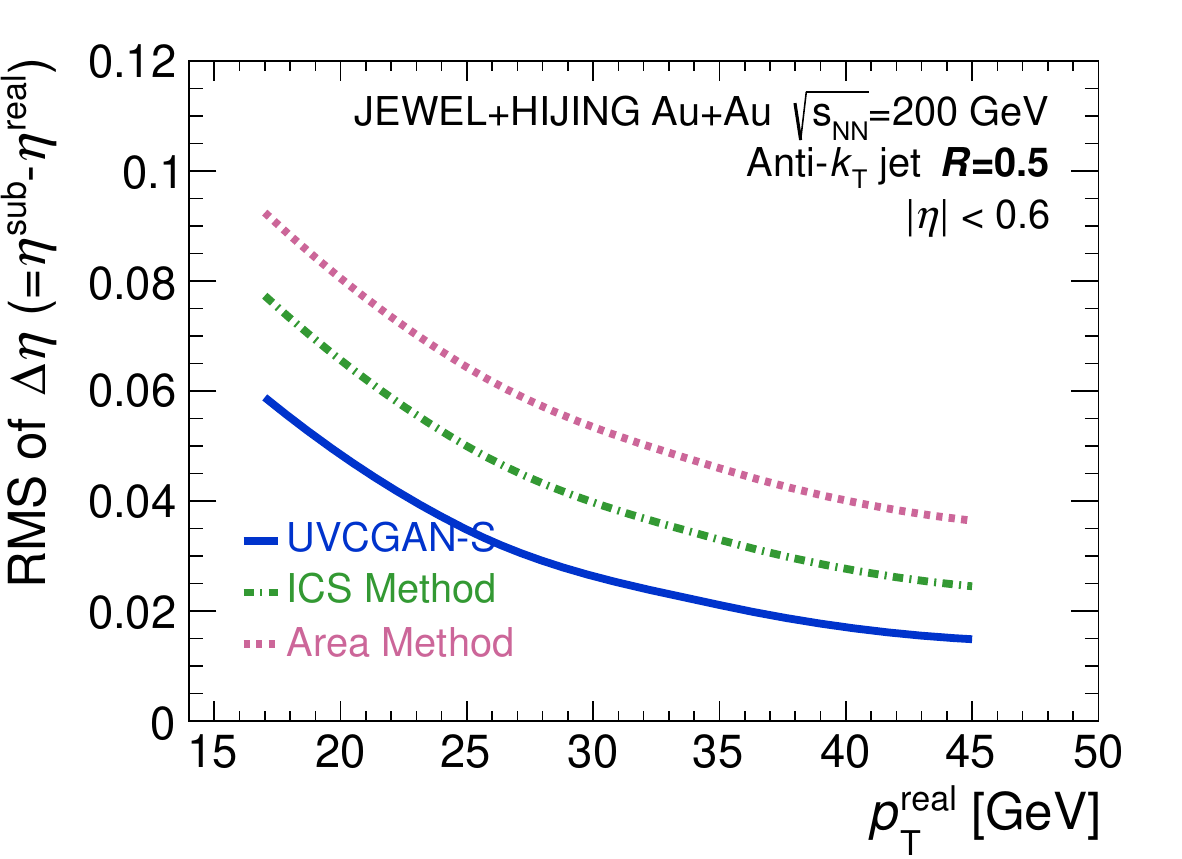}\\
    \includegraphics[width=0.43\textwidth]{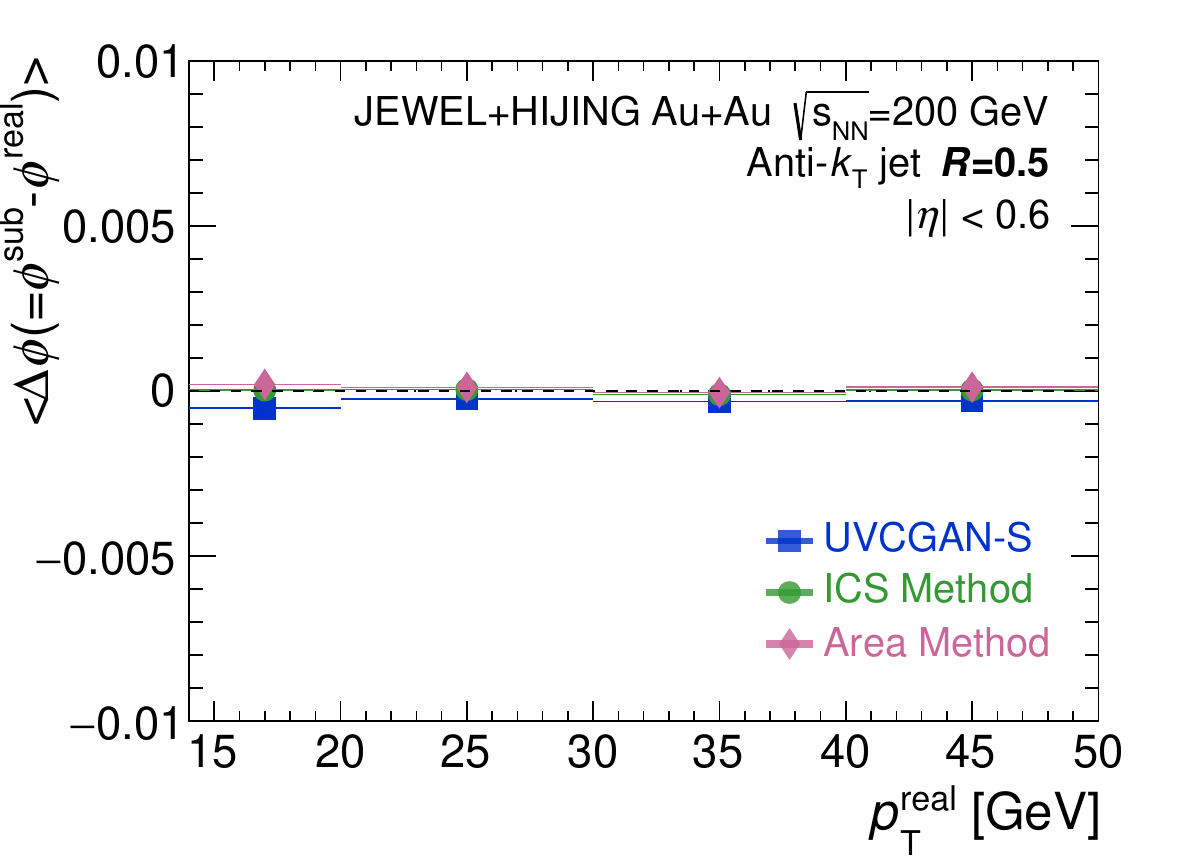}%
    \includegraphics[width=0.43\textwidth]{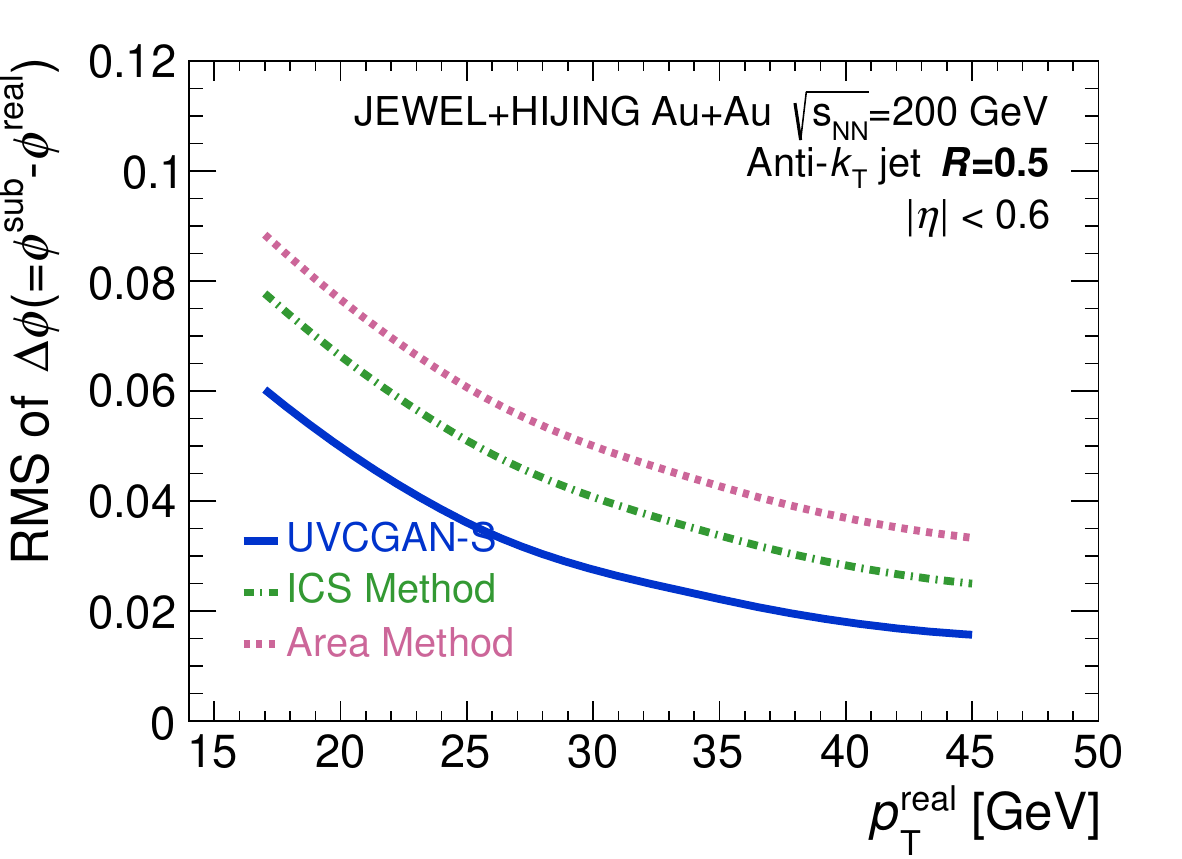}%
    \caption{
        Mean (Left) and RMS (Right) of jet $\eta$ position difference ($\Delta\eta=\eta^{\rm sub}-\eta^{\rm real}$) distributions as a function of jet \pt for $R=0.2$ (Top) and $R=0.5$ (Bottom) in \jewelhijing.}
    \label{fig:jewel_deta_dphi_app}
\end{figure}

\begin{figure}[ht!]
    \centering
    \includegraphics[width=0.43\textwidth]{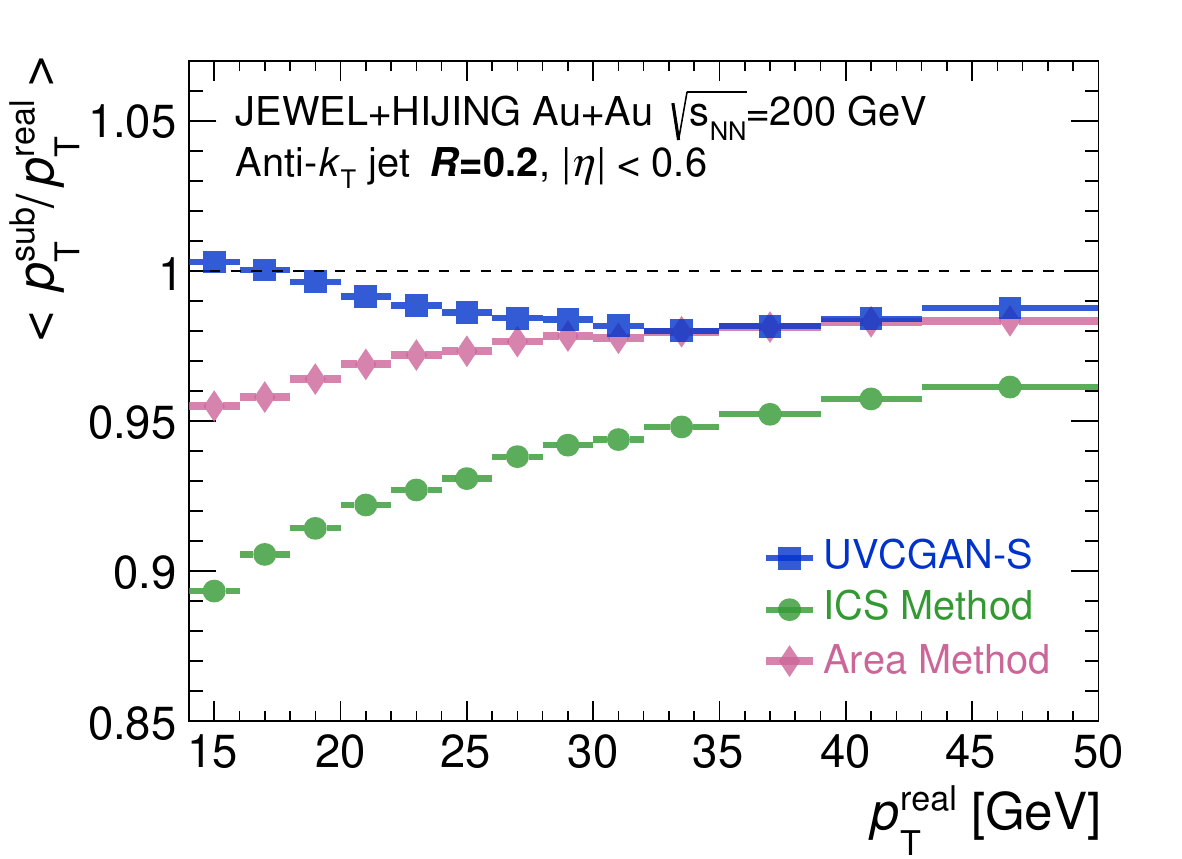}
    \includegraphics[width=0.43\textwidth]{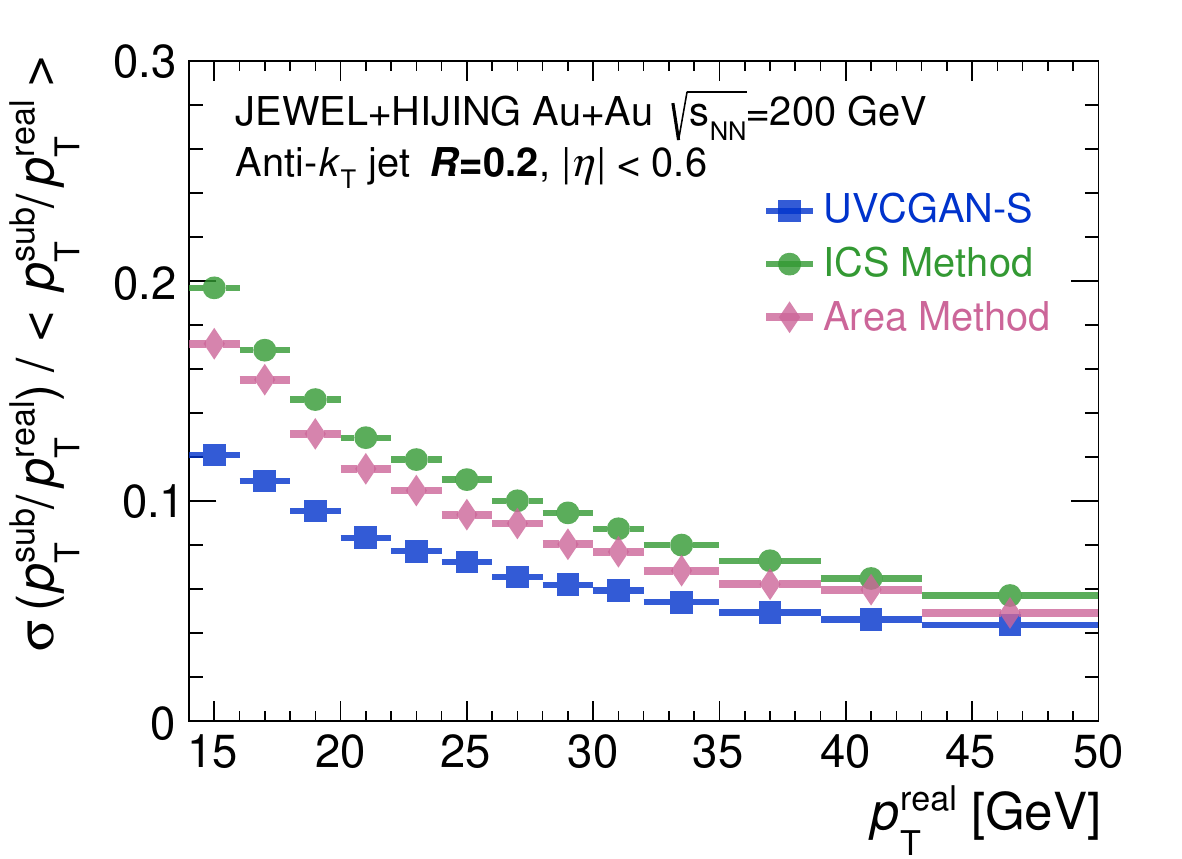}
    \includegraphics[width=0.43\textwidth]{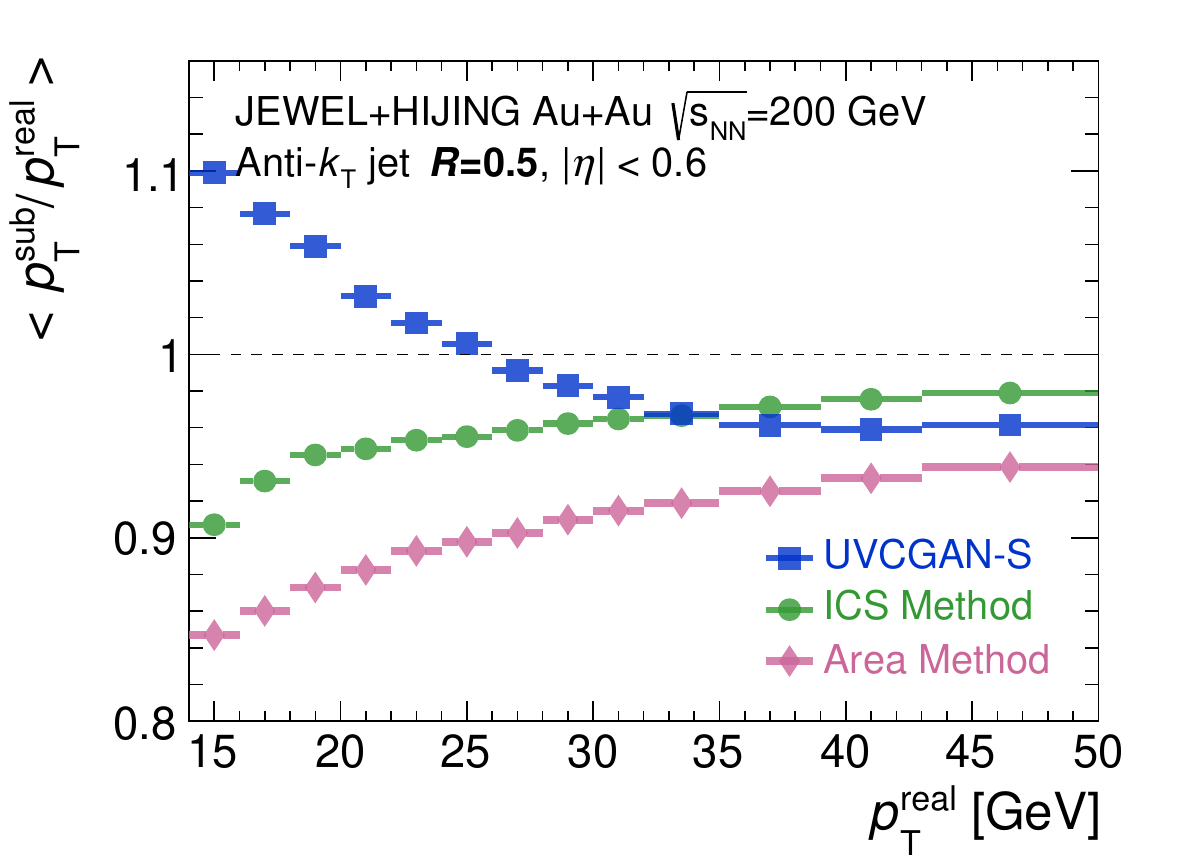}
    \includegraphics[width=0.43\textwidth]{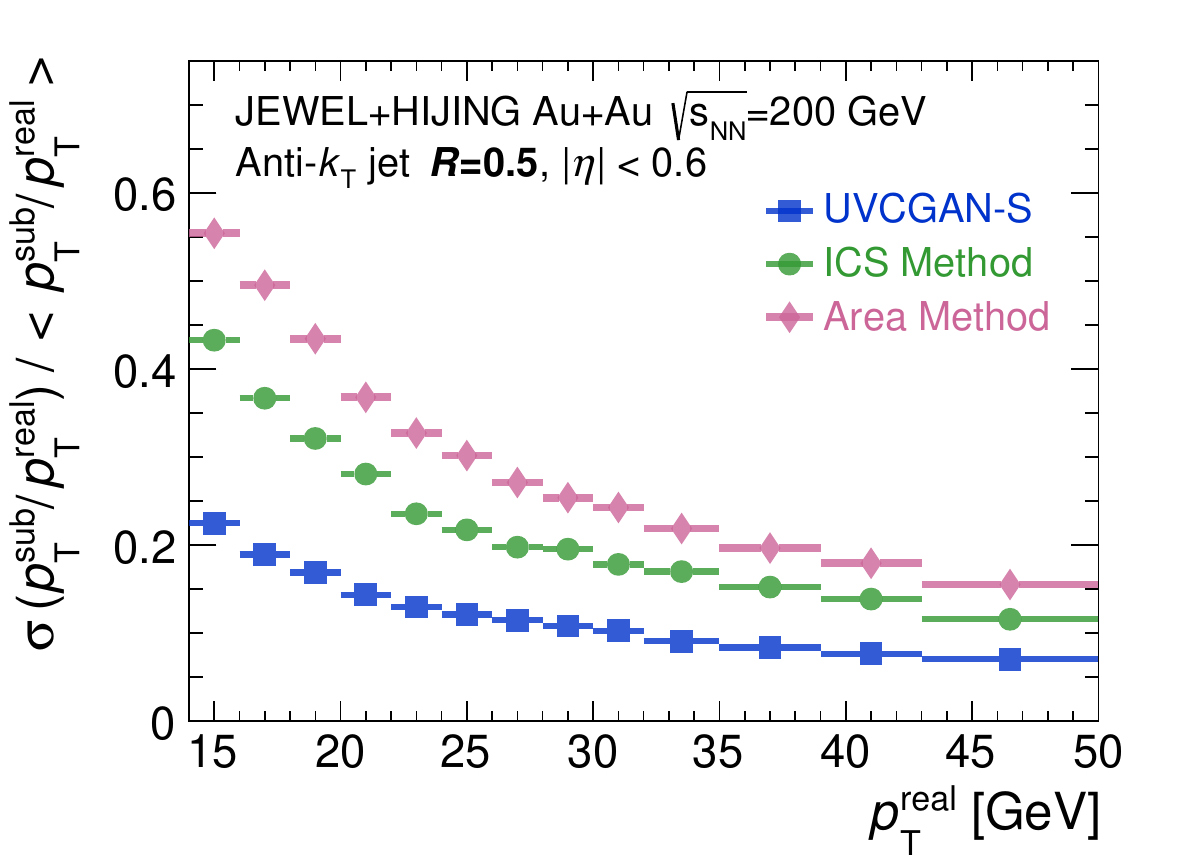}
    \caption{Jet transverse momentum mean (Left) and resolution (Right) for $R=0.2$ (Top) and $R=0.5$ (Bottom) in \jewelhijing.}
    \label{fig:jewel_JESJER}
\end{figure}

\autoref{fig:jewel_deta_dphi_app}-\ref{fig:jewel_JESJER} summarize the performance on jet position and momentum, respectively. The jet position performance, measured by the mean and RMS of $\Delta\eta$ and $\Delta\phi$, is very similar to the performance observed in the \pythiahijing case, demonstrating the method's robustness against signal modification. The momentum resolution (JER) achieved for \jewelhijing is also excellent, while the momentum scale (JES) is found to be better or comparable to conventional methods.

\begin{figure}[ht!]
    \centering
    \includegraphics[width=0.43\textwidth]{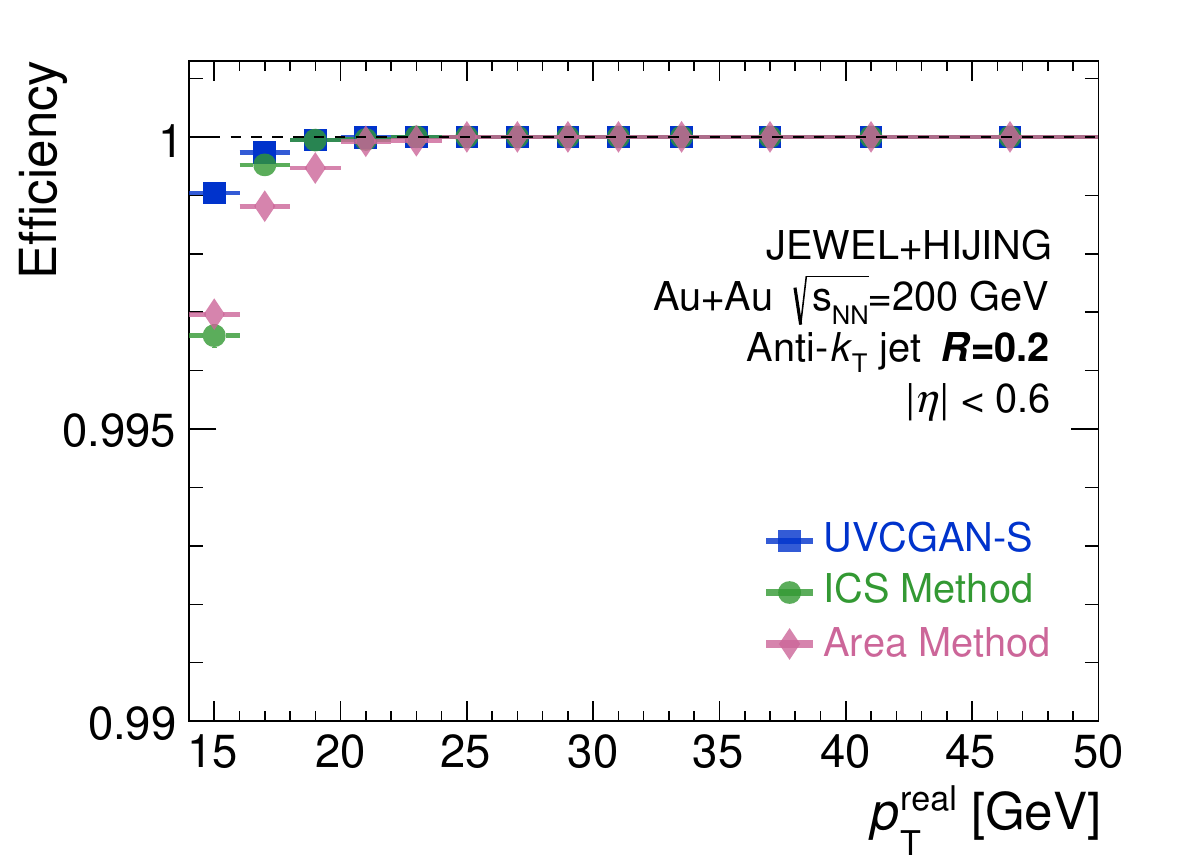}%
    \includegraphics[width=0.43\textwidth]{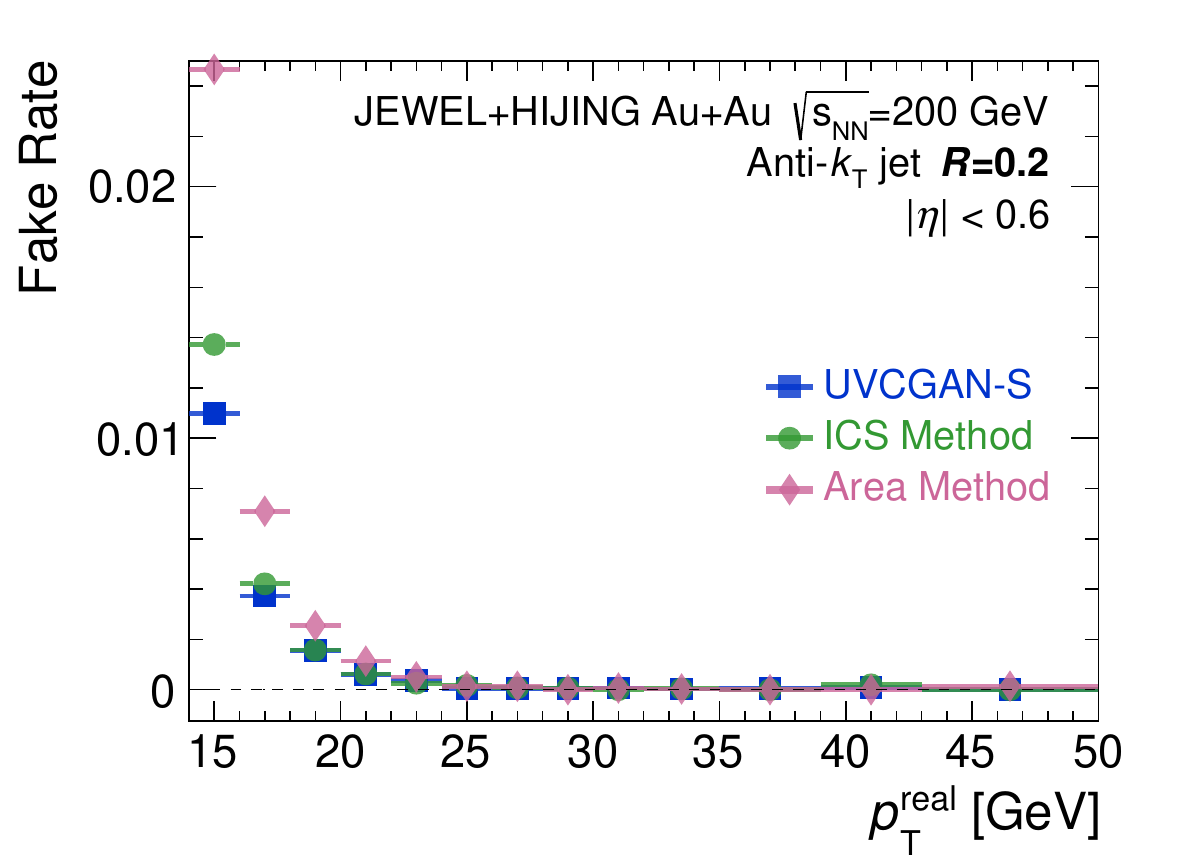}\\
    \includegraphics[width=0.43\textwidth]{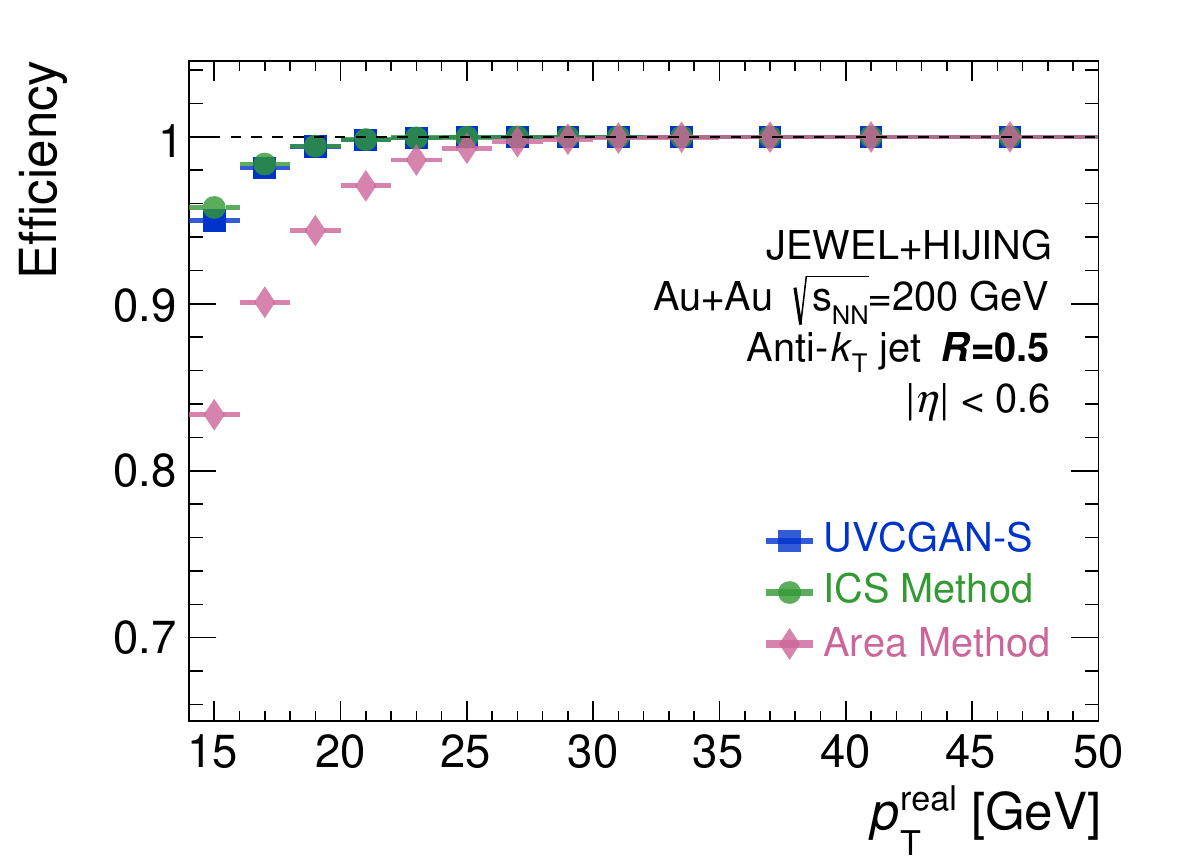}%
    \includegraphics[width=0.43\textwidth]{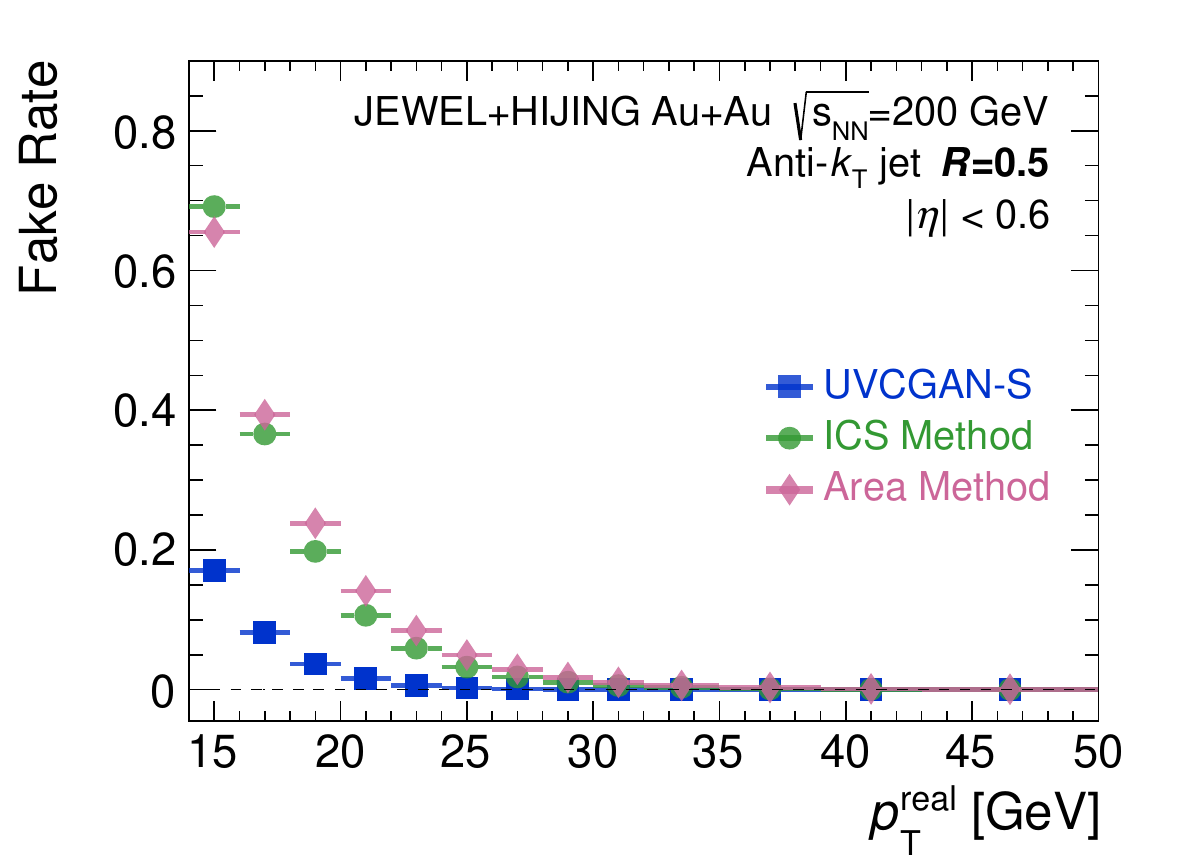}
    \caption{Jet reconstruction efficiency (Left) and fake rate (Right) as a function of the ground-truth jet \ptreal for $R=0.2$ (Top) and $R=0.5$ (Bottom) in \jewelhijing. The statistical uncertainties are smaller than the marker sizes. }
    \label{fig:jewel_eff_fake}
\end{figure}

\autoref{fig:jewel_eff_fake} compares the jet reconstruction efficiency and fake rates for the quenched \jewel sample. Similar to the unquenched case, \ourmodel exhibits significantly better efficiency and lower fake rates than conventional methods, particularly for larger jet radii.

\begin{figure}[ht!]
    \centering
    
     \tikzexternaldisable
    \begin{tikzpicture}
        \def\width{0.43\textwidth}
        \def\xs{0.02\textwidth}
        \def\ys{0.02\textwidth}
        \node[inner sep=0] (fig1) at (0, 0) {\includegraphics[width=\width]{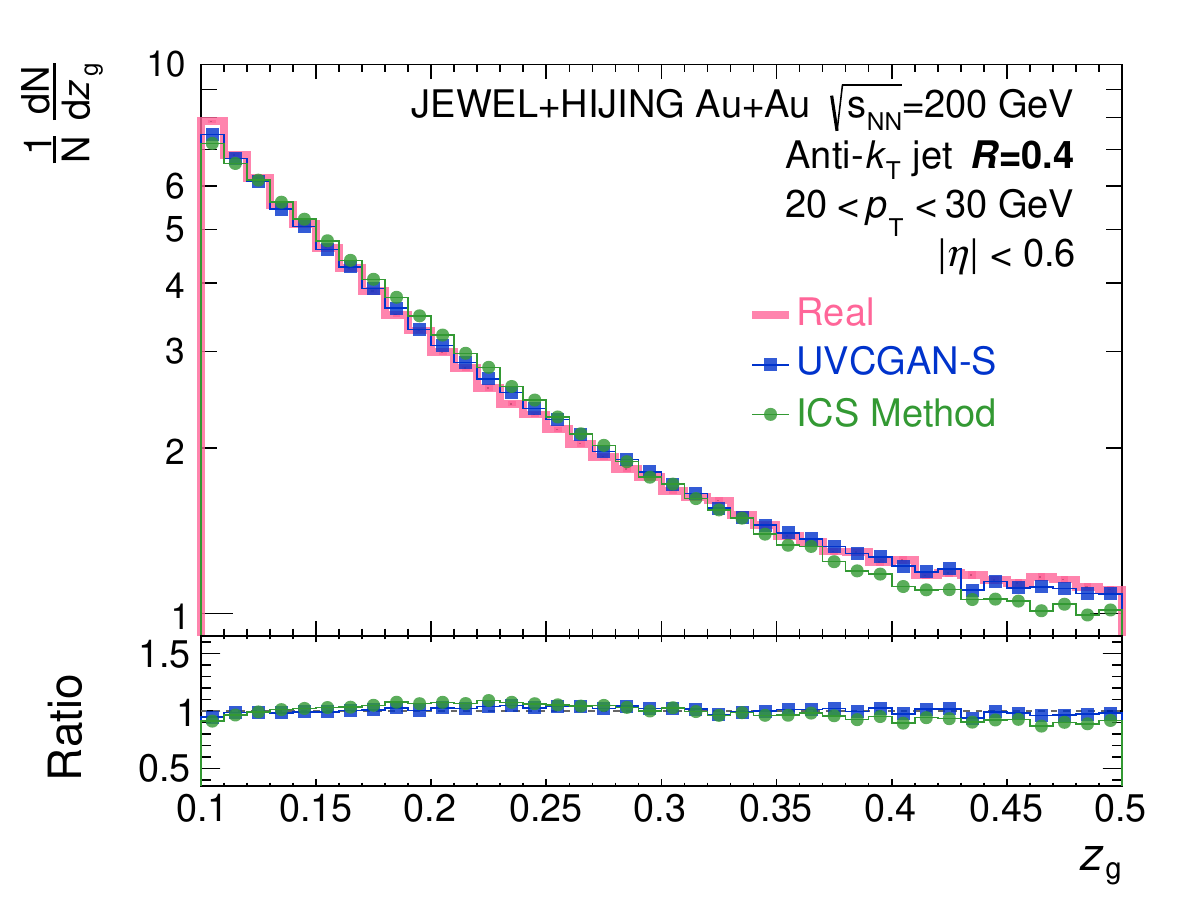}};
        \node[inner sep=0, anchor=west] (fig2) at ([xshift=\xs]fig1.east) {\includegraphics[width=\width]{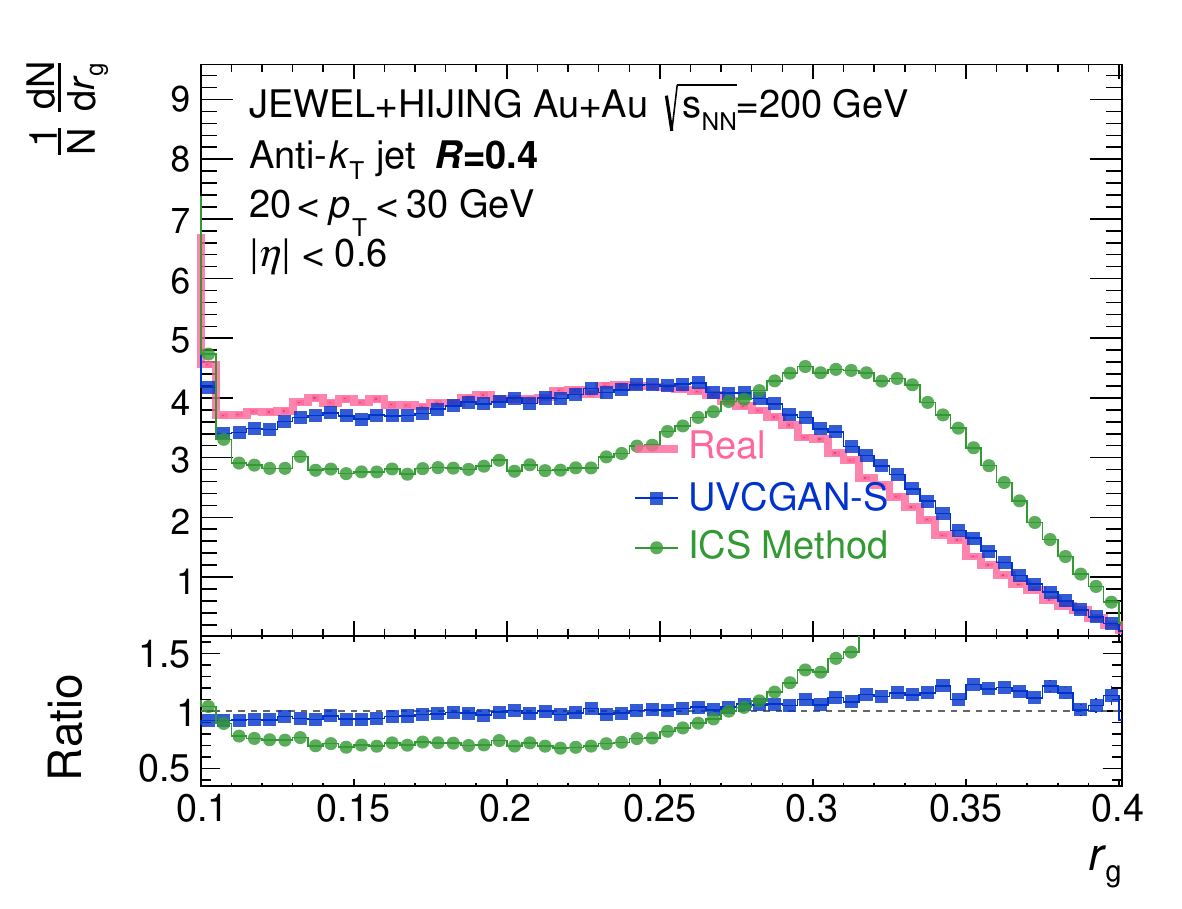}};
        \node[inner sep=0, anchor=north west] (fig3) at ([yshift=-\ys]fig1.south west) {\includegraphics[width=\width]{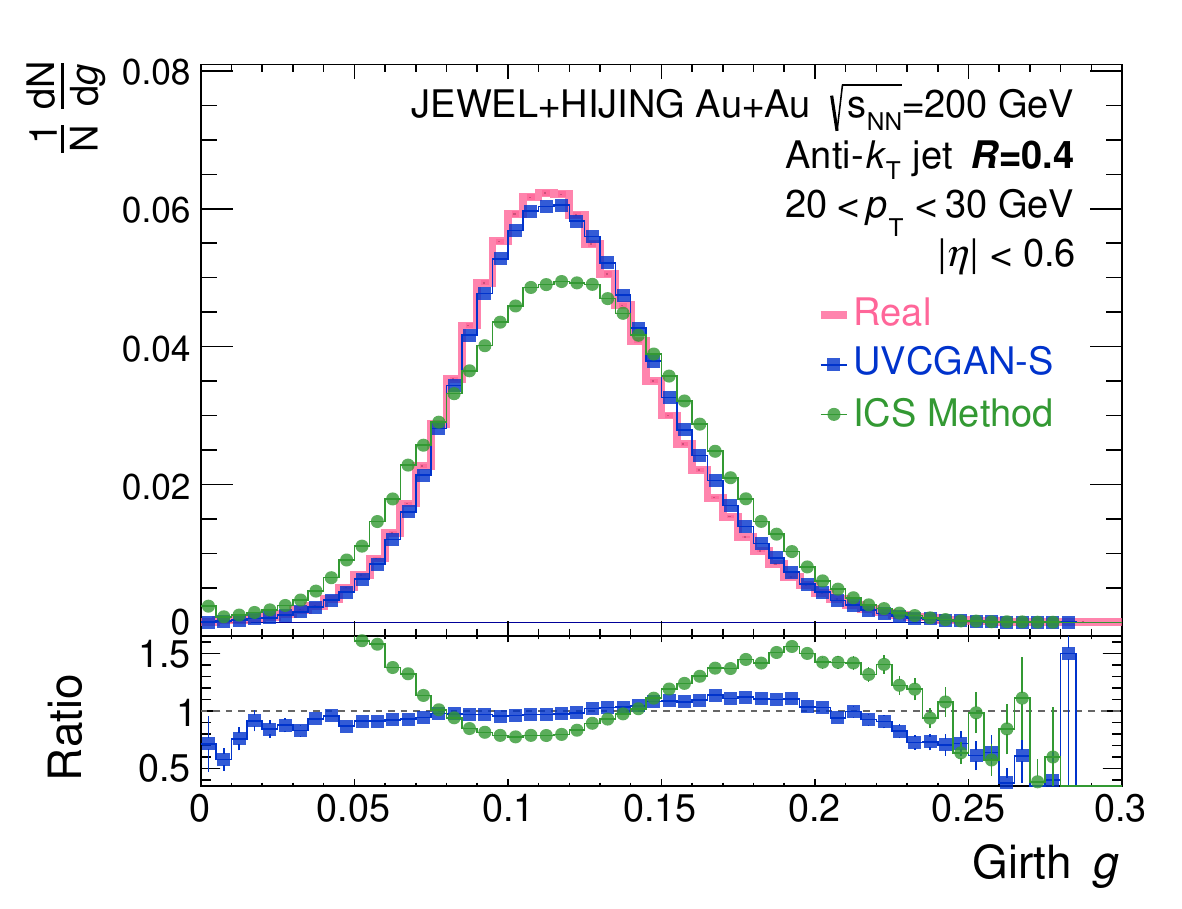}};
        \node[inner sep=0, anchor=west] (fig4) at ([xshift=\xs]fig3.east) {\includegraphics[width=\width]{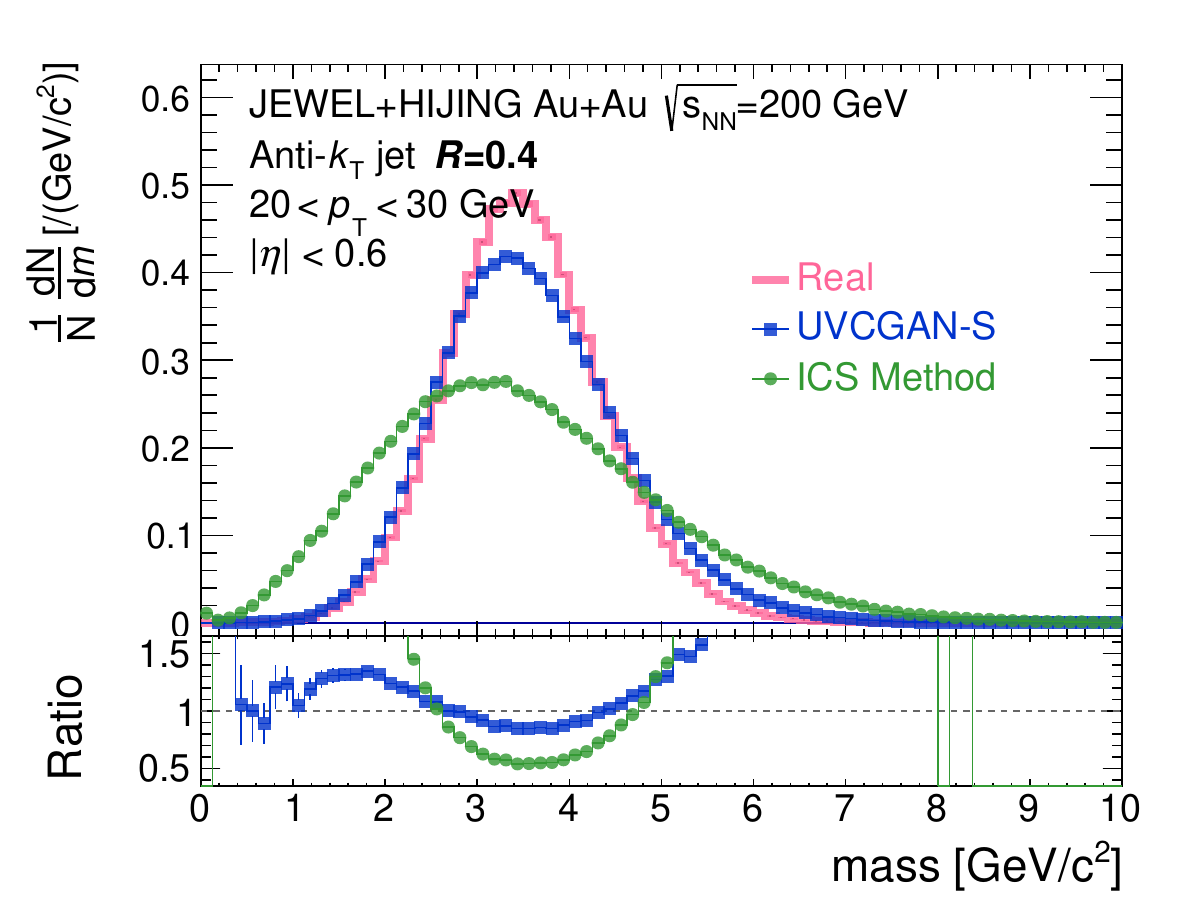}};
        \node[inner sep=0, anchor=north west] (fig5) at ([yshift=-\ys]fig3.south west) {\includegraphics[width=\width]{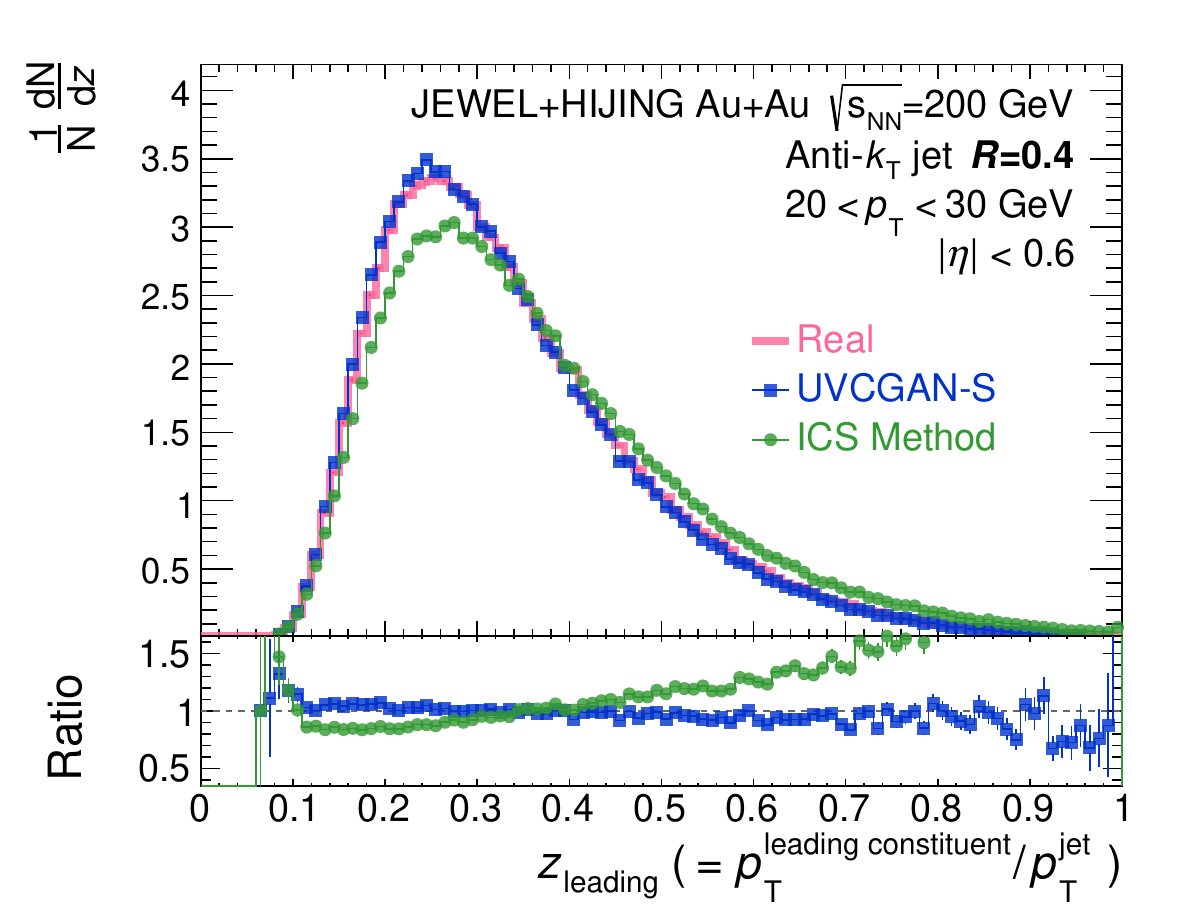}};
        \node[inner sep=0, anchor=west] (fig6) at ([xshift=\xs]fig5.east) {\includegraphics[width=\width]{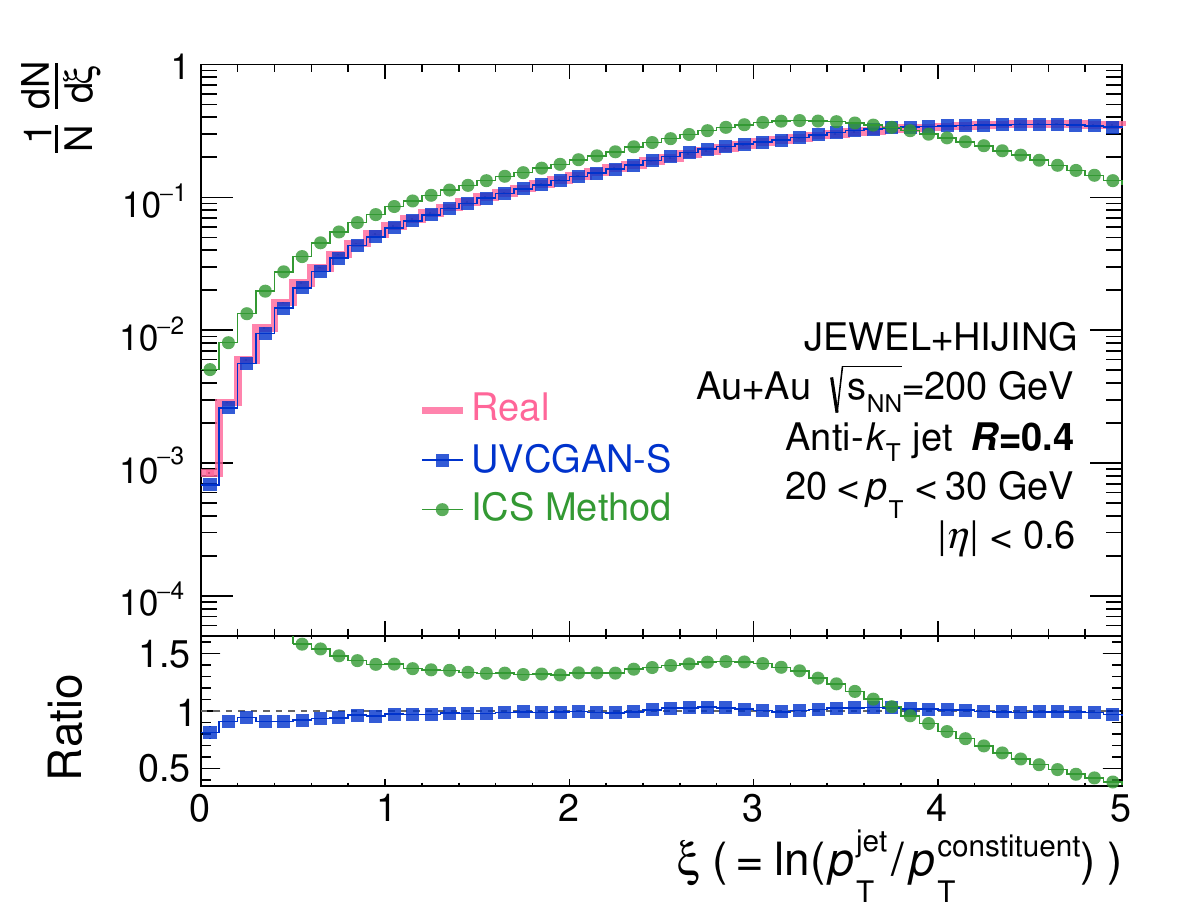}};
        \node[anchor=north west, font=\bfseries] at ([yshift=+10pt]fig1.north west) {(a)};
        \node[anchor=north west, font=\bfseries] at ([yshift=+10pt]fig2.north west) {(b)};
        \node[anchor=north west, font=\bfseries] at ([yshift=+10pt]fig3.north west) {(c)};
        \node[anchor=north west, font=\bfseries] at ([yshift=+10pt]fig4.north west) {(d)};
        \node[anchor=north west, font=\bfseries] at ([yshift=+10pt]fig5.north west) {(e)};
        \node[anchor=north west, font=\bfseries] at ([yshift=+10pt]fig6.north west) {(f)};
    \end{tikzpicture}
    \caption{Various jet substructure observable distributions (top panels) and their ratios to the ground truth (``Real'') distributions (bottom panels) in \jewelhijing: (a) Groomed momentum sharing fraction \zg, (b) Groomed jet radius \rg, (c) jet girth $g$, (d) jet mass, (e) momentum fraction of leading constituents \zleading, (f) jet fragmentation function $\xi$. All distributions are for jet with $R=0.4$ at $20<\pt<30$~GeV. The \ourmodel reconstruction is significantly closer to the ground truth (``Real'') distributions than the Iterative Method, demonstrating better preservation of jet substructure.
    }
    \label{fig:jewel_substructure}
\end{figure}

\autoref{fig:jewel_substructure} presents the reconstructed distributions for six key jet substructure observables ($\zg, \rg, g, m, \zleading, \xi$). The true quenched distributions (labeled ``Real (JEWEL)'') are statistically very different from the \pythia distribution (labeled ``Real (PYTHIA)''). The key finding is that the \ourmodel's reconstructed distributions (``UVCGAN-S (JEWEL)'') consistently track the ground truth \jewel distributions closely across all observables. This result is highly significant: it confirms that the model successfully learned the distinction between signal and background energy flow without tying itself to the specific internal structure of the unquenched training jets. Consequently, the network preserves the physical medium-induced modifications (the characteristic shift and broadening seen in the \jewel distributions) while simultaneously removing the background, demonstrating that the background subtraction is an emergent property that generalizes to out-of-distribution jet signals.

\clearpage

\input{src/app_train}

%% file: src/app_train.tex
\section{Training Details}\label{sec:training_details}

\subsection{Data Processing}

Our training data consists of sPHENIX calorimeter images with spatial resolution $24 \times 64$ pixels and single-channel format ($C_{a_0} = C_{a_1} = C_b = 1$).
During the training, we sample images from three datasets independently: background components ($985791$ images), isolated jet signals from simulation ($635582$ images), and embedded jets ($633000$ images). All datasets are stored in HDF5 format with unpaired sampling during training. Representative examples of the images (combined jet+background images, and separate signal and background images) are shown in Figure~\ref{fig:cyclegan}.

We found that logarithmically normalizing images before presenting them to the generators improves the signal extraction performance:
\begin{equation}
x_{\text{norm}} = \log(x_{\text{raw}} + b)
\end{equation}

where $b = 0.1$ prevents numerical issues with zero-valued pixels. This compresses the energy scale while preserving relative patterns needed for jet identification.

\subsection{Network Architecture and Training}

The generator architecture follows UVCGANv2's hybrid Vision Transformer~\cite{torbunov2023} design but we removed initial convolutional stage due to the small input size ($24 \times 64$ pixels). The final configuration uses 12 transformer blocks with 6 attention heads and 384 embedding features. The style modulated UNet network employs a 3-level feature hierarchy with [96, 192, 384] channels.

For discriminators, we switched from UVCGANv2's shallow design to a deeper ResNet-based architecture~\cite{he2016deep}. Our discriminator uses 4 levels with progressively increasing channels ($64 \to 128 \to 256 \to 512$) and 3 ResNet blocks per level. This modification significantly improved signal-background separation performance compared to the original shallow discriminators. Interestingly, we found that vanilla CycleGAN completely fails with these stronger discriminators - they are too powerful for the artifact-based cycle-consistency to work, but work well with our stratified approach.

Additionally, we changed the adversarial loss from LSGAN (used in UVCGANv2) to hinge loss~\cite{miyato2018spectral}, which further improved performance in our experiments.

The training was conducted for 400 epochs with batch size 4 and learning rate $5 \times 10^{-5}$ using Adam optimizer ($\beta_1=0.5, \beta_2=0.99$). Loss weights were $\lambda_\text{cycle}^{B\to A \to B} = 10, \lambda_\text{cycle}^{A_0\to B \to A_0} = 10, \lambda_\text{cycle}^{A_1\to B \to A_1} = 100$, and $\lambda_{idt} = 0.5 \lambda_\text{cycle}$. The higher cycle weight for signal components (100) emphasizes preserving jet structure during decomposition.

The model is implemented in PyTorch. The architecture contains approximately 32 million trainable parameters per generator and 20 million trainable parameters per discriminator, totaling approximately 124 million trainable parameters across two generators and three discriminators. Training for 400 epochs was performed on NVIDIA A6000 GPUs, requiring approximately 120 hours.